\documentclass[journal=chreay,manuscript=review,layout=traditional]{achemso}

\usepackage{mathpazo}
\usepackage[T1]{fontenc}
\usepackage{amssymb}
\usepackage[fleqn]{amsmath}
\usepackage{graphicx}
\usepackage{titlesec}
\usepackage{fncychap}
\usepackage[titletoc,title]{appendix}
\usepackage{bm}
\usepackage{color}
\usepackage{dsfont}

\usepackage{hyperref}
\hypersetup{
    colorlinks=true,
    linkcolor=black,
    filecolor=magenta,      
    urlcolor=cyan,
    }

\usepackage{xcolor}

\author{J\'er\'emy R. Rouxel}
\email{jeremy.rouxel@epfl.ch}
\affiliation{Univ Lyon, UJM-Saint-Etienne, CNRS, IOGS, Laboratoire Hubert Curien UMR 5516, Saint-Etienne F-42023, France}

\author{Shaul Mukamel}
\email{smukamel@uci.edu}
\affiliation{Department of Chemistry and Physics \& Astronomy, University of California, Irvine, California 92697-2025, USA}

\title{Molecular chirality and its monitoring by ultrafast X-ray pulses}

\date{\today}

\begin{document}

\begin{abstract}
Major advances in X-ray sources including the development of circularly polarized and orbital angular momentum pulses make it possible to probe matter chirality at unprecedented energy regimes and with {\AA}ngstr{\"o}m and femtosecond spatiotemporal resolutions.
We survey the theory of stationary and time-resolved nonlinear chiral measurements that can be carried out in the X-ray regime using tabletop X-ray sources or large scale (XFEL, synchrotron) facilities. A variety of possible signals and their information content are surveyed.
\end{abstract}

\maketitle

\textbf{
This document is the Accepted Manuscript version of a Published Work that appeared in final form in Chemical Reviews, copyright \copyright 2022 American Chemical Society after peer review and technical editing by the publisher. To access the final edited and published work see \url{https://doi.org/10.1021/acs.chemrev.2c00115}.
}

\tableofcontents

\section{Introduction}

Newly-developed X-ray sources such as free electron lasers (FEL) or High Harmonic Generation (HHG) offer most valuable insights on molecular structures and processes, with unprecedented spatiotemporal resolutions and atomistic sensitivity of molecular dynamical events.
Significant advances have been made in the control of the polarization and the spatial structure of X-ray beams which are required to probe chirality.
In the optical regime, such capabilities have allowed to probe chiral molecules efficiently. 
X-ray chiral techniques are now developing in a similar fashion.
This review surveys chiral techniques and their simulations with emphasis on the X-ray regime.

Chiral molecules are defined by their lack of mirror symmetry, and are ubiquitous and of high relevance to chemical and biological function\cite{hicks2002chirality} and to drug design.
Two mirror-symmetric molecules are known as enantiomers and possess some distinct physical and chemical properties.
The geometrical operation linking opposite enantiomers is an inversion, also known as a parity transformation.
The toxicity of enantiomers also differs.
For example, R-methadone has analgesic and respiratory effects while the S-methadone has no such effects\cite{kasprzyk2010pharmacologically,lin2009effects}. 
D-cocaine has higher activity and faster toxicokinetics than L-cocaine\cite{smith2009chiral}.
(-)-(S)-thalidomide has multiple positive effects but prescription of the racemate in the late 50s led to a pharmaceutical disaster by causing fetal malformations\cite{eriksson1995stereospecific}.
Historically, early studies were made by Pasteur in 1857 who had observed the selective fermentation by bacteria of D-tartaric acid only\cite{pasteur1858memoire}.
As can be seen from these selected examples, the determination of chirality of a compound is of vital importance for the pharmaceutical, chemical and material industries.

Enantiomeric pairs interact with light in a different manner. 
Molecular chirality was first discovered through the rotation of the light polarization induced by a chiral sample due to the different speed of left and right circularly polarized light\cite{cotton1896recherches,pasteur1848relations}, a technique now known as Optical Rotatory Dispersion (ORD).
This effect can lead to the use of chiral molecules, crystals or nanostructures to design optical devices that control the electromagnetic field polarization\cite{yu2012broadband,cui2014giant,wang2016optical,
schulz2017organic,dhbaibi2018exciton}.
Various light-matter interaction processes may lead to enantiomer-specific spectroscopic signals, thereby detecting e.g. enantiomeric excess in a solution.
A chiral signal has an opposite sign for the two enantiomers\cite{berova2000circular,barron2004molecular,berova2011comprehensive} due to the extra minus sign introduced by the parity inversion, and thus vanishes in a racemic mixture.
Chiral signals are directly linked to the molecular geometry, and thus provide valuable structural information.
Such signals can be obtained by, e.g., taking the difference of two spectroscopic observables for different polarizations. 
For example,  the circular dichroism (CD) technique measures the difference in the absorption of left and right circularly polarized light.
This technique usually involves the cancellation of contributions within the electric dipole approximation, and the observation of a pseudo-tensor quantity that becomes the leading term in the matter response.
Pseudo-tensors are quantities that do not change sign under a parity operation when their rank is odd, and reverse their sign when it is even.
A pseudo-tensor is crucial in chiral signals since it ensures that a parity operation (equivalent to a exchange of two enantiomers or of left and right polarization) provides the desired cancellation.
Unfortunately, these pseudo-tensors usually require higher order multipoles (magnetic dipoles and electric quadrupole) in the interaction. 
The interaction multipoles are the terms of a converging multipolar expansion of the molecule charge or polarization densities\cite{silver2013irreducible}.
Their contributions to the signals are much weaker than their non-chiral counterparts by a factor $a/\lambda$, typically $10^{-2}$ to $10^{-3}$ where $a$ is a molecular size and $\lambda$ is the wavelength of the incident light.

In some cases, chiral signals can be obtained even in the electric dipole approximation when considering unbounded states or nonlinear optical interactions \cite{ordonez2018generalized} or when using the minimal coupling interaction (avoiding the multipolar expansion altogether). Such signals are stronger since they do not contain the $a/\lambda$ factor and can lead to highly sensitive chiral discrimination, making them potentially interesting for various applications.

Molecular chirality offers a window onto fundamental questions such as the origin of biological homochirality
\cite{barron2008chirality,fujii2004homochirality,podlech1999new,
quack2002important}. 
L- amino-acids and D- sugars are exclusively present in biological molecules and it is believed that life could not exist with heterochiral systems\cite{cline2005physical,bailey1998circular}.
Competing theories attempt to explain the link between chirality and life.
First, parity violation exists at the fundamental level due to the electroweak interaction, resulting in a ground state energy difference between opposite enantiomers \cite{pelloni2013parity}. 
However, this parity-violating energy difference is extremely small ($\approx 10^{-19}$ eV) and has not been measured experimentally so far.
Some have proposed\cite{salam1991role,globus2020chiral} that, despite being very small, this energy difference could favour one enantiomer over another over extremely long timescales.
Others consider this small energy difference as insignificant and promote a statistical approach implying a spontaneous symmetry breaking\cite{sandars2003toy}.
Detecting extra terrestrial life could resolve this issue.

Detecting chirality with a high degree of accuracy and sensitivity is of interest for a broad range of applications. The main techniques routinely employed in the IR and visible regimes are circular dichroism (CD), optical activity (OA), optical rotatory dispersion (ORD) and Raman optical activity (ROA). 
In this review, we present extensions of these techniques as well as new possible X-ray techniques.

The first chiral X-ray magnetic circular dichroism (XMCD) measurements in crystals were carried out with synchrotron sources. 
In this technique, the difference in X-ray absorption spectra (XAS) of left and right incoming polarization is taken in the presence of a strong magnetic field\cite{stohr1999exploring,chen1990soft}.
This technique provides valuable information on atomic spins.
However, the parity breaking is induced by an external field and is not linked to an intrinsic molecular chirality which is the subject of this review.
With recent developments of X-ray sources and their polarization control, natural X-ray chirality can now be probed at large scale facilities such as synchrotrons and X-ray free-electron lasers (XFEL) as well as with tabletop sources (HHG).

X-rays offer a particularly useful probe of molecular chirality thanks their short wavelength. 
In essence, chirality is a molecular structural property and the atomic resolution of X-rays gives better structural information than possible in the optical regime. 
The chirality is typically localized around a chiral center, often a carbon atom. 
Alternatively, it may be associated with global structures such as helices\cite{hicks2002chirality}. 
By tuning the X-ray wavelength to be resonant with a core transition, one can get local insight on the chirality and map it across the molecule instead of merely considering it a global property.
It is also possible to probe the chirality of the surrounding atoms in order to measure the delocalization of chirality.
Already in the frequency domain, X-ray chiral signals offer numerous new insights but their capabilities are even more striking in the time domain.
Photoexcitation of chiral molecules can trigger ultrafast charge migration and nuclear dynamics in a few femtoseconds, and X-rays can resolve them while combining while adding its usual advantages in structure and element sensitivities.
Understanding these dynamics and controlling the photoexcitation of chiral molecules have great promise for photoinduced chiral purification.

Frequency-domain signals which probe molecular chirality in the ground state have been implemented in lower frequency (optical and infrared) regimes.
Time-resolved extensions of all these signals are straightforward by starting with a system in a non-stationary state and monitoring the chiral nuclear and electronic dynamics.

\begin{figure*}[!h]
  \centering
  \includegraphics[width=0.6\textwidth]{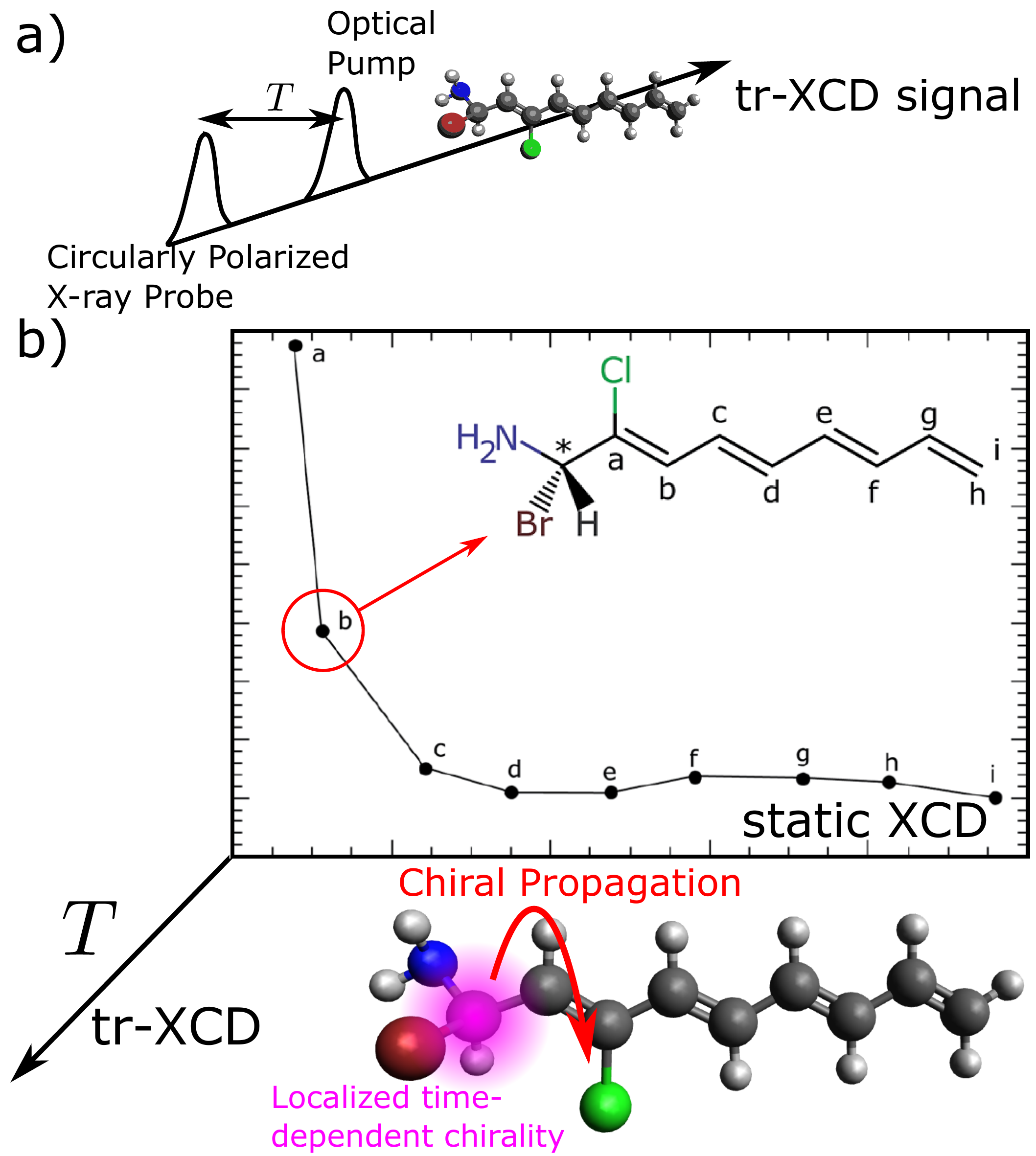}
  \caption{ a) Scheme for a tr-CD signal: a chiral dynamics is trigger by an actinic pump and subsequently probed by a chiral sensitive technique, here CD\cite{zhang2017x}. b). X-rays have been shown to probe the spatial extension of the molecular chiral density. Going into the time-domain will offer a window of how this quantity will propagate spatially within a molecule.
\label{fig:intro1}}
\end{figure*}

Chiral dynamics can provide most valuable information on a large variety of systems.
First, photo-triggered chiral dynamics in molecules that chiral in their ground states can be resolved.
Compared to time-resolved (tr) signals that are sensitive to the overall dynamics, core-resonant chiral dynamics provides structural information at the vicinity of the chiral center.
Thanks to the element selectivity of core transitions, one get two-point molecular information by exciting it at some position and detecting at another.
For example, in Fig. \ref{fig:intro1}, a chiral dynamics is triggered in the linear molecule, 1-bromo-1-amino-n-chloro-nonene. 
It has been shown that an X-ray chromophore, here chlorine, offers a sensitive window into the delocalization of chirality\cite{zhang2017x}, see section \ref{section:xcd}. 
By going into the time domain, it should be possible to probe how fast does a chiral dynamics is propagate to different locations within a molecule.
Second, achiral molecules can acquire chirality through photoexcited nuclear dynamics in the excited states. 
This is the case for example in formamide that is planar and achiral in its electronic ground state and becomes chiral in the excited state by out-of-plane bending\cite{rouxel2017photoinduced}.
Finally, chirality can be induced in an achiral molecule through interaction with another chiral molecule\cite{allenmark2003induced}.
This is relevant to biological systems where chiral molecules present in the protein structure can interact with a reaction center through dipole-dipole coupling.
The photosynthetic reaction center, that often contains X-ray chromophores, e.g. metallo-porphyrins, can acquire an optical activity through the presence of nearby chiral molecules.
The induced molecular chirality can thus measure of the time-dependent coupling between the chiral molecule and the probed achiral one.

In this review, we present expressions for time-resolved (tr-) techniques such as tr-XCD (X-ray Circular Dichroism), tr-XROA (X-ray Raman Optical Activity), tr-PECD (Photoelectron Circular Dichroism), tr-HD (Helical Dichroism).
To keep the presentation general, we do not specify the preparation process and simply assume a non-stationary initial molecular wavefuntion.
A standard pump-probe approach can be used where an actinic pump pulse triggers a desired dynamics that is subsequently probed after a delay using the chiral detection mode\cite{beaulieu2018photoexcitation}.
Alternatively, a stimulated Raman pump can induce a broadband ultrafast  excitation\cite{biggs2012two}. 
More elaborate multipulse preparation schemes can be envisioned as well.

Chiral HHG (cHHG) is a recent exciting development in time-resolved chiral X-ray experiments\cite{cireasa2015probing,kfir2015generation}. 
An intense mid-IR field\cite{1508.02890} is used to ionize a molecule. 
The released electron is then accelerated by the intense laser field until it recombines with the same molecule, emitting soft X-ray HHG light (up to 500 eV) in the process. Enantiomers were found to produce a different HHG spectrum depending on the incoming laser ellipticity\cite{cireasa2015probing}.
A precise control of the emitted light polarization in free electron laser (FEL) sources has been achieved through undulators\cite{allaria2014control}. 
In FEL, an electron bunch is propagated through a magnetic periodic structrure, the undulator, giving rise to an electromagnetic emission coupled to the bunch\cite{FEL1}. 
Fine details of the electron trajectory are imprinted in the generated light : a sinusoidal trajectory results in linearly polarized light, while an helicoidal trajectory produces a circularly polarized signal\cite{ferrari2015single,gluskin1991xuv}. 
The total control of the emitted polarization of FEL has been achieved recently\cite{ferrari2015single,allaria2014control}.
Synchrotrons constitute a third common source of X-ray light that have long produced circular polarized light \cite{melrose1971degree} for continuous or nanosecond samples.
Circularly polarized X-rays can be obtained by collecting them off-axis above and below the orbit plane. 
This source provides imperfect polarization states at a rather low flux.
Insertion devices are then used to amplify the beam\cite{dartyge1992essential}.
Each of these sources have their own merits. 
HHG sources offer the shortest (attosecond) pulse duration and are tabletop but are limited to soft X-rays and weak fluxes. 
Recent developments have reached the carbon K-edge at 284 eV\cite{li201753}. Synchrotrons are very stable sources over a broad frequency range (10 eV to 120 keV) and are widely available worldwide. 
The most common (multi-bunch) operation mode provides a continuous light source that can be used for frequency domain techniques. 
A single bunch can be singled out for time-resolved experiments with a resolution down to tens of picoseconds\cite{holldack2014single}.
X-ray FEL are less common but produce ultrashort high brilliance beams (few femtoseconds or even attosecond \cite{duris2020tunable}) particularly suitable for ultrafast time-resolved nonlinear experiments.
The improvement and development of dedicated beamlines with accurate polarization control of X-rays at synchrotrons\cite{yamamoto2014new, hussain2012circular, nahon2012desirs, ohresser2014deimos, zhi2015new} and FELs\cite{hartmann2016circular, roussel2017polarization, kubota2019polarization, schneidmiller2013obtaining} are making chiral techniques increasingly available to a broader range of experiments.
We further note that many beamlines built for XMCD at synchrotrons can be used for XCD measurements.

Two X-ray light characteristics make them particularly suitable for chirality measurements. 
First, due to its high photon energy, X-ray light can easily ionize the sample, and the emitted photoelectrons can be detected leading to X-ray photoelectron spectroscopy (XPS). 
Its chiral extension, PECD, is obtained by measuring the difference in XPS with left and right polarizations\cite{goetz2019quantum, goetz2019perfect, kastner2016enantiomeric}.
Second, the X-ray wavelength can be comparable to inter-atomic distance, leading to an extremely small Abbe diffraction limit.
Fields with varying spatial profile across the molecular charge density can be produced. 
The multipolar expansion of the light-matter coupling then  completely fails and the light-matter interaction must be described in the minimal coupling Hamiltonian that implicitly incorporates all multipoles. 
Such signals can further exploit the orbital angular momentum of light\cite{rouxel2016non}, potentially leading to a new family of chiral techniques.

The above survey of chiral signals suggests a natural classification scheme of signals according to the interaction Hamiltonian that should be used in their description: electric dipole, multipolar, or minimal coupling.
This classification will be adopted in this review. Multipolar coupling signals such as XCD, XROA and four wave mixing (4WM) signals are presented in Section \ref{multipolesPART}. 
These are direct extensions of the corresponding optical techniques but take advantage of several unique properties of X-ray pulses namely element specificity and localization of the interaction, ultrashort (down to the attosecond) pulses and large bandwidths. 
Signals that exist in the electric dipole approximation are discussed in Section \ref{electricPART}. 
These include signals involving complex electric dipoles, i.e. bound to unbound transitions, such as PECD as well as ones derived with even order susceptibilities such as sum frequency generation $\chi^{(2)}$ (SFG).
In Section \ref{minimalPART}, we discuss signals that require the full spatial variation of the incoming field. Helical dichroism (HD), the difference in absorption of Laguerre-Gauss beams with a specific transverse profile and opposite orbital momenta, is an example.

Ultrashort X-ray sources can look at time-resolved chiral signals in which an excited state is first prepared and is then probed through a chiral observable such as CD or ROA.
Time-resolved near UV CD (tr-CD) in a pump-probe set up has been reported recently\cite{dartigalongue2005observation}.
We have shown that tr-XCD involving a visible pump - X-ray probe can provide valuable information\cite{rouxel2017photoinduced}.
In a simplified picture, the molecular potential energy surface (PES) can be described by a double-well potential, where the minima correspond to the two enantiomers.
Commonly, the enantiomers are stable in their ground state at room temperature; the energy barrier between them is high and prevents interconversion dynamics.
This barrier can be small in the excited state and an asymmetric wavepacket in the excited PES can propagate back and forth between the two minima, giving rise to a time-evolving chiral signal.

In summary, X-rays offer many advantages to chiral techniques. Their use as a local probe of core transitions or the {\AA}ngstr{\"o}m-resolved diffraction provide valuable structural insights.
Additionally, the magnitude of chiral X-ray signals is favorable and can enhance intrinsically weak signals.
Finally, X-rays and EUV are at the forefront of extreme ultrafast measurements by reaching the attosecond regime.

\section{General considerations regarding X-ray chiral measurements}

\label{appendixGenCons}

\subsection{Pseudo-tensors, polarization vectors and rotational averaging}

We first introduce the notation necessary to describe chiral spectroscopic observables that will be used throughout this review.
We start with the parity operator $\mathcal P$ which reverses the sign of all coordinates. 
It spawns the parity group that has two irreducible representations, with characters $+1$ and $-1$.
Under a coordinate inversion, a true vector $\bm \mu$ becomes $\mathcal P \bm \mu = - \bm\mu$. A pseudo-vector $\bm m$, on the other hand, does not change sign, i.e. $\mathcal P \bm m = \bm m$. Examples of true vectors and tensors are the electric field $\bm E$ or vector potential $\bm A$, the electric dipole moment $\bm\mu$ and the electric quadrupole $\bm q$. 
The magnetic field $\bm B$ and the magnetic dipole moment $\bm m$ are examples of pseudo-vectors.
An arbitrary tensor, such as nonlinear response functions, can be decomposed into irreducible parts, that are symmetric and antisymmetric upon the action of the parity operator.
A rank $n$ tensor, also called true tensor, changes sign upon inversion according to $\mathcal P T^{(n)} = (-1)^n T^{(n)}$ while a rank $n$ pseudotensor changes sign as $\mathcal P T^{(n)} = (-1)^{n+1} T^{(n)}$.

Chiral signals originate from the lack of parity. Pseudo-tensorial response functions represent such quantities and spectroscopic observables sensitive to chirality are expressed from the antisymmetric part of response tensors.
One straightforward manner to do that is to cancel out the symmetric contribution by taking combinations of the same observable with different polarization configurations.
In particular, this can be done with circular polarizations given by
\begin{equation}
\label{basisdef}
\bm e_L = -\frac{1}{\sqrt 2} 
\begin{pmatrix}
1 \\ 
i \\ 
0
\end{pmatrix} 
\ \ \ \ \ \ \ \ \ \
\bm e_R = \frac{1}{\sqrt 2} 
\begin{pmatrix}
1 \\ 
-i \\ 
0
\end{pmatrix} 
\end{equation}
\noindent where $\bm e_L$ and $\bm e_R$ are the left and right-handed polarization vectors for a plane wave propagating along $z$.
The vectors $(\bm e_L, \bm e_R, \bm e_z)$ form an orthonormal basis with the following properties:
\begin{eqnarray}
&&\bm e_L^* = - \bm e_R \hspace{2.4cm} \bm e_R^* = - \bm e_L\\
&&\bm e_L\cdot \bm e_L = \bm e_R\cdot \bm e_R = 0 \hspace{0.5cm} \bm e_L\cdot \bm e_R = -1\label{basisdotprod}\\
&&\bm e_z \times \bm e_L = - i \bm e_L \hspace{1.4cm} \bm e_z \times \bm e_R =  i \bm e_R\label{basiscrossprod}
\end{eqnarray}
Eq. \ref{basisdotprod} shows that the basis obtained using circular polarization, known as the irreducible basis, does not have a cartesian metric but instead an antidiagonal one with components $(-1,1,-1)$ along the antidiagonal. One must then be careful when performing tensor contractions with this basis.
Eq. \ref{basiscrossprod} allows to get the polarization vectors $\bm b$ of the magnetic field 
\begin{equation}
\bm b_{L/R} = \frac{1}{c} \bm e_z \times \bm e_{L/R}
\end{equation}

Many chiral measurements rely on the cancellation of non-chiral contributions when the difference of the observable between left and right polarization is taken.
Thus, the following formulas will be routinely used:
\begin{eqnarray}
\bm e_L^a \bm e_L^{b*} - \bm e_R^a \bm e_R^{b*} &=& - i \epsilon^{ab}_c \hat k^c\label{eq:sumpolar}\\
\bm e_L^a \bm e_L^{b*} + \bm e_R^a \bm e_R^{b*} &=& \delta_{ab} - \hat k^a\hat k^b
\end{eqnarray}
\noindent where $\epsilon^{ab}_c$ and $\delta_{ab}$ are the Levi-Civita and Kronecker symbols.
This relationship is true for left and right plane wave propagating along a wavevector $\bm k$. $\hat k$ is the unit vector along $\bm k$. In the basis defined in Eq. \ref{basisdef}, we have $\hat k = \bm e_z$.

Finally, rotational averaging will be repeatedly carried out later on. Chiral signals are only defined for  rotationally averaged ensembles. This is because, in many cases, signals such as circular dichroism do not vanish in oriented achiral samples and are thus not chirality-specific in these cases.
Rotational averaging of a rank $N$ tensor $T$, $\langle T \rangle_\Omega$, is achieved by rotating the tensor $T$ using rotation matrices and integrating over all solid angle:
\begin{multline}
\langle T^{i_1...i_N}\rangle_\Omega = \int d\Omega R^{i_1}_{\lambda_1}(\Omega)...R^{i_N}_{\lambda_N}(\Omega) T^{\lambda_1...\lambda_N}
= (I^{(N)})_{\lambda_1...\lambda_N}^{i_1...i_N} T^{\lambda_1...\lambda_N}
\end{multline}
\noindent where the tensor $T^{\lambda_1...\lambda_N}$ is in the molecular frame and $\langle T^{i_1...i_N}\rangle_\Omega$ is in the laboratory frame. The averaging tensors $I^{(N)}$ are given in Appendix \ref{appendixROTAV}.

\subsection{Coupling of light to chiral matter}

Signals described by three radiation/matter coupling Hamiltonians will be considered in this review. We first introduce the electric dipole coupling Hamiltonian.
\begin{equation}
H_\text{int}(t) = - \bm \mu\cdot \bold E(t)
\label{hintEdip}
\end{equation}
\noindent where $\bm \mu$ is the electric dipole and $\bold E$ is the incoming electric field. This interaction does not generate pseudo-tensorial quantities for even order susceptibilities. However, if the electric dipole is complex, it is possible to obtain chiral signals\cite{ordonez2018generalized} as discussed in section \ref{electricPART}.

The second interaction is given by the multipolar Hamiltonian\cite{salam2009molecular,craig1998molecular} truncated at the electric quadrupole
\begin{equation}
H_\text{int}(t) = -\bm \mu\cdot \bold E(t)-\bm m\cdot \bold B(t)-\bm q\cdot \nabla\bold E(t)
\label{hintMULTI}
\end{equation}
\noindent where $\bm m$ and $\bm q$ are the magnetic dipole and the electric quadrupole respectively and $\bold B$ is the magnetic field.
Since $\bold m$ is a pseudo-vector and $\bold q$ is a second-rank tensor, this coupling can generate chiral observables.

Finally, the last, exact, interaction Hamiltonian discussed in section \ref{minimalPART} is
\begin{equation}
H_{\text{int}}(t) = - \int d\bold r \ \bold j(\bold r) \cdot \bold A(\bold r,t) + \frac{e}{2m} \int d\bold r \ \sigma (\bold r) \bold A^2(\bold r,t)
\label{hintminimal}
\end{equation}
where $\bold j(\bold r)$ and $\sigma(\bold r)$ are the current and charge density operators and $\bold A$ is the vector potential. $e$ and $m$ are the electron electric charge and mass.
This coupling is derived by the standard minimal coupling Hamiltonian $\sum_\alpha(\bold p_\alpha - q_\alpha \bold A(r_\alpha,t))^2/2m_\alpha$ where the sum is over elementary charges. Expanding the square and taking the continuous limit for the sum leads to Eq. \ref{hintminimal}. 
This description is convenient to describe extended structures. 
Standard electronic structure codes can provide the matrix elements of the charge and the current density operators
\begin{eqnarray}
\sigma(\bold r) &=& e \psi^\dagger(\bold r)\psi(\bold r)\\
\bold j(\bold r) &=& \frac{e}{2m} \big(|\bm r\rangle\langle \bm r| \bm p + \bm p^\dagger |\bm r\rangle\langle \bm r| \big)\nonumber\\
		 &=& \frac{e \hbar}{2mi}\big(\psi^\dagger(\bold r) \bold\nabla \psi(\bold r) - (\bold\nabla\psi(\bold r)^\dagger)\psi(\bold r)\big)
		 \label{eq:j}
\end{eqnarray}	
where  $\psi^\dagger(\bold r)$ and $\psi(\bold r)$ are the electron field fermion creation and annihilation operators at position $\bold r$ and  $\bm p$ is the momentum operator.
The $\sigma \bold A^2$ interaction can not induce chiral signals by varying the polarization since the matter quantity involved is a scalar, insensitive to the field polarization. However, the use of spatially varying fields with orbital momentum is still possible.
The current density is a vector field that need not have a well defined parity, i.e. it can have both odd and even contributions under a parity inversion. 
Chiral signals can thus arise by mixing these two contributions.

The three coupling Hamiltonian discussed in this section (electric dipole coupling, multipolar coupling, minimal coupling) naturally construct a way to classify chiral sensitive signals as summarized in Table \ref{tab:summary}.
Interestingly, this classification is not only formal but also encompass different experimental aspect.
The multipolar coupling signals rely usually on dichroic measurement in which the strong achiral background gets canceled out.
The electric dipole coupling signals use either nonlinear interactions in the incoming field or transition toward a continuum (photoionization).
Finally, the minimal coupling Hamiltonian is well suited for techniques using beams with large spatial variations.

\begin{table}[h!]
\begin{tabular}{ |p{3cm}||p{3cm}|p{3cm}|p{3cm}|  }
 \hline
  & Multipolar & Electric Dipole & Minimal Coupling\\
 \hline
 Static   		& XCD    	& SFG/DFG	&   HD\\
 				& XROA		& cHHG   	&	HROA\\
 		    		&  			& PECD	  	&   \\\hline
 Time-resolved	& tr-XCD  	& tr-PECD	& 	tr-HD\\
  				& tr-ROA  	& tr-cHHG   	& 	tr-HROA\\
 \hline
\end{tabular}
\caption{\label{tab:summary}
Chiral-sensitive signals discussed in this review. XCD: x-ray Circular Dichroism. XROA: x-ray Raman Optical Activity. PECD: Photoelectron Circular Dichroism. cHHG: chiral High Harmonic Generation. HD: Helical Dichroism. HROA: Helical Raman Optical Activity. tr: time-resolved.}
\end{table}

\section{Chiral signals involving odd-parity multipoles}
\label{multipolesPART}

We first discuss chiral signals that are expressed in terms of odd-parity multipoles, e.g. the magnetic dipole and the electric quadrupole at the lowest order.
These multipoles are usually not the first non-vanishing order in the multipolar expansion of the interaction Hamiltonian and the relevant signals are small chiral corrections to a strong achiral background.
These chiral contributions to the signal can be isolated by cancelling out the unwanted achiral terms through multiple measurements with opposite circular polarization configurations.
Despite their weakness, these signals have been most used experimentally.
This section presents first the frequency-domain signals and selected experimental and simulation results in the X-ray regime.

\subsection{X-ray circular dichroism}
\label{section:xcd}
\begin{figure*}[!h]
  \centering
  \includegraphics[width=0.6\textwidth]{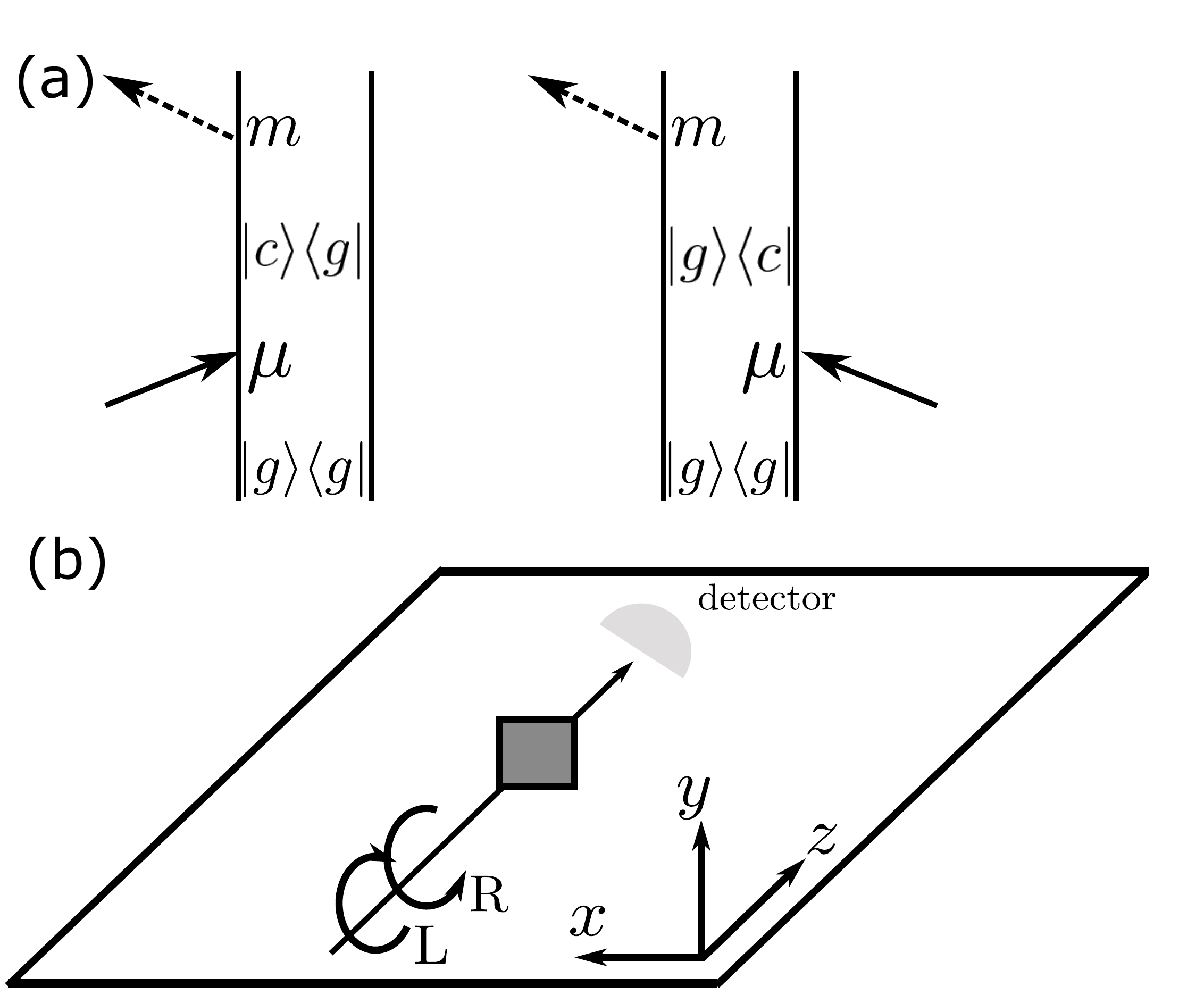}
  \caption{(a) Ladder diagrams of the CD signal (for diagram rules see appendix \ref{appendix:perturbation}). Permutation of the electric dipole $\bm \mu$ and magnetic dipole $\bm m$ must be included. (b) Geometry of a CD measurement.
\label{cd1}}
\end{figure*}

In this section, we consider signals described by the multipolar Hamiltonian, Eq. \ref{hintMULTI}.
We start with the simplest chiral signal, circular dichroism (CD): the absorption difference between left and right polarized incoming light. 
XCD signals can be measured in the frequency or in the time-domain. In the former, circularly polarized plane waves are used to measure absorption and their frequency is scanned, while in the latter, the free induction decay following an impulsive excitation is measured in time and is then Fourier transformed. 
The ladder diagrams contributing to the CD signal and the experimental geometry are given in Figs. \ref{cd1}(a) and (b) respectively.

The signal is usually normalized by the sum of the absorption spectra that has a purely electric dipole character.
\begin{equation}
S_\text{CD}(\omega) = \frac{A_L(\omega)-A_R(\omega)}{\frac{1}{2}(A_L(\omega)+A_R(\omega))}
\label{CDdef}
\end{equation}
\noindent where $A_p$ is the absorption signal of a $p$ polarized light ($p = L$ for left and $R$ for right) and $\omega$ is the incoming field frequency.
The electric quadrupole moment $\bm q$ contribution to the CD signal averages to zero in isotropic ensembles of  randomly oriented molecules since the electric quadrupole is a symmetric tensor and may be neglected\cite{berova2000circular}.
This is in contrast with anisotropic systems, e.g. crystals, whereby rotational averaging is not done and the  signal is dominated by the electric quadrupole - electric dipole response.
In the X-ray regime, the absorption baseline is routinely shifted to zero below the edge and the definition in Eq. \ref{CDdef} leads to a vanishing denominator and large uncertainties.
To avoid divergences, we thus use an alternate definition:
\begin{equation}
S_\text{CD}(\omega) = \frac{A_L(\omega)-A_R(\omega)}{\frac{1}{2}(\text{max}A_L(\omega)+\text{max}A_R(\omega))}
\label{CDdef2}
\end{equation}
\noindent where the $\text{max}A$ indicates the value of the spectrum at the main edge. This definition tends to underestimate the signal asymmetry ratio around the edge but prevents a divergence when the absorption signal is very small.

\begin{figure*}[!h]
  \centering
  \includegraphics[width=0.5\textwidth]{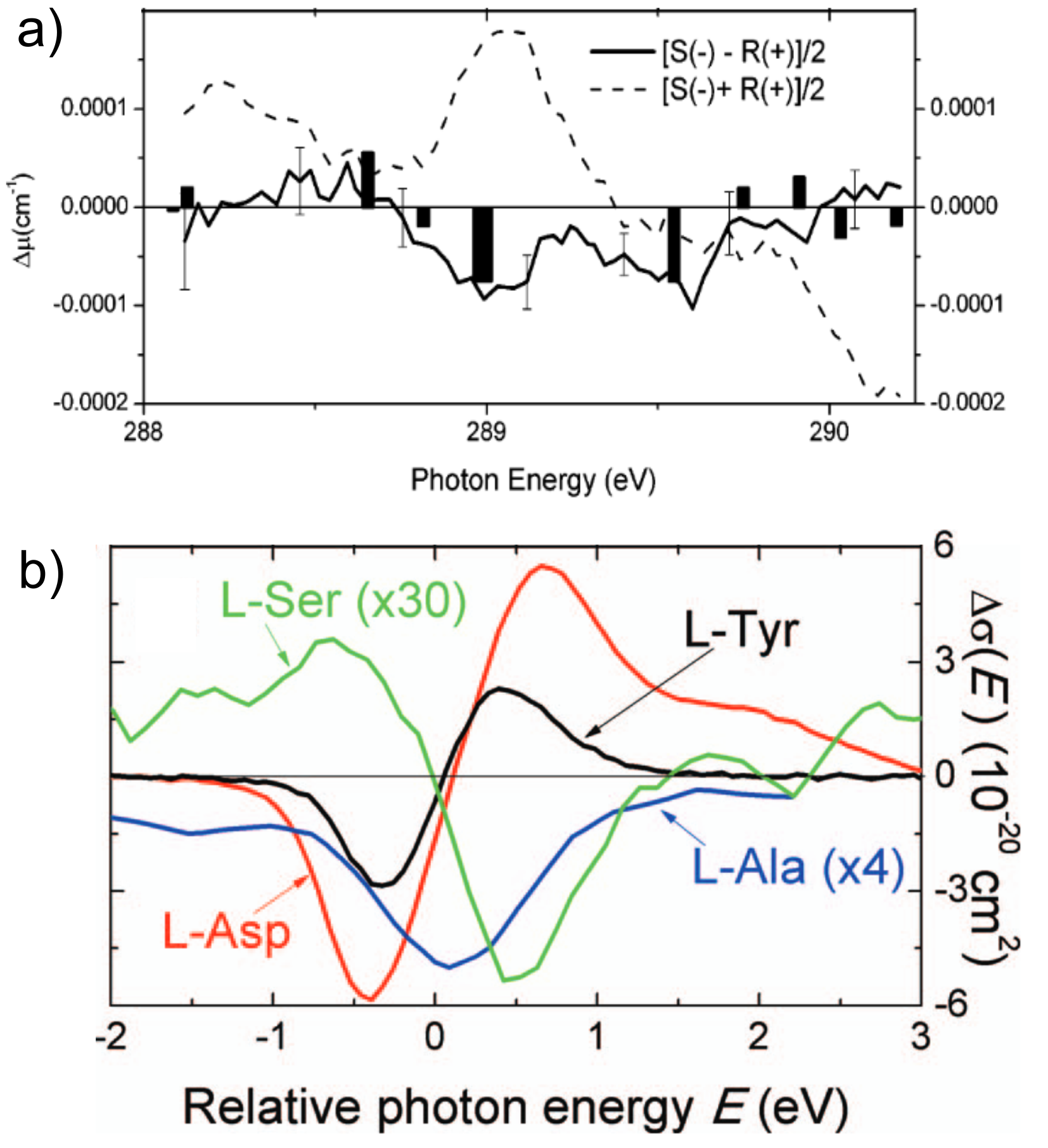}
  \caption{
a) CD spectrum at the C K-edge of S-(-)-methyloxirane measured by Turchini et al\cite{turchini2004core}. 
Reprinted with permission from JACS, 2004. Copyright 2004 American Chemical Society.
b) CD spectra at the O K-edge of L-Tyrosine (black), L-Aspartic acid (red), L-Serine (green), and L-Alanine (blue) films measured by Izumi et al\cite{izumi2013characteristic}. Reprinted from J. Chem. Phys., 2013. Copyright 2013 AIP Publishing.
\label{cd2}}
\end{figure*}

XCD in solids, also known as X-ray Natural Circular Dichroism (XNCD) as opposed to XMCD, has been first measured in 1998 by Alagna et al. on single crystals at the Nd L$_3$-edge\cite{alagna1998x}.
XCD has also been measured in molecular crystal of cobalt and neodymium  complex\cite{stewart1999circular,peacock2001natural} at their Co K- and Nd L-edges respectively.
Theoretical works have highlighted the importance of the electric quadrupole - electric dipole coupling for XCD\cite{natoli1998calculation,carra2000x,peacock2001natural}.
In the following, we focus on randomly oriented molecules where the quadrupolar term vanishes.
The first measurements of XCD on a chiral molecule at the carbon K-edge\cite{turchini2004core} were reported on methyloxirane in gaz phase, shown in Fig. \ref{cd2}a.
The experiment made use of the elliptical wiggler at the ELETTRA synchrotron to generate circularly polarized light and the measured asymmetry ratio was $\sim 10^{-3}$.
This molecule contains multiple chiral carbons and it was shown that tuning the photon energy can lead to a selection of the probed chiral carbon.

XCD have also been measured in serine and alanine thin films at the oxygen K-edge \cite{izumi2009measurement,izumi2013characteristic}, shown in Fig. \ref{cd2}b, and at the nitrogen K-edge on histidine\cite{izumi2014nitrogen}.
The O K-edge XCD experiments were conducted using helical undulators at SPring-8 on powder samples deposited on a SiN thin film by sublimation. 
Deposited samples can simplify the experimental setup but requires to cancel out the linear anisotropy component if it is present. Fig.\ref{cd2}b highlights the spectra around 1s $\rightarrow \pi^*$ transition of the oxygen atom in the C-OH group. 
The large variation of spectra in this area shows tha XCD can be very sensitive to small local structural variations.

Multiple XCD signals have been measured on chiral molecules, mostly in a crystallised form\cite{srinivasan2018enantiomeric} or embedded in a film \cite{tanaka2010chiroptical,izumi2013characteristic}. 
XCD measurements on oriented samples is not per se a measure of molecular chirality but are still an interesting probe of electric dipole - electric quadrupole mechanisms.
Such mechanisms also play an important role in other chiral signals such as X-ray Raman Optical Activity.

The absorption signal of a $p$ polarized weak probe is given by\cite{goulon2011x, mukamel1999principles}
\begin{equation}
A_p(\omega) = 2 \omega \Im \bold E^{p*}(\omega)\cdot\langle \bm \mu(\omega)\rangle+\bold B^{p*}(\omega)\cdot\langle\bold m^p(\omega)\rangle
\label{SigFreqDispAbs}
\end{equation}

This signal can be decomposed into $E-E$ (electric dipole - electric dipole) and $M-E$ (magnetic dipole - electric dipole) contributions:
{\small
\begin{eqnarray}
A_p^{E-E}(\omega) &=& - \frac{4}{3\hbar}  |E_0|^2 \text{Im}\sum_c \omega_{cg} \bm e_p^*\cdot \bm e_p \frac{\bm \mu_{gc}\cdot \bm \mu_{cg}}{\omega_{cg}^2 - (\omega-i\Gamma_{cg})^2}\\
A_p^{M-E}(\omega) &=& - \frac{4}{3\hbar} \frac{|E_0|^2}{c} \text{Im}\sum_c \omega_{cg}  \bm b_p^*\cdot \bm e_p \frac{\bm m_{gc}\cdot \bm \mu_{cg}}{\omega_{cg}^2 - (\omega-i\Gamma_{cg})^2}
\end{eqnarray}
}

Since $\bm e_L^*\cdot \bm e_L = \bm e_R^*\cdot \bm e_R = -1$, the $A_L^{E-E} - A_R^{E-E}$ vanishes.
On the other hand, $\bm b_L^*\cdot \bm e_L = -\bm b_R^*\cdot \bm e_R = -i$ and the $M-E$ contributions add up.
Introducing the lineshape function $f_{cg}(\omega)$ given by
\begin{equation}
f_{cg}(\omega) = \frac{\omega_{cg}}{\omega_{cg}^2 - (\omega-i\Gamma_{cg})^2},
\end{equation}
the left and right polarization difference and sum become:
\begin{eqnarray}
A_L(\omega)-A_R(\omega) &=& 2 \frac{4}{3\hbar}\frac{|E_0|^2}{c} \text{Re}\sum_c f_{cg}(\omega) \bm m_{gc}\cdot \bm \mu_{cg}\\
A_L(\omega)+A_R(\omega) &=& 2 \frac{4}{3\hbar}|E_0|^2 \text{Im}\sum_c f_{cg}(\omega) \bm \mu_{gc}\cdot \bm \mu_{cg}
\end{eqnarray}

Finally, the sum-over-states (SOS) expression of the CD signal defined in Eq. \ref{CDdef} reads:
\begin{equation}
S_\text{CD}(\omega) = \frac{2}{c} \frac{\text{Re} \sum_cf_{cg}(\omega) \bm m_{gc}\cdot \bm \mu_{cg}}{\text{Im}\sum_c f_{cg}(\omega) \bm \mu_{gc}\cdot \bm \mu_{cg}}
\label{CDfinal}
\end{equation}

\begin{figure}[!h]
  \centering
  \includegraphics[width=0.5\textwidth]{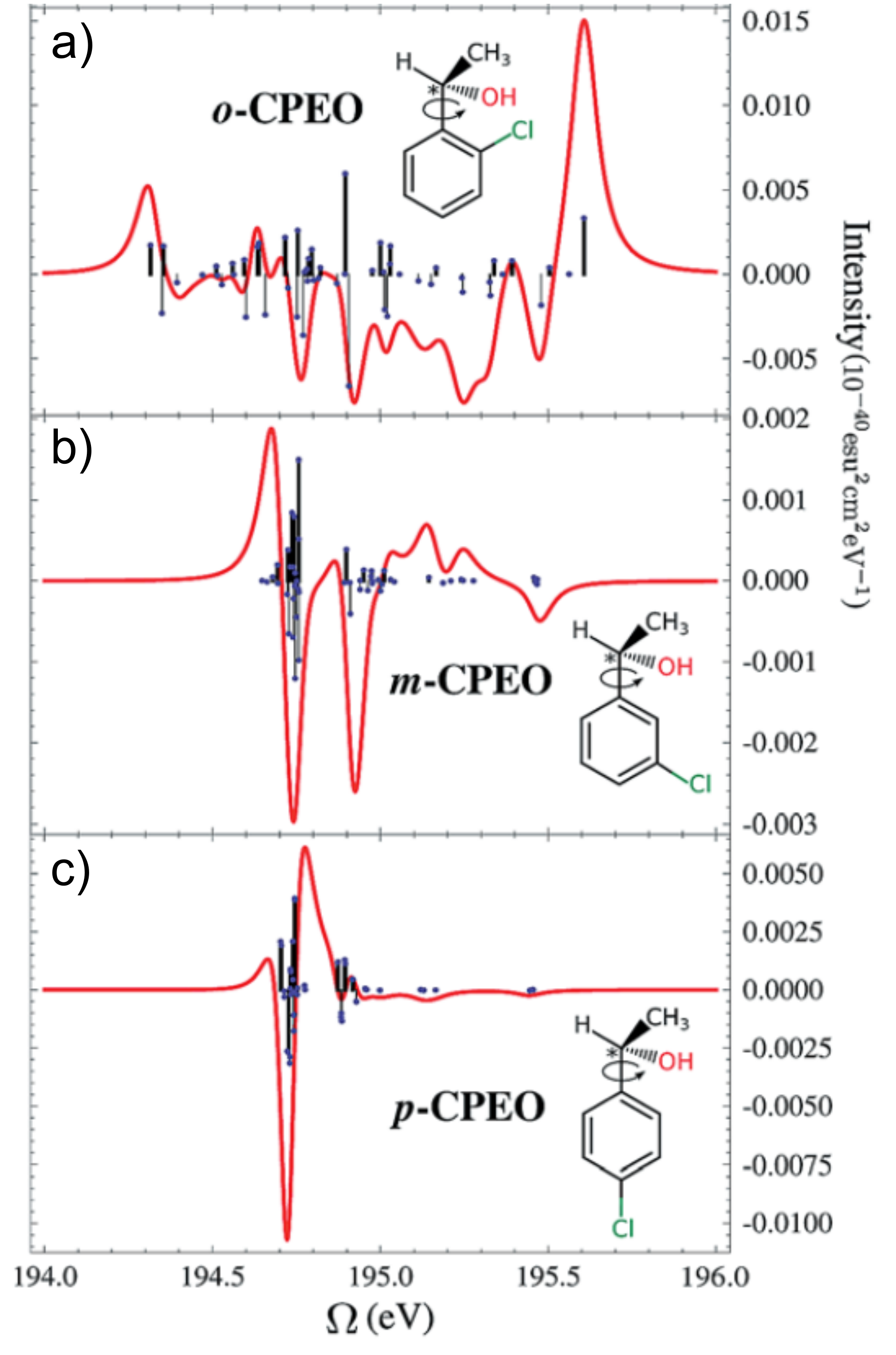}
  \caption{XCD (red) signals at the Cl L$_{2,3}$-edge on substituted $o,m,p$-chlorophenylethanol molecules, shown in panels a, b, and c respectively. The chemical structures are shown in insert with the chiral center marked by an asterisk\cite{zhang2017x}. 
\label{cd3}}
\end{figure}

Eq. \ref{CDfinal} is the standard expression for CD signals\cite{berova2000circular}. It is common to interpret it using the rotatory strength 
\begin{equation}
R_{gc} = \bm m_{gc}\cdot \bm \mu_{cg}
\label{rotatory_strength}
\end{equation}

CD signals have been routinely used in the IR/visible regimes due to their high sensitivity to molecular structure.
In the X-ray regime, one can combine the structural information with the element specificity of the X-ray excitation to probe the molecular chirality at the vicinity of a chosen atomic element within the molecule. 

Fig. \ref{cd3} depicts simulated the X-ray CD spectra of ortho-, meta- and para- chlorophenylethanol (CPEO) molecules at the Cl L$_{2,3}$-edge. The simulations were carried out using the restricted excitation window TDDFT (REW-TDDFT) under the Tamm-Dancoff approximation.
The chlorine atom which serves as the X-ray chromophore is located at the various positions (ortho, meta and para) provides a local probe of molecular chirality.
The CD spectra depicted in Fig. \ref{cd3} show that the maximum chiral signal strength follows the order $o>p>m$ for the different substituted molecules.
The meta substituted molecule has the weakest electronic coupling with the chiral center.
This trend is in agreement with other optical reactivity and conductivity properties\cite{tretiak2002density,hansen2009interfering}.

\begin{figure}[!h]
  \centering
  \includegraphics[width=0.5\textwidth]{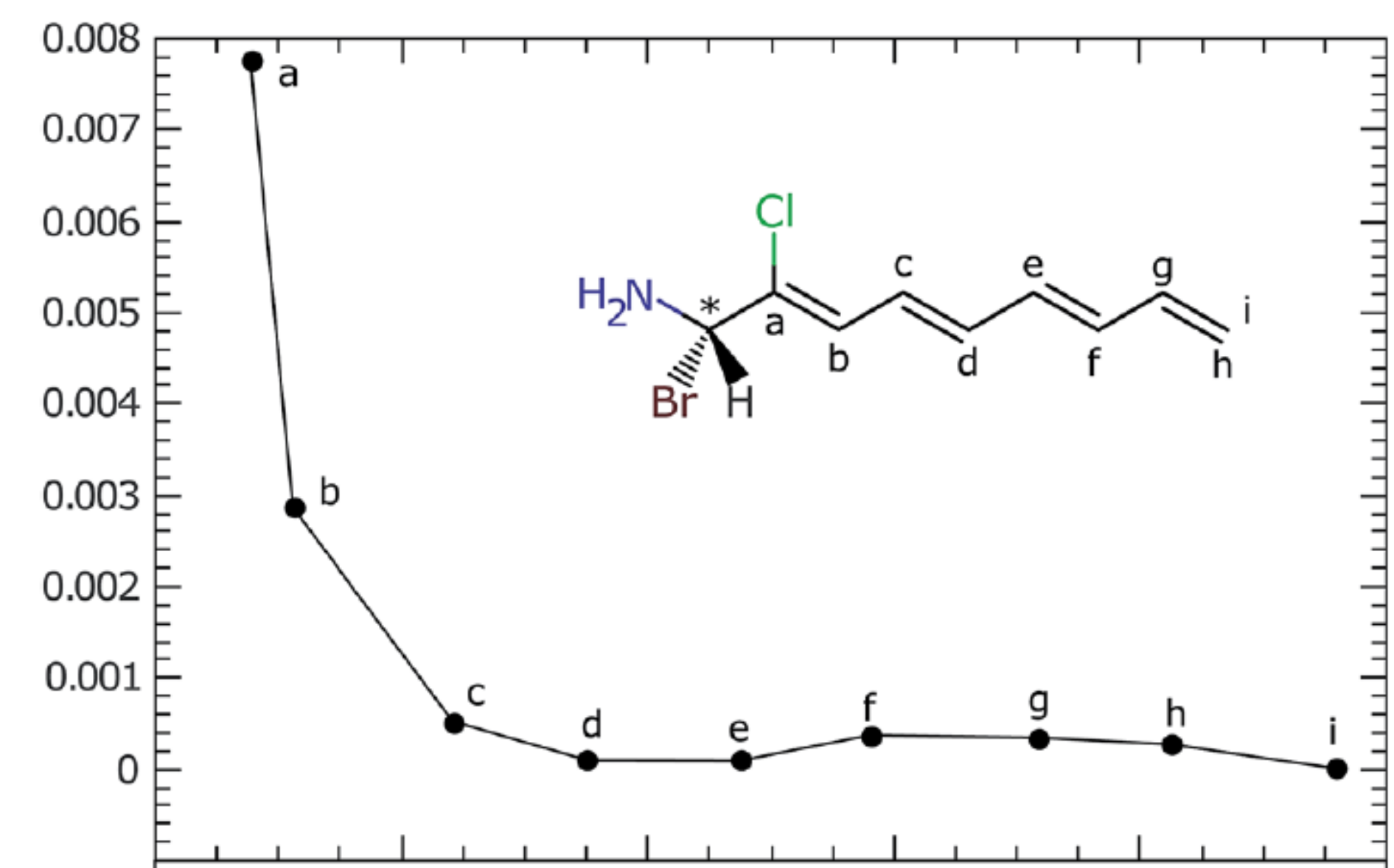}
  \caption{Magnitude of the rotatory strength, Eq. \ref{rotatory_strength}, as a function of the distance between the X-ray chromophore and the chiral center at the
one of the strongest XAS peaks. The amplitude of the rotatory strength decays when the distance between the chiral center and the atom probed by X-rays increases. Reprint from Chem. Sci., 2017\cite{zhang2017x}.
\label{cd4}}
\end{figure}

Fig. \ref{cd4} shows another example of X-ray CD at the Chlorine L$_{2,3}$-edge in 1-bromo-n-chloronona-2,4,6,8-tetraen-1-amine chains.
The rotatory strength at the strongest XAS peak is displayed vs of the X-ray chromophore position.
We see a decrease with the distance between the X-ray chromophore and the chiral center, revealing again the sensitivity of the XCD to the local molecular chirality.

\begin{figure}[!h]
  \centering
  \includegraphics[width=0.5\textwidth]{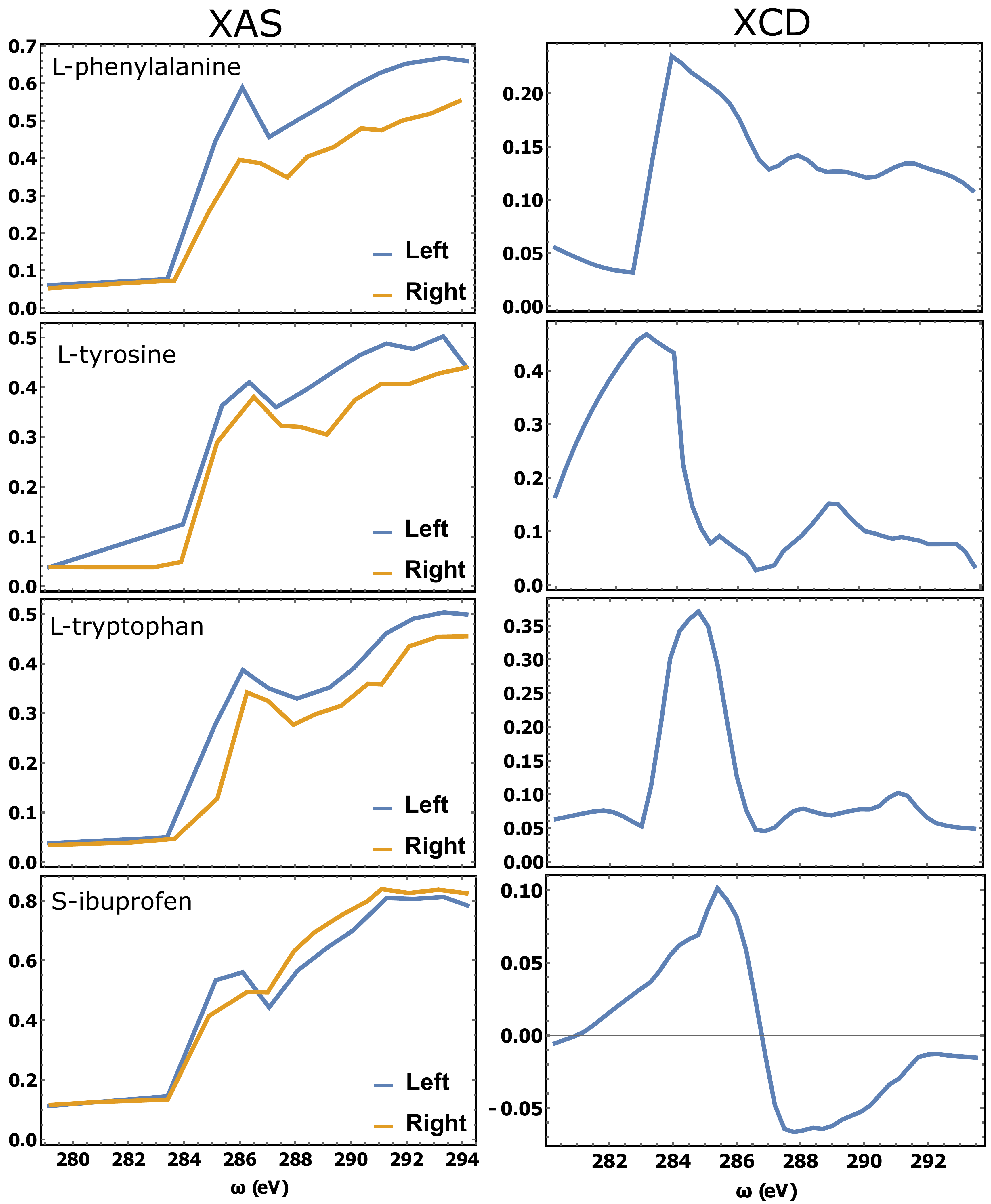}
  \caption{Carbon K-edge XCD spectra of three left-handed amino-acids, L-phenylalanine, L-tryptophan, L-tyrosine, and of S-ibuprofen (unpublished data, courtesy of R. Mincigrucci).
\label{cd5}}
\end{figure}
 
Experimental XCD signals at the C K-edge of three amino-acids (L-phenylalanine, L-tyrosine, L-tryptophan) and S-ibuprofen deposited on a SiN substrate measured at Elettra, Trieste\cite{} are shown in Fig. \ref{cd5}.
The differences of the XCD spectra of the three amino-acids illustrate the sensitivity of XCD to the molecular geometry over a large distance from the chiral center, despite the localized nature of the core orbitals.
Comparison with ab initio calculations of the XCD spectra shows that the contribution from carbon atoms localized at inequivalent sites in the molcules can be spectrally separated in the XANES spectrum.
Thus, probing with a narrowband X-ray pulse allows to select a location within the molecule.

We now turn to the time-resolved extension of XCD, tr-XCD.
The most common setup is to pump a chiral molecule or an achiral molecule that becomes chiral in its excited state with an actinic pulse.
A delayed X-ray pulse then probes the molecular dynamics.
The X-ray probe is finally frequency-dispersed on a detector.

As in the frequency-domain, this signal is measured by taking the difference in the absorption of left and right Circularly Polarized Light (CPL) of a delayed probe with respect to an initial actinic pump.
Eq. \ref{SigFreqDispAbs} can be extended to time-dependent polarization and magnetization:
\begin{eqnarray}
A_{\text{p}}(\omega, T) &=& 2 \omega \Im\Big( \bold E^{p*}(\omega)\cdot \langle\bm \mu^{p}(\omega,T)\rangle + \bold B^{p*}(\omega)\cdot\langle\bm m^{p}(\omega,T)\rangle\Big)
\label{sigdef1}
\end{eqnarray}
\noindent where $T$ is the time delay between the actinic pulse and the probe and $p = L, R$ is the polarization.
The time and frequency resolved tr-XCD signal is defined by :
\begin{equation}
S_{\text{tr-XCD}}(\omega,\tau) = 
\frac{A_{\text{L}}(\omega, \tau)-A_{\text{R}}(\omega, \tau)}
{\frac{1}{2}(A_{\text{L}}(\omega, \tau)+A_{\text{R}}(\omega, \tau))}
\label{trcddef}
\end{equation}

The time-resolved (frequency-integrated) tr-XCD can also  provides interesting information:
\begin{equation}
S_\text{tr-XCD}(\tau) = \int d\omega S_\text{tr-XCD}(\omega,\tau)
\label{finalSt}
\end{equation}

The tr-XCD is obtained by expanding to one more order into the CPL probe. 
Chiral contributions appear through pseudo-scalars and we shall limit the discussion to terms mixing electric and magnetic dipoles (electric quadrupoles vanishes upon rotational averaging):
\begin{eqnarray}
\langle\bm \mu^{p}_\text{chir}(\omega,T)\rangle &=& 
\int dt \ e^{-i\omega t} \langle\bm \mu^p_\text{chir}(t,T)\rangle \\
&=& 
\int dt d\tau_1 \ e^{-i\omega t} \langle\bm \mu^p(t) (\bm m^p(\tau_1)\cdot \bm B^p(\tau_1,T))_-\rangle\\
&=& \int dt d\tau_1 \ e^{-i\omega t} \langle\langle \bm \mu^p | \mathcal G(t-\tau_1)(\bm m^p\cdot \bm B^p(\tau_1,T))_- \mathcal G(\tau_1-t_0)|\rho(t_0)\rangle\rangle
\end{eqnarray}
A similar expression is obtained for the magnetic dipole expectation value:
\begin{eqnarray}
\langle\bm m^{p}_\text{chir}(\omega,T)\rangle &=& 
\int dt d\tau_1 \ e^{-i\omega t} \langle\langle \bm m^p | \mathcal G(t-\tau_1)(\bm \mu^p\cdot \bm E^p(\tau_1,T))_- \mathcal G(\tau_1-t_0)|\rho(t_0)\rangle\rangle
\end{eqnarray}

\begin{figure*}[!h]
  \centering
  \includegraphics[width=0.8\textwidth]{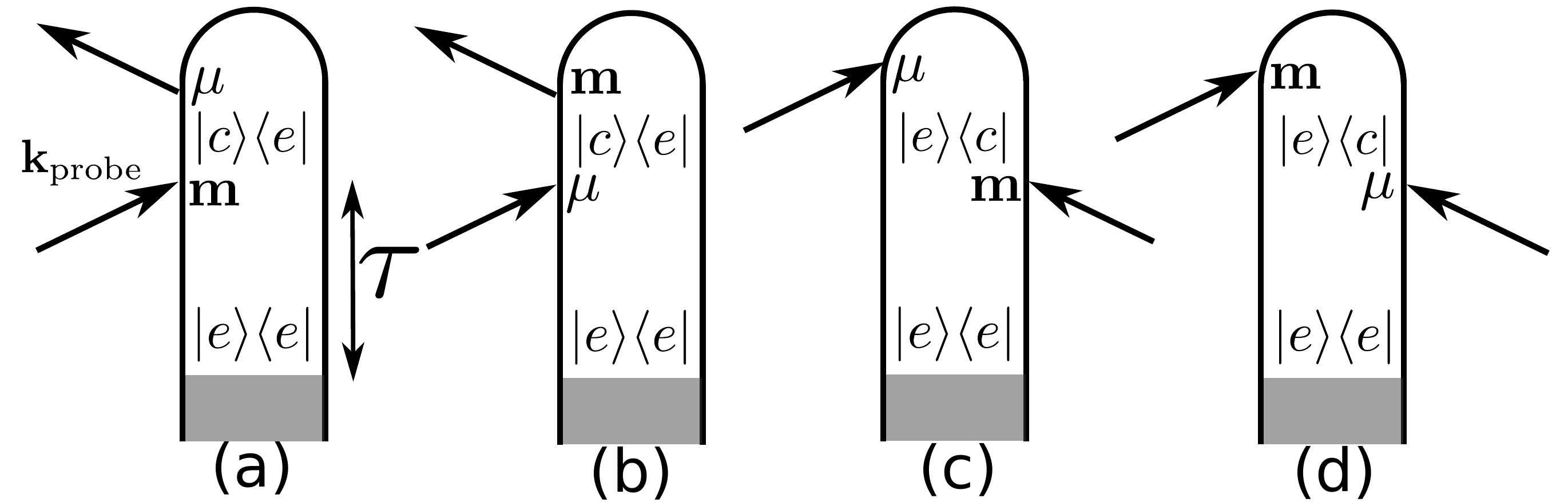}
  \caption{Loop diagrams for time-resolved circular dichroism. $e$ and $c$ represent the valence- and the core-excited states respectively.
\label{trcd1}}
\end{figure*}

These expressions are given in Liouville space and can be expanded into sum over state expressions. 
However, most interesting chiral dynamics involve both nuclear and electronic dynamics and are thus not simply expanded over states since the propagator $\mathcal G$ contains a numerical propagation of the nuclear wavepacket.
The expectation values are conveniently defined using the wavefunction in Hilbert space as:
\begin{multline}
\langle\bm \mu^{p}_\text{chir}(\omega,T)\rangle = \int dt d\tau_1 \ e^{-i \omega t}
\Big(\langle \Psi(t_0) | G^\dagger(t,t_0)\bm\mu^\dagger G(t,\tau_1)\bold m\cdot\bold B^p(\tau_1,T) G(\tau_1,t_0)|\Psi(t_0)\rangle \\
- \langle \Psi(t_0) | G^\dagger(\tau_1,t_0)\bm m^\dagger\cdot\bm B^{p*}(\tau_1,T) G^\dagger(t,\tau_1)\bm \mu G(t,t_0)|\Psi(t_0)\rangle\Big)
\label{hilbertTRCD1}
\end{multline}
\begin{multline}
\langle\bm m^{p}_\text{chir}(\omega,T)\rangle = \int dt d\tau_1 \ e^{-i \omega t}
\Big(\langle \Psi(t_0) | G^\dagger(t,t_0)\bm m^\dagger G(t,\tau_1)\bm \mu\cdot\bold E^p(\tau_1,T) G(\tau_1,t_0)|\Psi(t_0)\rangle \\
- \langle \Psi(t_0) | G^\dagger(\tau_1,t_0)\bm \mu^\dagger\cdot\bm E^{p*}(\tau_1,T) G^\dagger(t,\tau_1)\bm m G(t,t_0)|\Psi(t_0)\rangle\Big)
\label{hilbertTRCD2}
\end{multline}
\noindent where the two terms in Eq. \ref{hilbertTRCD1} are shown in Fig. \ref{trcd1}(a) and (c) and two in Eq. \ref{hilbertTRCD2} in Fig. \ref{trcd1}(b) and (d).

Assuming that the wavefunction is a product of a nuclear and an electronic wavefunction $|\Psi(t_0)\rangle = \sum_e |\Phi(t_0)\rangle|\varphi\rangle$ where $|\Phi(t_0)\rangle$ is the nuclear wavepacket and $|\varphi\rangle$ is the electronic eigenfunction, the sum over the electronic states gives:
\begin{multline}
\langle\bm \mu^{p}_\text{chir}(\omega,T)\rangle = \sum_{ec}\int dt d\tau_1 \ e^{-i \omega t}
\Big(\\
e^{-i\omega_{ce}(t-\tau_1)}
\langle \Phi(t_0)_e | U_e^\dagger(t,t_0)\bm\mu_{ec}^\dagger U_c(t,\tau_1)\bold m_{ce}\cdot\bold B^p(\tau_1,T) U_e(\tau_1,t_0)|\Phi_e(t_0)\rangle \\
-e^{i\omega_{ce}(t-\tau_1)} 
\langle \Phi_e(t_0) | U_e^\dagger(\tau_1,t_0)\bm m_{ec}^\dagger\cdot\bm B^{p*}(\tau_1,T) U_c^\dagger(t,\tau_1)\bm \mu_{ce} U_e(t,t_0)|\Phi_e(t_0)\rangle\Big)
\label{hilbertTRCD3}
\end{multline}
\begin{multline}
\langle\bm m^{p}_\text{chir}(\omega,T)\rangle = \sum_{ec}\int dt d\tau_1 \ e^{-i \omega t}
\Big(\\
e^{-i\omega_{ce}(t-\tau_1)}
\langle \Phi_e(t_0) | U_e^\dagger(t,t_0)\bm m_{ec}^\dagger U_c(t,\tau_1)\bm \mu_{ce}\cdot\bold E^p(\tau_1,T) U_e(\tau_1,t_0)|\Phi_e(t_0)\rangle \\
-e^{i\omega_{ce}(t-\tau_1)} 
\langle \Phi_e(t_0) | U_e^\dagger(\tau_1,t_0)\bm \mu_{ec}^\dagger\cdot\bm E^{p*}(\tau_1,T) U_c^\dagger(t,\tau_1)\bm m_{ce} U_e(t,t_0)|\Phi_e(t_0)\rangle\Big)
\label{hilbertTRCD4}
\end{multline}
\noindent where $U_e$ is the propagator of the nuclear wavepacket of the state $e$.
Since the core-excited coherence is extremely short-lived compared to standard nuclear dynamics, we make the approximation $U_c^\dagger(t,\tau_1) = \mathbb{1}$ in the following.
Using that $\bm E^L(\omega) = \mathcal E(\omega) \bm e_L$, $\bm B^L(\omega) = \mathcal E(\omega) \bm b_L$ and that $e_{Li}^*b_{Lj}-e_{Ri}^* b_{Rj} = (-i/c)(e_{Li}^* e_{Lj}-e_{Ri}^* e_{Rj})= (-i/c)(\delta_{ij}-\hat k_i \hat k_j)$ (similarly $e_{Li}^*b_{Lj}-e_{Ri}^* b_{Rj}= (i/c)(\delta_{ij}-\hat k_i \hat k_j)$), the numerator of expression Eq. \ref{trcddef} becomes:
\begin{eqnarray}
A_{L}(\omega, T) &-& A_{R}(\omega, T)  = -\frac{2}{c} \omega \Re \sum_{ec} \int dt d\tau_1 \ e^{-i \omega t} \mathcal E(\omega) \mathcal E(\tau_1,T)(\delta_{ij}-\hat k_i \hat k_j)
\Big(\nonumber\\
&&e^{-i\omega_{ce}(t-\tau_1)}
\langle \Phi(t_0)_e | U_e^\dagger(t,t_0)\mu_{ec,i}^\dagger m_{ce,j} U_e(\tau_1,t_0)|\Phi_e(t_0)\rangle \nonumber\\
&+& e^{i\omega_{ce}(t-\tau_1)} 
\langle \Phi_e(t_0) | U_e^\dagger(\tau_1,t_0) m_{ec,i}^\dagger \mu_{ce,j} U_e(t,t_0)|\Phi_e(t_0)\rangle \nonumber\\
&-&e^{-i\omega_{ce}(t-\tau_1)}
\langle \Phi(t_0)_e | U_e^\dagger(t,t_0)m_{ec,i}^\dagger \mu_{ce,j} U_e(\tau_1,t_0)|\Phi_e(t_0)\rangle \nonumber\\
&-& e^{i\omega_{ce}(t-\tau_1)} 
\langle \Phi_e(t_0) | U_e^\dagger(\tau_1,t_0) \mu_{ec,i}^\dagger m_{ce,j} U_e(t,t_0)|\Phi_e(t_0)\rangle
\Big)
\label{TRCD5}
\end{eqnarray}

The denominator is given by:
\begin{eqnarray}
&&A_{L}(\omega, T) + A_{R}(\omega, T)  = 2 \omega \Re \sum_{ec} \int dt d\tau_1 \ e^{-i \omega t} \mathcal E(\omega) \mathcal E(\tau_1,T)(\delta_{ij}-\hat k_i \hat k_j)\Big(\nonumber\\
&&e^{-i\omega_{ce}(t-\tau_1)}
\langle \Phi(t_0)_e | U_e^\dagger(t,t_0)\mu_{ec,i}^\dagger \mu_{ce,j} U_e(\tau_1,t_0)|\Phi_e(t_0)\rangle \nonumber\\
&+& e^{i\omega_{ce}(t-\tau_1)} 
\langle \Phi_e(t_0) | U_e^\dagger(\tau_1,t_0) \mu_{ec,i}^\dagger \mu_{ce,j} U_e(t,t_0)|\Phi_e(t_0)\rangle\Big)
\label{TRCD6}
\end{eqnarray}

The tr-XCD signal is finally obtained by rotational averaging of Eqs. \ref{TRCD5} and \ref{TRCD6}. 
This operation is not trivial when nuclear dynamics are included because the propagation of the nuclear wavepacket is usually carried out numerically for a given molecular orientation in the laboratory frame.
Under these conditions, one can not use the simple tensor averaging procedures of Appendix \ref{appendixROTAV} and must rely on a numerical averaging: the full solid angle is discretized and wavepacket propagation must be carried out for each orientation. The matter correlation functions are computed for each orientation and averaged out.
When nuclear coordinates are neglected and only wavepacket dynamics is considered, the numerator, Eq. \ref{TRCD5} simplifies considerably and rotational averaging gives:
\begin{multline}
\langle A_{L}(\omega, T) - A_{R}(\omega, T)\rangle_\Omega  = -\frac{4}{c} \omega \Im \sum_{ec} \int dt d\tau_1 \ e^{-i \omega t} \mathcal E(\omega) \mathcal E(\tau_1,T)\sin(\omega_{ce}(t-\tau_1))
\\
\times (\delta_{ij}-\hat k_i \hat k_j)\langle(\mu_{ec,i}^\dagger m_{ce,j}-m_{ec,i}^\dagger \mu_{ce,j}) \rho_{ee}(t_0)\rangle_\Omega
\label{TRCD7}
\end{multline}

\noindent where $\rho_{ee}(t_0)$ are the excited state populations after the actinic pulse. For example, if the actinic pulse interacts twice, the averaging tensor $I^{(4)}$ may be used to average the signal.

\begin{figure*}[!h]
  \centering
  \includegraphics[width=1.\textwidth]{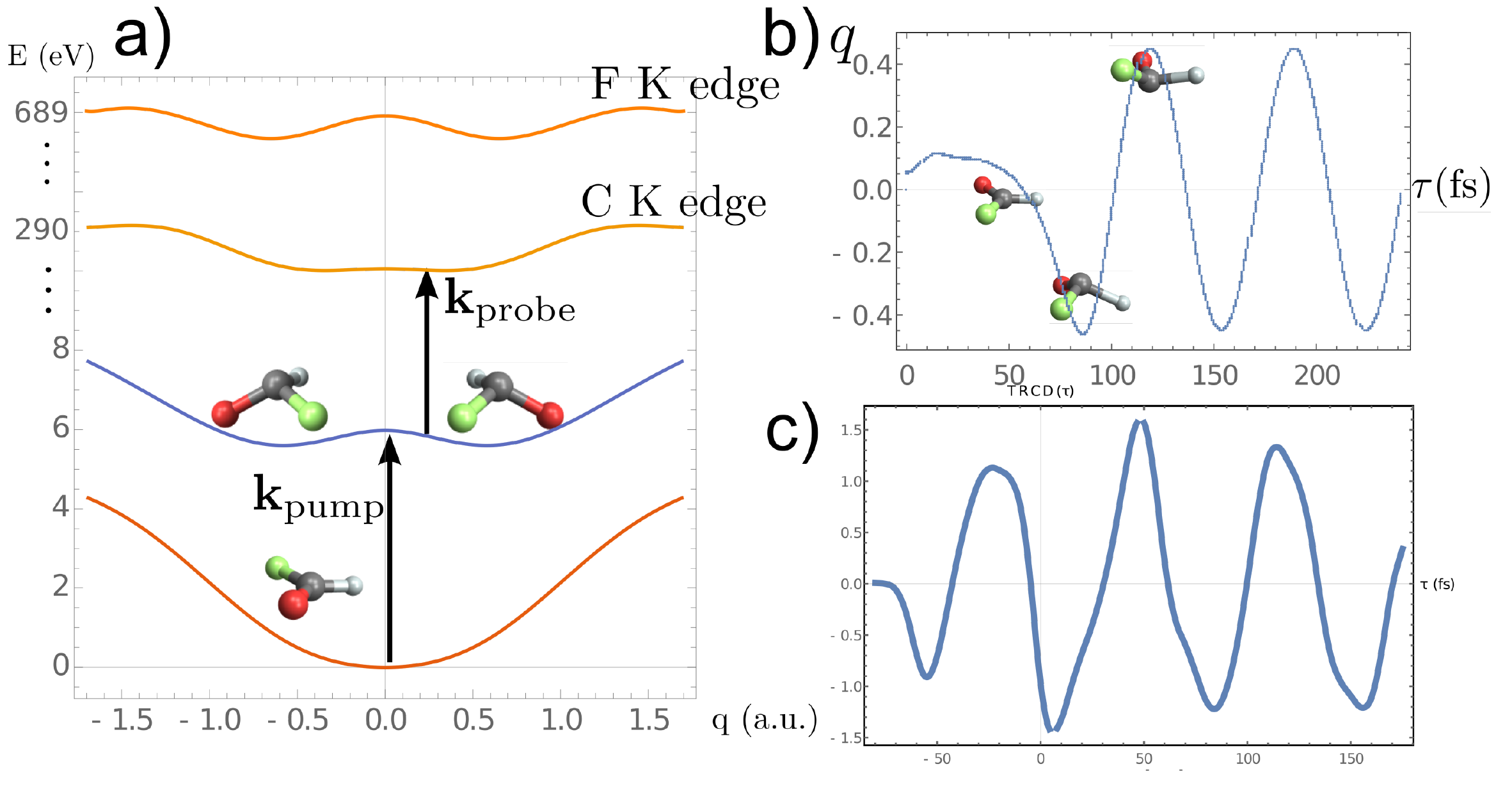}
  \caption{a) PES for an achiral molecule in its ground state that becomes chiral in its excited state. The two minima indicate the two enantiomers and a symmetry-breaking pump can generate a wavepacket with an excess probability of one of them whose average dynamics is displayed in panel b). A delayed X-ray probe can then selectively probe atoms in the molecule and lead to a tr-XCD following closely the molecular dynamics. Reprint from Rouxel at al\cite{rouxel2017photoinduced}.
\label{trcd2}}
\end{figure*}

A simulated tr-XCD trace on formamide is shown in Fig. \ref{trcd2}c. 
This molecule is chiral and trigonal planar in its ground state and  is trigonal pyramidal in its first excited state at approximately 6eV. 
The lone pair in the excited geometry completes the four groups needed to have a chiral carbon and the molecule thus becomes chiral in its excited state.
The potentiel energy surfaces (PES) along the bending mode shown in Fig. \ref{trcd2}a have two minima in the excited state corresponding to the two enantiomers (molecule bended upward or downward).
Photoexcitation with a circularly polarized UV beam launches an asymmetric wavepacket in the excited PES and the molecule starts oscillating back and forth as shown by the expectation value of the bending coordinate depicted in Fig. \ref{trcd2}b.
Finally, the integrated tr-XCD is displayed in Fig. \ref{trcd2}c with the probe tuned at the C K-edge.
As expected, this signal follows closely the chiral dynamics of the molecule\cite{rouxel2017photoinduced}.

\subsection{X-ray Raman optical activity}
\label{sectionXROA}

Raman optical activity (ROA) is widely used  in the IR and the visible regimes to measure the vibrational or electronic optical activity\cite{barron2004molecular,polavarapu1990ab,barron2018raman}.
It has provided a clear signature  of the absolute configuration of small chiral molecules which is important for drug synthesis and in the secondary and tertiary structure determination of proteins.

Spontaneous Raman signals can be seen as a scattering event in which an incoming photon is scattered into an empty mode of the electromagnetic field\cite{mukamel1999principles}.
In order to create a photon population in the scattered mode, two orders in the perturbation by the interaction Hamiltonian are necessary (one with the ket and one with the bra) and the matter Raman response function are thus given by 4 point correlation functions.
To first order in the multipolar expansion, the spontaneous Raman signal contains four perturbations in the electric dipole coupling.
This contribution gets cancelled upon taking the difference between opposite polarizations, and the ROA requires the magnetic dipole and electric quadrupole coupling. 
The theory and practical implementation of ROA has been developed to a large extent by Barron\cite{barron2018raman}.

X-ray ROA (XROA) is the X-ray regime extension of the ROA.
So far, theoretical works on XROA are scarce\cite{rouxel2019x} and no experimental realization has been reported yet.
On the other hand, spontaneous Raman signals have been measured routinely\cite{sahle2015planning, schulke2007electron,iwashita2017seeing}. 
Stimulated and ultrafast extensions are under development\cite{weninger2013stimulated, rohringer2019x, o2020electronic}.
In the X-ray community, spontaneous Raman is often referred to as Inelastic X-ray Scattering (IXS).
The signal recorded by frequency dispersing the scattered photons of a single incoming light beam is named X-ray Emission Spectroscopy (XES) while the 2D maps recorded at resonance by scanning the incoming light and dispersing the scattered photons are know as Resonant Inelastic X-ray Scattering (RIXS)\cite{vanko2013spin}.
The increased availability of polarization control at synchrotron and FEL has enabled the chiral-sensitive version of these technique, XROA, in principle achievable at many facilities.
The main challenge remains the scattering cross-sections of these signals which are usually weak, and thus the even weaker XROA requires long acquisition time and high stability.

While the formal expressions of the signal reviewed in the section remain the same as in the lower frequency regime, XROA exploits the specific characteristics of the X-ray regime.

\begin{figure*}[!h]
  \centering
  \includegraphics[width=0.7\textwidth]{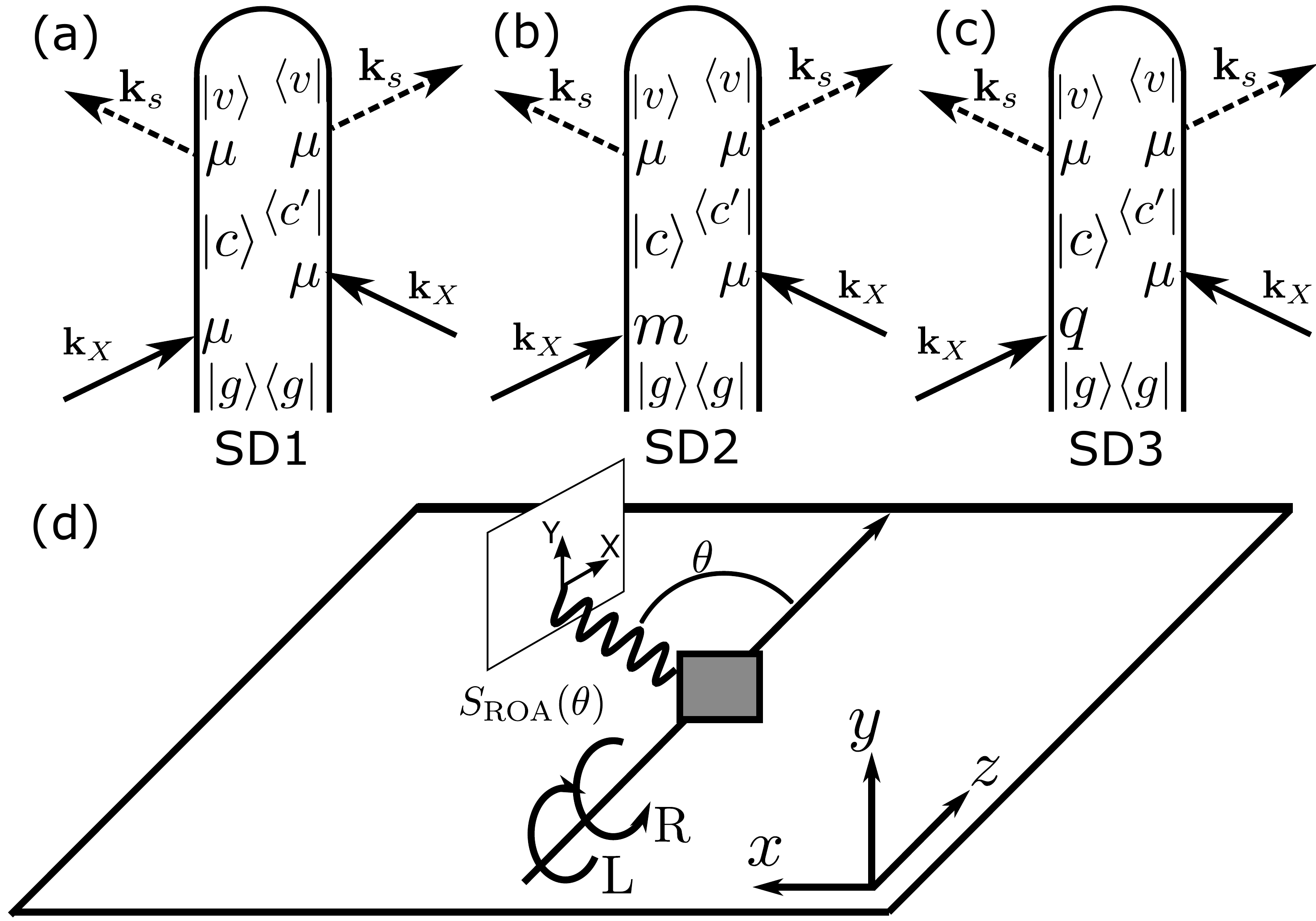}
  \caption{(a) Loop diagrams of the XROA signals. Permutation of the magnetic dipole and electric quadrupole have to be summed over to calculate the signal. (b) Geometry of a ROA measurement. This correspond to a incident circular polarization (ICP) setup. Alternatively, it is possible to excite the sample with linear polarization and detect left and right polarization on the detector or to use mixed schemes with both incident and detected circular polarized light.
\label{roa1}}
\end{figure*}

ROA being an incoherent spontaneous signal, Eq.\ref{homoINCmulti} in appendix \ref{appendixOBS} can be used as a starting point. Circularly polarized light passes through the sample and generates a Raman signal recorded as a function of the scattering angle, see Fig.\ref{roa1}(d). The ROA signal, Eq.\ref{XROAdef} is given by the small difference between Raman scattering of left and right polarized light.
\begin{multline}
\Delta_{\text{RAM}}(\omega_X,\omega_s,\theta,\bold e_s) = S_{\text{RAM}}(\omega_X,\omega_s,\theta,\bold e_{L},\bold e_s)
-S_{\text{RAM}}(\omega_X,\omega_s,\theta,\bold e_{R},\bold e_s)
\label{deltaRAM}
\end{multline}
\noindent where $S_{\text{RAM}}$ is the Raman scattering signal given by the diagrams shown in \ref{roa1}(a) and $\bm e_{L/R}$ and $\bm e_s$ are the polarization vectors of the incoming and spontaneously emitted photons respectively.
Similar to XCD, Eq.\ref{CDdef}, the XROA signal is defined by normalizing the difference, Eq. \ref{deltaRAM} with the sum of the signals with opposite polarizations\cite{barron2004raman}.
\begin{equation}
S_\text{XROA}(\omega_X,\omega_s,\theta,\bold e_s) = \frac{S_{\text{RAM}}(\omega_X,\omega_s,\theta,\bold e_{L},\bold e_s)-
S_{\text{RAM}}(\omega_X,\omega_s,\theta,\bold e_{R},\bold e_s)}{\frac{1}{2}(S_{\text{RAM}}(\omega_X,\omega_s,\theta,\bold e_{L},\bold e_s)+
S_{\text{RAM}}(\omega_X,\omega_s,\theta,\bold e_{R},\bold e_s))}
\label{XROAdef}
\end{equation}

\noindent Unlike in the usual ROA literature\cite{barron2004raman}, we have divided the denominator by 2 to normalize by the average of the achiral signals and to keep consistence with other signals in this review.

The XROA signal depends on multiple parameters: $\omega_\text{X}$ is the frequency of the incident X-rays, $\omega_s$ is the frequency of the detected scattered photons, $\theta$ is the scattering angle, defined in Fig.\ref{roa1}(d), and $\bm e_s$ is the detected light polarization.
Often, no polarizer is placed in front of the detector.
This simpler XROA signal $S_{\text{XROA}}(\omega_X,\omega_s,\theta)$ is obtained by
summing over detected polarizations and does not depend on the emitted polarization $\bm e_s$.
Alternatively, it is possible to use incoming linear polarizations and detect the amount of left and right polarization in the scattered light.
These two configurations are often labelled as Incident Circular Polarization (ICP) and Scattered Circular Polarization (SCP).
When both incident and scattered circular polarization are used, this leads to the Dual Circular Polarization (DCP) setup.
These setups result in different linear combination of the same observables\cite{barron2004molecular}. 
Below, we shall only consider the ICP.

We start by reviewing the XROA expressions.
We assume that the incident pulse propagates along the $z$ axis and that the scattered Raman signal is measured in the $xz$ plane as shown in Fig.\ref{roa1}(d). 
In an isotropic medium, the entire  matter+field system is cylindrically symmetric and out-of-plane detection does not carry extra information.
Following appendix \ref{appendixOBS}, the spontaneous Raman signal defined as the change of photon number emitted in the detected direction reads
\begin{multline}
S_{\text{RAM}}(\omega_s,\theta,\bm e_s) = -\frac{2}{\hbar} \text{Im} \int dt \langle\bold E_s^\dagger(t)\cdot \bm\mu(t)
+\bold B_s^\dagger(t)\cdot \bm m(t)+\nabla\bold E_s^\dagger(t)\cdot \bm q(t)\rangle
\end{multline}
\noindent where expectation values are defined by
\begin{equation}
\langle A(t) \rangle = \langle \Psi(t) | A | \Psi(t) \rangle
\end{equation} 
Perturbative expansion of the bra and the ket wavefunctions contributing to the signal are  depicted by the loop diagrams in Fig. \ref{roa1} a, b and c.
The molecular wavefunction is expanded to first order in the spontaneous field and to second order in the incident field.
Three terms contribute to the signals, as shown in Fig.\ref{roa1}(a-c). 
Each of these contributions can be written as a four-point correlation function:
\begin{multline}
S_{\text{RAM}}(\omega_X,\omega_s,\theta,\bm e_{L/R},\bm e_s) = \frac{2}{\hbar^4}\frac{\hbar\omega_s}{2 \epsilon_0} \text{Re}\\
\langle 
X\cdot F_{L/R}^* G^\dagger(\omega_X +\omega_g)
X^\dagger\cdot F_s G^\dagger(\omega_s -\omega_X + \omega_g)
X\cdot F_s^* G(\omega_X+\omega_g)
X^\dagger\cdot F_{L/R} \rangle_\Omega
\end{multline}
\noindent where $X\cdot F$ stands for $\bm \mu \cdot \bm e$, $\bm m \cdot \bm b$ or $\bm q\cdot (i \bm k \bm e)$ if the interaction is dipolar electric, dipolar magnetic or quadrupolar electric.
The leading term of the Raman signal given by interaction with the electric dipoles only (diagram (a)) does not contribute to ROA signal since it vanishes upon rotational averaging.
Assuming that the X-ray beam is resonant with the core-excited manifold, we obtain for the three diagrams in Fig.\ref{roa1}(a-c):
\begin{multline}
S_{D1}(\omega_X,\omega_s,\theta,\bm e_{L/R},\bm e_s) =
\frac{\omega_s}{\epsilon_0 \hbar^3}|\mathcal E_X|^2 \times\\
\text{Re} \sum_{cc'v}
\frac{\langle\bm \mu_{gc}\bm \mu_{cv}^\dagger\bm \mu_{vc'}\bm \mu_{c'g}^\dagger \rangle_\Omega \cdot \bm e_X^* \bm e_s \bm e_s^* \bm e_X}{(\omega_X-\omega_{cg}-i\epsilon)(\omega_s-\omega_X-\omega_{vg}-i\epsilon)(\omega_X-\omega_{c'g}+i\epsilon)}
\label{eq:roasd1}
\end{multline}
\begin{multline}
S_{D2}(\omega_X,\omega_s,\theta,\bm e_{L/R},\bm e_s) =
\frac{\omega_s}{\epsilon_0 \hbar^3}|\mathcal E_X|^2 \times\\
\text{Re} \sum_{cc'v}
\frac{\langle\bm \mu_{gc}\bm \mu_{cv}^\dagger\bm \mu_{vc'}\bm m_{c'g}^\dagger \rangle_\Omega \cdot \bm e_X^* \bm e_s \bm e_s^* \bm b_X}{(\omega_X-\omega_{cg}-i\epsilon)(\omega_s-\omega_X-\omega_{vg}-i\epsilon)(\omega_X-\omega_{c'g}+i\epsilon)}
\end{multline}
\begin{multline}
S_{D3}(\omega_X,\omega_s,\theta,\bm e_{L/R},\bm e_s) =
\frac{\omega_s}{\epsilon_0 \hbar^3}|\mathcal E_X|^2 \times\\
\text{Re} \sum_{cc'v}
\frac{\langle\bm \mu_{gc}\bm \mu_{cv}^\dagger\bm \mu_{vc'}\bm q_{c'g}^\dagger \rangle_\Omega \cdot \bm e_X^* \bm e_s \bm e_s^* (i\bm k_X\bm e_X)}{(\omega_X-\omega_{cg}-i\epsilon)(\omega_s-\omega_X-\omega_{vg}-i\epsilon)(\omega_X-\omega_{c'g}+i\epsilon)}
\label{eq:roasd3}
\end{multline}
The total Raman scattering signal including interactions to second order in the multipolar Hamiltonian is obtained by summing over these three terms as well as the ones obtained by permuting the magnetic dipole and electric quadrupole interactions with the electric dipole ones in diagrams $D_2$ and $D_3$.
Orientational averaging of the matter tensor is done with the $I^{(4)}$ tensor for $S_{D_1}$ and $S_{D_2}$. $I^{(5)}$ is used to average $S_{D_3}$.
Finally, the XROA signal is obtained by taking the difference between left and right polarizations of the incoming light.

We start by showing how the electric dipole contributions cancel out. When summing over states, we obtain:
\begin{multline}
\langle\bm \mu_{gc}\bm \mu_{cv}^\dagger\bm \mu_{vc'}\bm \mu_{c'g}^\dagger \rangle_\Omega \cdot (\bm e_{XL}^* \bm e_s \bm e_s^* \bm e_{XL}-\bm e_{XR}^* \bm e_s \bm e_s^* \bm e_{XR})= I^{(4)abcd}_{ijkl} \mu_{gc}^i \mu_{cv}^{j\dagger} \mu_{vc'}^k \mu_{c'g}^{l\dagger} i \epsilon_{adz} (e_s)_b  (e_s^*)_c
\label{Eq:ROAelec}
\end{multline}
Einstein summation convention is used for the sum over repeated cartesian indices.
Using the definition of $I^{(4)abcd}_{ijkl}$ in appendix A, Eq. \ref{Eq:ROAelec} contains three terms which are proportional to $\delta_{ab}\delta_{cd}\epsilon_{adz}(\bm e_s)_b  (\bm e_s^*)_c$, 
$\delta_{ac}\delta_{bd}\epsilon_{adz}(\bm e_s)_b  (\bm e_s^*)_c$ and $\delta_{ad}\delta_{bc}\epsilon_{adz}(\bm e_s)_b  (\bm e_s^*)_c$ respectively. The first two terms are $(\bm e_s \wedge \bm e_s^*)_z$ which are zero and the last term is trivially vanishing because and the anti-symmetry of $\epsilon_{adz}$.

We now turn to the chiral terms. 
The XROA signal can be separated into the magnetic dipole and the electric quadrupole contributions:
\begin{equation}
\Delta_{\text{RAM}}(\omega_X, \omega_s, \bm e_s, \theta) = S_{\text{XROA}}^{\text{mag}}(\omega_X, \omega_s, \bm e_s, \theta) 
+ S_{\text{ROA}}^{\text{quad}}(\omega_X, \omega_s, \bm e_s, \theta)
\end{equation}
where $S_{ROA}^{\text{mag}}$ and  $S_{\text{XROA}}^{\text{quad}}$ are 
defined as
\begin{multline}
S_{\text{XROA}}^{\text{mag}}(\omega_X, \omega_s, \bm e_s, \theta) = \frac{\omega_s}{\epsilon_0 \hbar^3}|\mathcal E_X|^2 \text{Re} \sum_{cc'v} G_{cg}(\omega_X)G_{vg}(\omega_s-\omega_X)G_{c'g}^*(\omega_X) I^{(4)abcd}_{ijkl}\\
\times\Big(\mu_{gc}^i \mu_{cv}^{\dagger j}\mu_{vc'}^k m_{c'g}^{\dagger l} (e_{XL}^{a*} b_{XL}^{d} - e_{XR}^{a*} b_{XR}^{d}) e_s^b e_s^{*c}
+ \mu_{gc}^i \mu_{cv}^{\dagger j}m_{vc'}^k \mu_{c'g}^{\dagger l}(e_{XL}^{a*} e_{XL}^{d} - e_{XR}^{a*} e_{XR}^{d}) e_s^b b_s^{*c}\\
+ \mu_{gc}^i m_{cv}^{\dagger j}\mu_{vc'}^k \mu_{c'g}^{\dagger l}(e_{XL}^{a*} e_{XL}^{d} - e_{XR}^{a*} e_{XR}^{d}) b_s^b e_s^{*c}
+ m_{gc}^i \mu_{cv}^{\dagger j}\mu_{vc'}^k \mu_{c'g}^{\dagger l}(b_{XL}^{a*} e_{XL}^{d} - b_{XR}^{a*} e_{XR}^{d}) e_s^b e_s^{*c}
\Big)
\label{sigXROAmag}
\end{multline}
\begin{multline}
S_{\text{XROA}}^{\text{quad}}(\omega_X, \omega_s, \bm e_s, \theta) = \frac{\omega_s}{\epsilon_0 \hbar^3}|\mathcal E_X|^2 \text{Re} \sum_{cc'v} G_{cg}(\omega_X)G_{vg}(\omega_s-\omega_X)G_{c'g}^*(\omega_X) I^{(5)abcdf}_{ijklm}\\
\times\Big(\mu_{gc}^i \mu_{cv}^{\dagger j}\mu_{vc'}^k q_{c'g}^{\dagger l} (e_{XL}^{a*} i k_{X}^{d}e_{XL}^{f} - e_{XR}^{a*} i k_{X}^{d}e_{XR}^{f}) e_s^b e_s^{*c}\\
+ \mu_{gc}^i \mu_{cv}^{\dagger j}q_{vc'}^k \mu_{c'g}^{\dagger l}(e_{XL}^{a*} e_{XL}^{f} - e_{XR}^{a*} e_{XR}^{f}) e_s^b (-i k_s^{c}e_s^{*d})\\
+ \mu_{gc}^i q_{cv}^{\dagger j}\mu_{vc'}^k \mu_{c'g}^{\dagger l}(e_{XL}^{a*} e_{XL}^{f} - e_{XR}^{a*} e_{XR}^{f}) (i k_s^{b}e_s^{c}) e_s^{*d}\\
+ q_{gc}^i \mu_{cv}^{\dagger j}\mu_{vc'}^k \mu_{c'g}^{\dagger l}(-ik_{X}^{a}e_{XL}^{b*} e_{XL}^{f} - (-ik_{X}^{a}e_{XR}^{b*}) e_{XR}^{f}) e_s^c e_s^{*d}
\Big)
\label{sigXROAquad}
\end{multline}
\noindent where $G(\omega)$ are the Green's functions, see Appendix \ref{appendix:perturbation} and $I^{(n)}$ is the averaging tensor, see Appendix \ref{appendixROTAV}.
The four terms in each of Eq. \ref{sigXROAmag} and \ref{sigXROAquad} correspond to the possible permutations of the chiral interaction in diagrams (b) and (c), Fig.\ref{roa1}, obtained by permuting the position of the magnetic dipole or electric quadrupole interactions.
The signal calculation can be done in the following steps: calculation of the field polarization tensor, rotational averaging by contraction with the $I^{(n)}$ tensors, calculation of the final observable by contraction with the matter response tensor.
The four field polarization tensors for $S_{\text{XROA}}^\text{mag}$ in Eq. \ref{sigXROAmag} can be simplified using:
\begin{eqnarray}
cF^{m1}_{abcd} = c(e_{XL}^{a*} b_{XL}^{d} - e_{XR}^{a*} b_{XR}^{d}) e_s^b e_s^{*c} &=& i(\delta_{ad} - \delta_{az}\delta_{dz}) e_s^b e_s^{*c}\\
cF^{m2}_{abcd} = c(e_{XL}^{a*} e_{XL}^{d} - e_{XR}^{a*} e_{XR}^{d}) e_s^b b_s^{*c} &=& c e_s^b b_s^{*c} i \epsilon_{adz} \\
cF^{m3}_{abcd} = c(e_{XL}^{a*} e_{XL}^{d} - e_{XR}^{a*} e_{XR}^{d}) b_s^b e_s^{*c} &=& cb_s^b e_s^{*c} i \epsilon_{adz}\\
cF^{m4}_{abcd} = c(b_{XL}^{a*} e_{XL}^{d} - b_{XR}^{a*} e_{XR}^{d}) e_s^b e_s^{*c} &=& i(\delta_{ad} - \delta_{az}\delta_{dz}) e_s^b e_s^{*c}
\end{eqnarray}

Each of the four terms in Eq. \ref{sigXROAmag} are then obtained by contracting the field polarization tensors with the rotationally-averaged molecular response tensor. For exemple, for the first term, we get:
\begin{multline}
I^{(4)abcd}_{ijkl} \mu_{gc}^i \mu_{cv}^{j\dagger} \mu_{vc'}^k m_{c'g}^{l\dagger} F^{m1}_{abcd} 
=\frac{i}{c}\frac{1}{30}
\Big(\\
\delta_{ab}\delta_{cd}(\delta_{ad} - \delta_{az}\delta_{dz})e_s^b e_s^{*c})(4 \bm \mu_{gc}^i\bm \mu_{cv}^{i\dagger} \bm \mu_{vc'}^j \bm m_{c'g}^{j\dagger}- \bm \mu_{gc}^i\bm \mu_{cv}^{j\dagger} \bm \mu_{vc'}^i \bm m_{c'g}^{j\dagger}- \bm \mu_{gc}^i\bm \mu_{cv}^{j\dagger} \bm \mu_{vc'}^j \bm m_{c'g}^{i\dagger}) \\
+\delta_{ac}\delta_{bd}(\delta_{ad} - \delta_{az}\delta_{dz})e_s^b e_s^{*c})(- \bm \mu_{gc}^i\bm \mu_{cv}^{i\dagger} \bm \mu_{vc'}^j \bm m_{c'g}^{j\dagger}+4 \bm \mu_{gc}^i\bm \mu_{cv}^{j\dagger} \bm \mu_{vc'}^i \bm m_{c'g}^{j\dagger}- \bm \mu_{gc}^i\bm \mu_{cv}^{j\dagger} \bm \mu_{vc'}^j \bm m_{c'g}^{i\dagger}) \\
+\delta_{ad}\delta_{bc}(\delta_{ad} - \delta_{az}\delta_{dz})e_s^b e_s^{*c})(- \bm \mu_{gc}^i\bm \mu_{cv}^{i\dagger} \bm \mu_{vc'}^j \bm m_{c'g}^{j\dagger}- \bm \mu_{gc}^i\bm \mu_{cv}^{j\dagger} \bm \mu_{vc'}^i \bm m_{c'g}^{j\dagger}+4 \bm \mu_{gc}^i\bm \mu_{cv}^{j\dagger} \bm \mu_{vc'}^j \bm m_{c'g}^{i\dagger})\Big)
\end{multline}

For detection along the $Y$ axis, $\bold e_s = \bold e_Y$ (see Fig.\ref{roa1}(d)), this contribution simplifies to:
\begin{multline}
\frac{1}{c}I^{(4)abcd}_{ijkl} \mu_{gc}^i \mu_{cv}^{j\dagger} \mu_{vc'}^k m_{c'g}^{l\dagger} F^{m1}_{abcd} = \frac{i}{c}\frac{1}{30}
(3  \mu_{gc}^i \mu_{cv}^{i\dagger}  \mu_{vc'}^j  m_{c'g}^{j\dagger}
+3  \mu_{gc}^i \mu_{cv}^{j\dagger}  \mu_{vc'}^i  m_{c'g}^{j\dagger}
-2  \mu_{gc}^i \mu_{cv}^{j\dagger}  \mu_{vc'}^j  m_{c'g}^{i\dagger})
\end{multline}

Detection along $X$ gives:
\begin{multline}
\frac{1}{c}I^{(4)abcd}_{ijkl} \mu_{gc}^i \mu_{cv}^{j\dagger} \mu_{vc'}^k m_{c'g}^{l\dagger} F^{m1}_{abcd} = \frac{i}{c}\frac{1}{30}\cos^2(\theta)
(3  \mu_{gc}^i \mu_{cv}^{i\dagger}  \mu_{vc'}^j  m_{c'g}^{j\dagger}
+3  \mu_{gc}^i \mu_{cv}^{j\dagger}  \mu_{vc'}^i  m_{c'g}^{j\dagger}
-2  \mu_{gc}^i \mu_{cv}^{j\dagger}  \mu_{vc'}^j  m_{c'g}^{i\dagger})
\end{multline}

The three other magnetic contributions and the four quadrupole contributions can be calculated similarly for an arbitrary scattering angle and detection polarization.
These lengthy expressions are not given in this review and are more adequately derived using computer algebra systems.

It is also convenient to express the signals using the polarizabilities defined as\cite{craig1998molecular}:
\begin{eqnarray}
\alpha_{vg}^{ij}(\omega_X, -\omega_s) &=& \sum_c \frac{\mu_{vc}^i \mu_{cg}^{j\dagger}}{\omega_{cg}-\omega_X-i\epsilon} 
+ \frac{\mu_{vc}^j \mu_{cg}^{i\dagger}}{\omega_{cg}+\omega_s+i\epsilon}
\label{eq:alphap}\\
G_{vg}^{ij}(\omega_X, -\omega_s) &=& \sum_c \frac{\mu_{vc}^i m_{cg}^{j\dagger}}{\omega_{cg}-\omega_X-i\epsilon} 
+ \frac{m_{vc}^j \mu_{cg}^{i\dagger}}{\omega_{cg}+\omega_s+i\epsilon}
\label{eq:Gchirrep}\\
A_{vg}^{ijk}(\omega_X, -\omega_s) &=& \sum_c \frac{\mu_{vc}^i A_{cg}^{jk\dagger}}{\omega_{cg}-\omega_X-i\epsilon} 
+ \frac{A_{vc}^{jk} \mu_{cg}^{i\dagger}}{\omega_{cg}+\omega_s+i\epsilon}
\label{eq:Ap}
\end{eqnarray}
\noindent where $\alpha$ is the ordinary polarizability and $G$ and $A$ are the mixed electric-magnetic and mixed electric dipole-quadrupole polarizabilities.
One should not mistake the mixed electric-magnetic polarizability $G_{vg}^{ij}(\omega_X, -\omega_s)$ and the Green's function $G_{vg}(\omega_s - \omega_x)$ that often share the same symbol in the literature. 
The second term in each definition correspond to cases in which the scattered photon is  emitted before the absorption of the incoming one.
Since its denominator can not cancel out, it does not contribute near resonance and can be neglected (this is the rotating wave approximation).
Eqs. \ref{sigXROAmag} and \ref{sigXROAquad} simplifies into
\begin{multline}
S_{\text{XROA}}^{\text{mag}}(\omega_X, \omega_s, \bm e_s, \theta) = 
\frac{1}{c}\frac{2\omega_s}{\epsilon_0 \hbar^3}|\mathcal E_X|^2 \text{Re} 
\sum_v G_{vg}(\omega_s - \omega_x) I^{(4)abcd}_{ijkl}\\
\times \Big(\alpha^{ij}_{gv}(\omega_X,-\omega_s) G^{kl*}_{vg}(\omega_X,-\omega_s) F^{m4}_{cdba} + 
\alpha^{ij}_{gv}(\omega_X,-\omega_s) G^{lk}_{vg}(\omega_X,-\omega_s) F^{m3}_{dcab}\Big)
\label{eq:xroamag}
\end{multline}
\begin{multline}
S_{\text{XROA}}^{\text{quad}}(\omega_X, \omega_s, \bm e_s, \theta) = 
\frac{1}{c}\frac{2\omega_s}{\epsilon_0 \hbar^3}|\mathcal E_X|^2 \text{Re} 
\sum_v G_{vg}(\omega_s - \omega_x) I^{(5)abcde}_{ijklm}\\
\times \Big(\alpha^{ij}_{gv}(\omega_X,-\omega_s) A^{klm*}_{vg}(\omega_X,-\omega_s) F^{q1}_{xx} + 
\alpha^{ij}_{gv}(\omega_X,-\omega_s) A^{mlk}_{vg}(\omega_X,-\omega_s) F^{q1}_{xx}\Big)
\end{multline}
\noindent with
\begin{eqnarray}
c F^{q1}_{abcde} &=& c(e_{XL}^{e*} e_{XL}^{c} - e_{XR}^{e*} e_{XR}^{c})e_s^{a*} e_s^b ik_X^d = 
c\epsilon_{ecf}k_X^f k_X^d e_s^{a*} e_s^b\\
c F^{q2}_{abcde} &=& c(e_{XL}^{e*} e_{XL}^{c} - e_{XR}^{e*} e_{XR}^{c})e_s^{a*} e_s^b ik_s^d
= c\epsilon_{ecf}k_X^f k_s^d e_s^{a*} e_s^b
\end{eqnarray}

\noindent where we have used Eq. \ref{eq:sumpolar}.
The XROA signal $S_\text{XROA}(\omega_X,\omega_s,\theta,\bold e_s)$, Eq. \ref{XROAdef}, is obtained by normalizing by the achiral contribution:
\begin{eqnarray}
&&S_\text{RAM}^\text{achir}(\omega_X,\omega_s,\theta,\bold e_s) = \frac{1}{2}(S_{\text{RAM}}(\omega_X,\omega_s,\theta,\bold e_{L},\bold e_s)+
S_{\text{RAM}}(\omega_X,\omega_s,\theta,\bold e_{R},\bold e_s))\\
=&& \frac{1}{2} \frac{2\omega_s}{\epsilon_0 \hbar^3}|\mathcal E_X|^2 
I^{(4)abcd}_{ijkl} \alpha_{vg}^{ij}(\omega_X,-\omega_s) G_{vg}(\omega_s - \omega_x) \alpha_{vg}^{kl\dagger}(\omega_X,-\omega_s) e_s^{a*} (\delta_{bc} - \hat k_X^b\hat k_X^c) e_s^d
\label{eq:xroaelecd}
\end{eqnarray}

Eqs. \ref{eq:xroamag} to \ref{eq:xroaelecd} may be used to calculate $S_\text{XROA}(\omega_X,\omega_s,\theta,\bold e_s)$, Eq. \ref{XROAdef}.
For non-resonant ROA ($\epsilon = 0$ in Eqs. \ref{eq:alphap} to \ref{eq:Ap}) and for the specific values of the scattering angle, $\theta = 0^\circ$ (forward scattering), $180^\circ$ (backward scattering) and $90^\circ$, Barron et al.\cite{barron2004molecular, barron2004raman} have given simple closed form expressions given by:
\begin{eqnarray}
S_\text{XROA}(\omega_X,\omega_s, 0^\circ)
&=& 2 \frac{4\big(45(\alpha G)^{(0)} + (\alpha G)^{(2)}- (\alpha A)^{(1)}\big)}
{c(45(\alpha\alpha)^{(0)}+ 7 (\alpha\alpha)^{(2)})}\\
S_\text{XROA}(\omega_X,\omega_s, 180^\circ)
&=& 2 \frac{24\big((\alpha G)^{(2)}+ (\alpha A)^{(1)}/3\big)}{c(45(\alpha\alpha)^{(0)}+ 7 (\alpha\alpha)^{(2)})}\\
S_\text{XROA}(\omega_X,\omega_s, 90^\circ,\bold e_s = \bold e_x)
&=& 2\frac{12\big(45 (\alpha G)^{(0)} + 7(\alpha G)^{(2)}+(\alpha A)^{(1)}\big)}{c(45(\alpha\alpha)^{(0)}+ 7 (\alpha\alpha)^{(2)})}\\
S_\text{XROA}(\omega_X,\omega_s, 180^\circ,\bold e_s  = \bold e_y)
&=& 2 \frac{2\big((\alpha G)^{(2)}-(\alpha A)^{(1)}/3\big)}{c (\alpha\alpha)^{(2)}}
\end{eqnarray}
\noindent with 
\begin{eqnarray}
(\alpha\alpha)^{(0)} &=& \sum_v \frac{1}{3} \alpha^{ii}_{vg}G_{vg}(\omega_s - \omega_x)\frac{1}{3} \alpha^{jj}_{vg}\\
(\alpha G)^{(0)} &=& \sum_v \frac{1}{3} \alpha^{ii}_{vg}G_{vg}(\omega_s - \omega_x)\frac{1}{3} G^{jj}_{vg}\\
(\alpha\alpha)^{(2)} &=& \frac{1}{2}\sum_vG_{vg}(\omega_s - \omega_x)(3 \alpha^{ij}_{vg}\alpha^{ij}_{vg} - \alpha^{ii}_{vg}\alpha^{jj}_{vg})\\
(\alpha G)^{(2)} &=& \frac{1}{2}\sum_vG_{vg}(\omega_s - \omega_x)(3 \alpha^{ij}_{vg}G^{ij}_{vg} - \alpha^{ii}_{vg} G^{jj}_{vg})\\
(\alpha A)^{(1)} &=& \frac{1}{2}\sum_v G_{vg}(\omega_s - \omega_x)\omega_X \ \epsilon_{ikl} \alpha^{ij}_{vg} A^{klj}_{vg}
\end{eqnarray}

\begin{figure}[!h]
  \centering
  \includegraphics[width=0.5\textwidth]{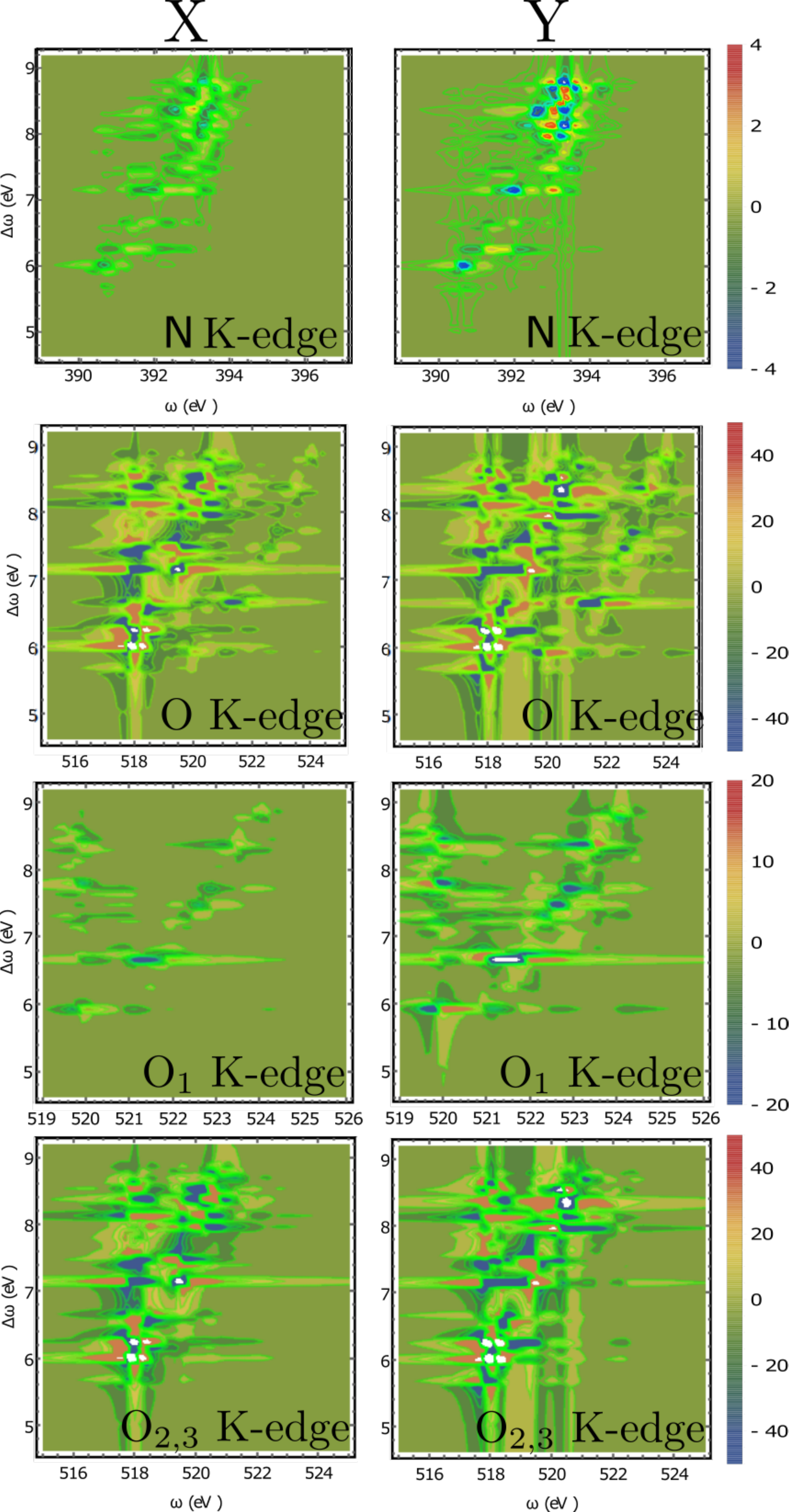}
  \caption{Simulations of ROA spectra at the N and O K-edge on D-tyrosine. Left column: X polarized detection. Right column: Y polarized detection. Reprint from Rouxel et al\cite{rouxel2019x}.
\label{roa2}}
\end{figure}

In the optical regime\cite{hecht1999raman}, a back reflection geometry can conveniently be used to collect the ROA signal.
In the X-ray, multiple instruments exist to measure inelastic scattering with a variety of choice regarding the scattering angle. For example, there exist fixed geometries with a 90$^\circ$ scattering angle\cite{yin2017endstation} or movable spectrometer scanning a range of scanning angles \cite{abela2019swissfel}.

This geometry eliminates the strong background of the intense incoming X-ray beam.
XROA can be complementary to standard electronic CD:
in electronic CD, the valence excited states are prepared by a direct dipole coupling involving the ground and the valence states but, for XROA they are prepared via two successive interactions involving the core excited manifold.
This is reminiscent of the complementarity due to selection rules of IR absorption and Raman spectra.
XROA is thus able to measure the optical activity of states that are dark for valence excitations.
Also, unlike XCD, XROA depends on the electric quadrupole interactions, even for randomly oriented molecules, and can probe molecular chirality when magnetic dipoles are weak, as has been demonstrated for tris(ethylenediamine) cobalt (III) ion (Co(en)$^{3+}_3$)\cite{stewart1999circular}.

We now turn to time-resolved XROA (tr-XROA).
First, we consider a stimulated process in which a delayed probe pulse is overlapped in the direction of the emitted photon. 
This is similar to time-domain CARS\cite{yampolsky2014seeing} and a chiral-sensitive version of Stimulated X-ray Raman Spectroscopy (SXRS) \cite{tanaka2002coherent}.
Simulations of SXRS signals have shown that the broad bandwidth of the pump pulse allows to simultaneously probe many valence states.
For frozen nuclei, a chiral extension to SXRS could then nicely complement XCD by probing the valence states chirality over a broad frequency range with the added benefits of elements selectivity and an alternate set of selection rules.

\begin{figure*}[!h]
  \centering
  \includegraphics[width=0.9\textwidth]{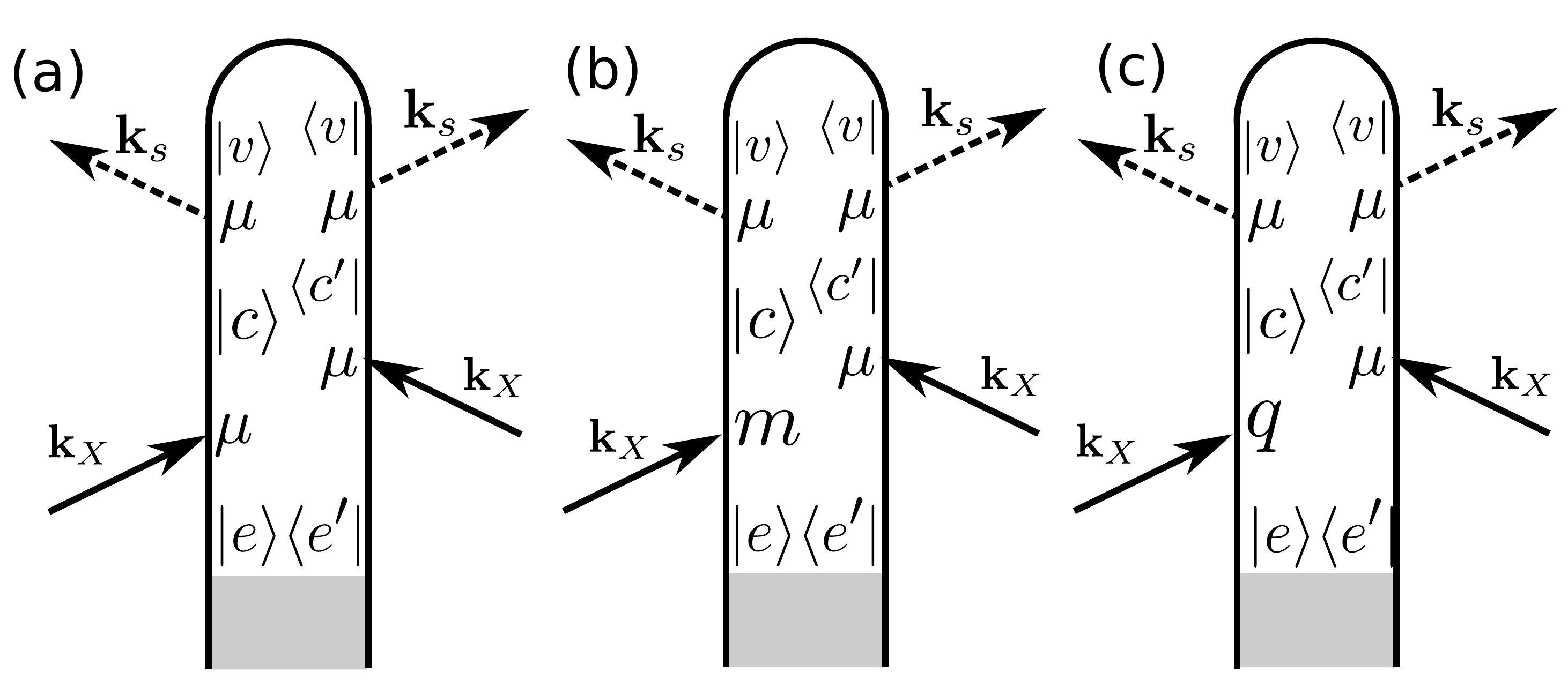}
  \caption{Loop diagrams of the tr-XROA signals. Diagram (a) is the achiral contribution while diagrams (b) and (c) are the chiral magnetic dipole and electric quadrupole ones. 
\label{roa3}}
\end{figure*}

Alternatively, it is possible to consider the full XROA process described in this section as a detection of a molecule prepared in a non-equilibrium state by an actinic pulse. 
The signal derivation follows the same steps of the previous section on tr-XCD and is not discussed in this review.
This signal is named tr-XROA and is defined by:
\begin{equation}
S_\text{tr-XROA}(\Gamma ,T) = \frac{S_{\text{tr-RAM}}(\Gamma,\bold e_{L},T)-
S_{\text{tr-RAM}}(\Gamma,\bold e_{R},T)}{\frac{1}{2}(S_{\text{tr-RAM}}(\Gamma,\bold e_{L},T)+
S_{\text{tr-RAM}}(\Gamma,\bold e_{R},T))}
\end{equation}
\noindent where $T$ is the delay between the actinic pulse that triggers the dynamics (shaded area in Fig. \ref{roa3}
) and the incoming X-ray pulse. 
$\Gamma$ represents the other parameters of the signal (pulse durations, central frequencies, scattering angle, polarized detection).

As for frequency-domain XROA, the signal can be represented by 9 diagrams: an achiral term (Fig. \ref{roa3}a), 4 magnetic dipole terms (Fig. \ref{roa3}b and its permutations) and 4 electric quadrupole terms (Fig. \ref{roa3}c and its permutations).
The interaction with the actinic pulse can be treated pertubatively or numerically.
Similarly to Eqs. \ref{eq:roasd1} to \ref{eq:roasd3}, we now give the expression of the three diagrams in Fig. \ref{roa3}, in the time domain.
\begin{multline}
S_\text{tr-RAM}^{(a)}(\Gamma, \bold e_X, T) = \frac{2}{\hbar^4}N \mathcal E_s\text{Re}\int dt ds_3 ds_2 ds_1 \ e^{i\omega_s s_2}
\mathcal E_X(t-s_1)\mathcal E_X^*(t-s_2-s_3)\\
\times\langle \Psi(T) |
\mu_R\cdot\bold e_X^* G^\dagger(s_3)
\mu_R^\dagger\cdot \bold e_s G^\dagger(s_2)
\mu_L\cdot \bold e_s^* G(s_1) 
\mu_L^\dagger\cdot \bold e_X| \Psi(T)\rangle_\Omega
\end{multline}
\begin{multline}
S_\text{tr-RAM}^{(b)}(\Gamma, \bold e_X, T) = \frac{2}{\hbar^4}N \mathcal E_s\text{Re}\int dt ds_3 ds_2 ds_1 \ e^{i\omega_s s_2}
\mathcal E_X(t-s_1)\mathcal E_X^*(t-s_2-s_3)\\
\times\langle \Psi(T) |
\mu_R\cdot\bold e_X^* G^\dagger(s_3)
\mu_R^\dagger\cdot \bold e_s G^\dagger(s_2)
\mu_L\cdot \bold e_s^* G(s_1) 
m_L^\dagger\cdot \bold b_X| \Psi(T)\rangle_\Omega
\end{multline}
\begin{multline}
S_\text{tr-RAM}^{(c)}(\Gamma, \bold e_X, T) = \frac{2}{\hbar^4}N \mathcal E_s\text{Re}\int dt ds_3 ds_2 ds_1 \ e^{i\omega_s s_2}
\mathcal E_X(t-s_1)\mathcal E_X^*(t-s_2-s_3)\\
\times\langle \Psi(T) |
\mu_R\cdot\bold e_X^* G^\dagger(s_3)
\mu_R^\dagger\cdot \bold e_s G^\dagger(s_2)
\mu_L\cdot \bold e_s^* G(s_1) 
q_L^\dagger\cdot (i k_x\bold e_X)| \Psi(T)\rangle_\Omega
\end{multline}
These contributions and their permutations of the chiral interactions can be used to calculate the time-resolved Raman scattering truncated at the magnetic dipole - electric quadrupole order of the multipolar expansion.
If the molecular wavefunction $|\Psi(T)\rangle$ can be expanded in an eigenbasis, diagrammatic rules from Appendix \ref{appendix:perturbation} can be used to calculate sum-over-states expressions.
Otherwise, a numerical wavepacket propagation may be used to compute the multipoint correlation function, for example in the presence of nuclear dynamics.

\subsection{Chiral X-ray four-wave-mixing}

In concluding this section on chiral-signals within the multipolar coupling Hamiltonian, we now discuss the general chiral four-wave-mixing (4WM) signals.
Achiral 4WM signals have proven extremely useful in the optical regime by spectroscopic techniques such as the photon echo \cite{de1998ultrafast}, transient grating\cite{hofmann2019transient} and other multidimensional spectroscopy signals\cite{cundiff2013optical}.
The recently developed of ultrafast temporally-coherent X-ray sources have triggered extensive theoretical\cite{mukamel2005multiple, mukamel2013multidimensional,  chen2019photon} and experimental\cite{marcus2014free,bencivenga2015four,foglia2018first,rouxel2021hardtg} efforts aimed at extending 4WM techniques to the X-ray regime.
All-X-ray 4WM is challenging to implement while hybrid optical/X-rays techniques have been realized more readily\cite{cao2016noncollinear,lin2021coupled,marroux2018multidimensional}.

Chiral 4WM (c4WM) are third-order $\chi^{(3)}$ techniques that employ circularly polarized light to generate signals that only exist in chiral samples\cite{cao2016noncollinear,abramavicius2009coherent}.
These are described by four-point correlation functions that include one interaction with either a magnetic dipole or an electric quadrupole transition.
Each pulse in an X-ray 4WM process can be resonant with a different core orbital and, which combined with the other structural information gained from chiral sensitive techniques, can provide unique information on matter.
Since 4WM involves four pulses, the number of possible polarization configurations greatly increases compared to lower order techniques.

Given that X-ray 4WM techniques are in their infancy, X-ray c4WM have not been implemented so far.
Lower frequency c4WM could serve as an inspiration\cite{nunes1993sensitive, hache1999nonlinear,fischer2005nonlinear,
mesnil2000experimental,belkin2005non}.
Optical c4WM offers increased sensitivity to chirality and allow to follow the time-evolution of chirality.

XROA discussed in the previous section also fits in the framework on c4WM by having two interaction with the incoming beam and two with photon modes initially in the vacuum state\cite{mukamel1999principles}. 
The XROA formalism demonstrated that multipolar signal expressions  become complex as the interaction order increases.
A systematic classification of c4WMs signals is thus needed to describe the numerous possibilities.

\begin{figure}[!h]
  \centering
  \includegraphics[width=0.5\textwidth]{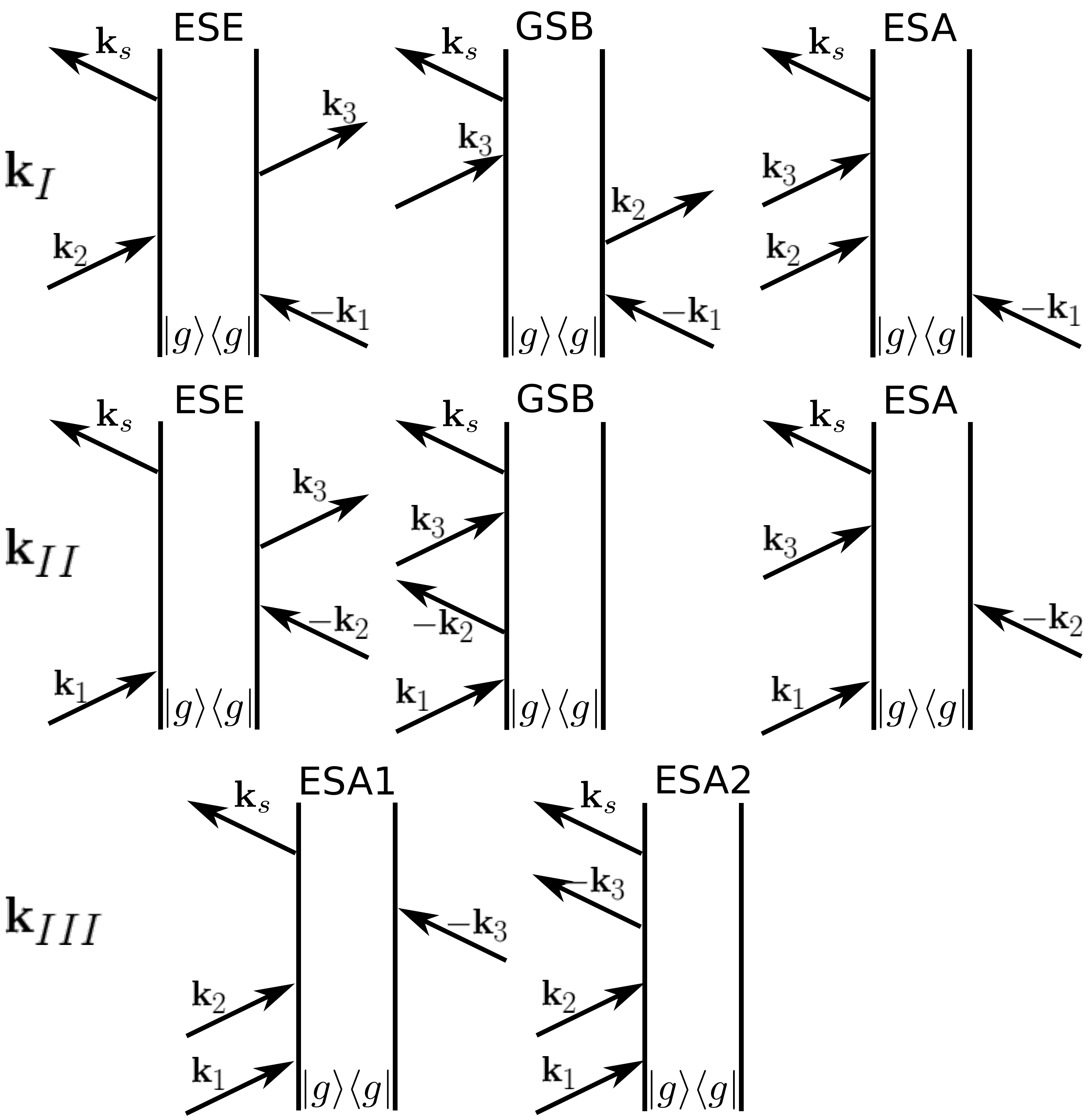}
  \caption{Ladder diagrams for the $\bold k_I, \bold k_{II}$ and $\bold k_{III}$ Four-wave mixing signals. Each interaction can be either with an electric dipole, a magnetic dipole or an electric quadrupole coupling, for a total of 9 diagrams for each of the ones displayed.
\label{4wm1}}
\end{figure}

First, c4WM, like its achiral counterparts, is emitted in a specific phase matching direction. 
This allows to select a subset of contributing diagrams to the detected signals (see Fig. \ref{4wm1}) and leads to three families of techniques that are labelled according to their phase matching condition as:
\begin{eqnarray}
\bold k_I &=& - \bold k_1 +\bold k_2 + \bold k_3\\
\bold k_{II} &=& + \bold k_1 -\bold k_2 + \bold k_3\\
\bold k_{III} &=& + \bold k_1 +\bold k_2 - \bold k_3
\end{eqnarray}
Fig. \ref{4wm1} shows the ladder diagrams corresponding to each technique.
When the multipolar interaction Hamiltonian (Eq. \ref{hintMULTI}) is used, each vertex can represent either the electric dipole, the magnetic dipole or the electric quadrupole.
Each of the diagrams in Fig. \ref{4wm1} appears 9 times in c4WM by including all possible permutations of a single chiral interaction for each pathway.
Diagrams with only electric dipole $\bm \mu\cdot \bm E$ interactions represent the achiral contribution to the signal.
In addition, there are 4 diagrams with a single magnetic dipole coupling $\bm m\cdot \bm B$ within the interaction pathway and similarly 4 diagrams with a single electric quadrupole coupling $\bm q\cdot \nabla E$.

Abramavicius et al. \cite{abramavicius2009coherent} had surveyed the possible configurations for linearly polarized pulses and different phase matching conditions.
In the following, we develop this classification using irreducible tensors\cite{rouxel2020chiral}.
This allows to clearly identify the pseudo-scalar component of the tensor that contribute to c4WM and to carry out the rotational averaging of the response tensors.

A generic 4WM signal can be split into an achiral and a chiral contribution:
\begin{equation}
S_{\text{4WM}}(\Gamma) = S_{\text{achir}}(\Gamma) + S_{\text{chir}}(\Gamma)
\end{equation}
\noindent where $\Gamma$ represents the different relevant pulse parameters, typically central frequencies, polarizations and bandwidths and where $S_{\text{chir}}(\Gamma)$ is the c4WM signal.
As customary with multipolar chiral signals, the c4WM signal can be obtained by measuring the differential signal between left and right polarized light.
We restrict the following discussion to time-domain, heterodyne detected 4WM signals. The achiral contribution $S_\text{achir}^\text{het}$ is given by a four-point correlation function of the electric dipole operator:\cite{mukamel1999principles}
\begin{multline}
S_{\text{achir}}^\text{het}(\Gamma) = -\frac{2}{\hbar} \Im\int dt dt_3 dt_2 dt_1 \bold R_{\mu\mu\mu\mu}(t_3,t_2,t_1)  \bold E_s(t) \bold E_3(t-t_3)  \bold E_2(t-t_3-t_2)
 \bold E_1(t-t_3-t_2-t_1)
\label{heterodynenonchir}
\end{multline}
\noindent with 
\begin{equation}
\bold R_{\mu\mu\mu\mu}(t_3,t_2,t_1) = \Big(-\frac{i}{\hbar}\Big)^3 \langle \bm\mu_\text{left}\mathcal G(t_3)\bm\mu_-\mathcal G(t_2)\bm\mu_-\mathcal G(t_1)\bm\mu_-\rangle_\Omega
\end{equation}
\noindent where the superoperators $\mu_-$ and $\mu_\text{left}$ are defined in Appendix \ref{appendix:perturbation}. 
For brevity, we now drop the arguments of the response functions and of the incoming fields and focus on the tensorial nature of the response. Eq. \ref{heterodynenonchir} then becomes:
\begin{equation}
S_\text{achir}^\text{het}(\Gamma) = -\frac{2}{\hbar} \Im\int dt dt_3 dt_2 dt_1 \bold R_{\mu\mu\mu\mu}\bullet (\bold E_s \otimes \bold E_3 \otimes \bold E_2 \otimes \bold E_1)
\label{eq:heterodynenonchir_short}
\end{equation}
\noindent where $\otimes$ indicates the direct product and $\bullet$ is the fully contracted product.
The chiral contribution $S_\text{chiral}^\text{het}$ is given by 
\begin{eqnarray}
&& S_\text{chiral}^\text{het}(\Gamma) =  -\frac{2}{\hbar} \Im\int dt dt_3 dt_2 dt_1 \nonumber\\
&&( \bm R_{m\mu\mu\mu} \bullet (\bold B_s\otimes \bold E_3 \otimes \bold E_2 \otimes \bold E_1) + \bm R_{q\mu\mu\mu}\bullet ( \nabla \bold E_s\otimes \bold E_3 \otimes \bold E_2 \otimes \bold E_1)\nonumber\\
&&+ \bm R_{\mu m\mu\mu} \bullet (\bold E_s\otimes \bold B_3 \otimes \bold E_2 \otimes \bold E_1) + \bm R_{\mu q\mu\mu}\bullet (\bold  E_s\otimes \nabla \bold E_3 \otimes \bold E_2 \otimes \bold E_1)\nonumber\\
&&+\bm R_{\mu \mu m\mu} \bullet (\bold E_s\otimes \bold E_3 \otimes \bold B_2 \otimes \bold E_1) + \bm R_{\mu\mu q\mu}\bullet (\bold  E_s\otimes \bold E_3 \otimes\nabla \bold E_2 \otimes \bold E_1)\nonumber\\
&&+\bm R_{\mu\mu\mu m} \bullet (\bold E_s\otimes \bold E_3 \otimes \bold E_2 \otimes \bold B_1) + \bm R_{\mu\mu\mu q}\bullet (\bold E_s\otimes \bold E_3 \otimes \bold E_2 \otimes \nabla \bold E_1))
\label{chiraltot1}
\end{eqnarray}
\noindent where we have omitted the time variable for brevity.
The matter correlation function $\bold R_{m\mu\mu\mu}$ is given by:
\begin{equation}
\bold R_{m\mu\mu\mu}(t_3,t_2,t_1) = \Big(-\frac{i}{\hbar}\Big)^3 \langle \bm m_\text{left}\mathcal G(t_3)\bm\mu_-\mathcal G(t_2)\bm\mu_-\mathcal G(t_1)\bm\mu_-\rangle
\end{equation}
\noindent with similar contributions for the other terms (magnetic dipole and the electric quadrupole).
The chiral contribution can be greatly simplified by making the slowly varying envelope approximation.
\begin{eqnarray}
\bm E_i(t) &=& \mathcal E_i(t) \ \bm \epsilon_i\\
\bm B_i(t) &=& \mathcal E_i(t) \ \frac{1}{c} u_i \bm{\hat k}_i \wedge \bm \epsilon_i = \mathcal E_i(t)\frac{u_i}{c} \bm b_i\\
\bm \nabla E_i(t) &=& \mathcal E_i(t) \ i u_i \frac{\omega_i}{c}\bm{\hat k}_i \otimes \bm \epsilon_i
\end{eqnarray}
\noindent where $\bm \epsilon_i$ is the polarization of the $i$th pulse and where $u_i = \pm 1$ depends on the chosen phase matching direction. The vectors $(u_1,u_2,u_3) = (-1,1,1), \ (1,-1,1), \ (1,1,-1)$ represent $\bold k_I, \bold k_{II}$ and $\bold k_{III}$ techniques respectively.
The chiral contribution to the signal then becomes:
\begin{multline}
S_\text{chiral}^\text{het}(\Gamma) =  -\frac{2}{\hbar c} \Im\int dt dt_3 dt_2 dt_1 
\mathcal E_s(t)\mathcal E_3(t-t_3)\mathcal E_2(t-t_3-t_2)\mathcal E_1(t-t_3-t_2-t_1)\\
\times \Big( -\bm R_{m\mu\mu\mu} \bullet (\bm F_{\bm b_s \bm \epsilon_3 \bm \epsilon_2 \bm \epsilon_1}) -i\omega_s \bm R_{q\mu\mu\mu}\bullet (\bm F_{(\bm k_s \otimes\bm \epsilon_s) \bm \epsilon_3 \bm \epsilon_2 \bm \epsilon_1})
+u_3 \bm R_{\mu m\mu\mu} \bullet (\bm F_{\bm \epsilon_s \bm b_3 \bm \epsilon_2 \bm \epsilon_1})\\
 +i\omega_3 u_3\bm R_{\mu q\mu\mu}\bullet (\bm F_{\bm  \epsilon_s (\bm k_3 \otimes \bm \epsilon_3) \bm \epsilon_2 \bm \epsilon_1})
+u_2\bm R_{\mu \mu m\mu} \bullet (\bm F_{\bm \epsilon_s \bm \epsilon_3  \bm B_2  \bm \epsilon_1}) +i\omega_2 u_2 \bm R_{\mu\mu q\mu}\bullet (\bm F_{\bm  \epsilon_s \bm \epsilon_3 (\bm k_2 \otimes \bm \epsilon_2) \bm \epsilon_1})\\
+u_1\bm R_{\mu\mu\mu m} \bullet (\bm F_{\bm \epsilon_s \bm \epsilon_3  \bm \epsilon_2  \bm b_1}) +i\omega_1 u_1\bm R_{\mu\mu\mu q}\bullet (\bm F_{\bm \epsilon_s \bm \epsilon_3  \bm \epsilon_2 (\bm k_1 \otimes \bm \epsilon_1)})\Big)
\label{chiraltot1}
\end{multline}
\noindent where we have defined $\bm F_{\bm t \bm u \bm v \bm w} = (\bm t \otimes \bm u \otimes\bm v \otimes\bm w)$.

To single out the chiral response, we consider the rotationally-averaged tensor and define cancellation schemes that eliminate the achiral contributions.
Irreducible tensor algebra is a powerful tool for accomplishing this goal, and the contraction between irreducible tensors can then be written as:
\begin{equation}
T\bullet U = \sum_{\tau JM} (-1)^M \ _\tau T^{JM} \ _\tau U^{J-M}
\label{eq:irrtensprod}
\end{equation}
\noindent where $J$ is the irreducible tensor rank and $\tau$ is the seniority index\cite{jerphagnon1978description}.
Only the $J=0$ rotational invariants survive the rotational averaging and the contraction reduces to a sum over the $J=0$ contributions.
For a rank 4 tensor constructed from direct products of rank 1 tensors, there are three rotational invariants:
\begin{eqnarray}
_0 T_{\bm A\bm B\bm C\bm D}^{J=0} &=& \{\{\bm A\otimes \bm B\}_0 \otimes \{\bm C\otimes \bm D\}_0\}_0 = \frac{1}{3}(\bm A\cdot \bm B)(\bm C\cdot \bm D) \\
_1 T_{\bm A\bm B\bm C\bm D}^{J=0} &=& \{\{\bm A\otimes \bm B\}_1 \otimes \{\bm C\otimes \bm D\}_1\}_0 = \frac{1}{\sqrt 3}(\bm A\wedge \bm B)\cdot(\bm C\wedge \bm D)\\
_2 T_{\bm A\bm B\bm C\bm D}^{J=0} &=& \{\{\bm A\otimes \bm B\}_2 \otimes \{\bm C\otimes \bm D\}_2\}_0 \nonumber\\
&=& \frac{1}{\sqrt 5}(\frac{1}{2}(\bm A\cdot \bm C)(\bm B\cdot \bm D)-\frac{1}{3}(\bm A\cdot \bm B)(\bm C\cdot \bm D)+\frac{1}{2}(\bm A\cdot \bm D)(\bm B\cdot \bm C))
\end{eqnarray}
\noindent where $A, B, C$ and $D$ can be either an electric dipole $\bm \mu$ or a magnetic dipole $\bm m$ interaction for the response function or the electric and magnetic field polarizations for the field tensor.
It follows from Eq.\ref{eq:irrtensprod} that the contraction with field tensor of the rotationally average response implies only the $J=0$ components.
For example, the first term in Eq. \ref{chiraltot1}  reduces to 
\begin{equation}
S_{\bm m\bm\mu\bm\mu\bm\mu}^\text{het}(\Gamma) =  -\frac{2}{\hbar c} \Im\int dt dt_3 dt_2 dt_1 
\Big( {_0 R_{\bm m\bm \mu\bm \mu\bm \mu}^{J=0}} \ {_0 F_{\bm b_s\bm \epsilon_3 \bm \epsilon_2\bm \epsilon_1}^{J=0}} + 
{_1 R_{\bm m\bm \mu\bm \mu\bm \mu}^{J=0}} \ {_1 F_{\bm b_s\bm \epsilon_3 \bm \epsilon_2\bm \epsilon_1}^{J=0}}
+ {_2 R_{\bm m\bm \mu\bm \mu\bm \mu}^{J=0}} \ {_2 F_{\bm b_s\bm \epsilon_3 \bm \epsilon_2\bm \epsilon_1}^{J=0}}\Big)
\end{equation}
\noindent and similar terms exist for the other permutations $\bm \mu\bm m\bm \mu\bm \mu$, $\bm \mu\bm \mu\bm m\bm \mu$ and $\bm \mu\bm \mu\bm \mu\bm m$.

The quadrupolar response involves rank 5 tensors which can only have two irreducible scalars:
\begin{eqnarray}
_1 T_{\bm A\bm B\bm C\bm Q}^{J=0} &=& \{\{\bm A\otimes \bm B\}_0 \otimes \{\bm C\otimes \bm Q\}_1\}_0 \\
_2 T_{\bm A\bm B\bm C\bm Q}^{J=0} &=& \{\{\bm A\otimes \bm B\}_1 \otimes \{\bm C\otimes \bm Q\}_2\}_0
\end{eqnarray}
The second term of Eq. \ref{chiraltot1} becomes
\begin{equation}
S_{\bm q\bm\mu\bm\mu\bm\mu}^\text{het}(\Gamma) =  \frac{2\omega_s}{\hbar c} \Re\int dt dt_3 dt_2 dt_1 
\Big({_0 R_{\bm q\bm \mu\bm \mu\bm \mu}^{J=0}} \ _0 F_{(\bm k_s\otimes\bm \epsilon_s)\bm \epsilon_3 \bm \epsilon_2\bm \epsilon_1}^{J=0} + 
_1 R_{\bm q\bm \mu\bm \mu\bm \mu}^{J=0} \ _1 F_{(\bm k_s\otimes\bm \epsilon_s)\bm \epsilon_3 \bm \epsilon_2\bm \epsilon_1}^{J=0}
\Big)
\end{equation}
Calculation of the $R^{J=0}$ and $F^{J=0}$ tensors allows to single out all the independent possible c4WM techniques and give them a clear algebraic meaning.

c4WM signals can be extracted from the total 4WM signal by cancellation of the achiral component.
CD and ROA only involved a single incoming beam so a single cancellation option was possible.
For c4WM, each of the four beams can be either left or right circularly polarized, leading to four types of c4WM signals denoted $\alpha, \beta, \gamma$ and $\delta$ for each phase matching direction ($\bold k_i = \bold k_{I}$, $\bold k_{II}$ or $\bold k_{III}$). 
Altogether, we get 12 possible schemes, see Eqs. \ref{chirtech1}-\ref{chirtech4}\cite{rouxel2020chiral}. 

The four polarization schemes are given by
\begin{eqnarray}
S_\text{chir}(\alpha) &=& S_\text{4WM}(L,L,L,L) - S_\text{4WM}(R,R,R,R)\label{chirtech1}\\
S_\text{chir}(\beta) &=& S_\text{4WM}(L,R,L,R) - S_\text{4WM}(R,L,R,L)\\
S_\text{chir}(\gamma) &=& S_\text{4WM}(L,L,R,R) - S_\text{4WM}(R,R,L,L)\\
S_\text{chir}(\delta) &=& S_\text{4WM}(L,R,R,L) - S_\text{4WM}(R,L,L,R)\label{chirtech4}
\end{eqnarray}
\noindent where the arguments of $S_\text{4WM}(e_s,e_3,e_2,e_1)$ indicate the polarization of pulses $\bm E_s$, $\bm E_3$, $\bm E_2$ and $\bm E_1$.
The other parameters (phase matching choice, beam parameters and delays, etc) are kept implicit.

X-ray 4WM techniques are still in their infancy and so are chiral 4WM techniques across the whole electromagnetic spectrum.
Current technical developments make them feasible and they will offer new ways to probe molecular chirality.
Two-dimensional spectroscopy in the IR or the optical regime has offered unique time-domain probes of energy transfer by coherence coupling between molecular excitons.
Its chiral extension will be able to follow the flux of chirality  within a molecular system, with the notable advantage of using an site- and element-specific probe for X-rays.

\section{Chiral signals in the electric dipole approximation}
\label{electricPART}

The chirality-sensitive techniques presented so far rely on the multipolar expansion truncated at the magnetic dipole / electric quadrupole level to generate pseudo-scalar quantities.
As a consequence, these signals constitute a small contribution on top of a strong achiral signal.
We now discuss chiral signals that exist even at the electric dipole level and are thus stronger.
Such signals generally involve a higher order perturbation in the incoming fields (SFG) or complex-valued electric dipoles (ionization-related signals).

\subsection{Sum-frequency-generations (XSFG)}
\label{sec:xsfg}
Even-order susceptibilities $\chi^{(2n)}$ usually vanish in centrosymmetric media such as ensembles of randomly oriented nonchiral molecules.
Most applications of these techniques use therefore molecules in anisotropic environments\cite{tang2020molecular} where the signal originates only from location where the centrosymmetry is broken, making it an sensitive probe of molecules at interfaces.
Additionally, $\chi^{(2n)}$ does not vanish in ensembles of chiral molecules which lack inversion symmetry, even upon rotational averaging.
Since the parity operation interchanges the two enantiomers, the molecular ensemble is not inversion-invariant\cite{belkin2001sum} and even-order signals can be observed in the bulk.
$\chi^{(2)}$ is finite and has been measured in randomly oriented ensembles of chiral molecules\cite{belkin2001sum}. 

Second order chiral signals such as Second Harmonic Generation (SHG), Sum Frequency Generation (SFG) and Difference Frequency Generation (DFG) do not vanish within the electric dipole approximation and do not require the magnetic dipole and electric quadrupole. 
This makes these chiral signals easier to detect since they are comparable in magnitude to their non chiral counterparts.

In the following, we discuss the SFG process obtained by an X-ray probe.
The discussion can be easily adapted to DFG or SHG.
A typical SFG pulse configuration uses an infrared pulse to excite a molecular vibrational coherence followed by an off-resonant broadband  visible pulse that stimulates the SFG process.
A similar configuration involving a visible pump and an X-ray probe to study valence electronic coherence can be introduced.
An all X-ray technique would involve very short lived double-core holes that may prove difficult to detect.
The change in the transmitted X-ray pulse intensity is recorded vs the delay between the two pulses. 

\begin{figure}[!h]
  \centering
  \includegraphics[width=1.\textwidth]{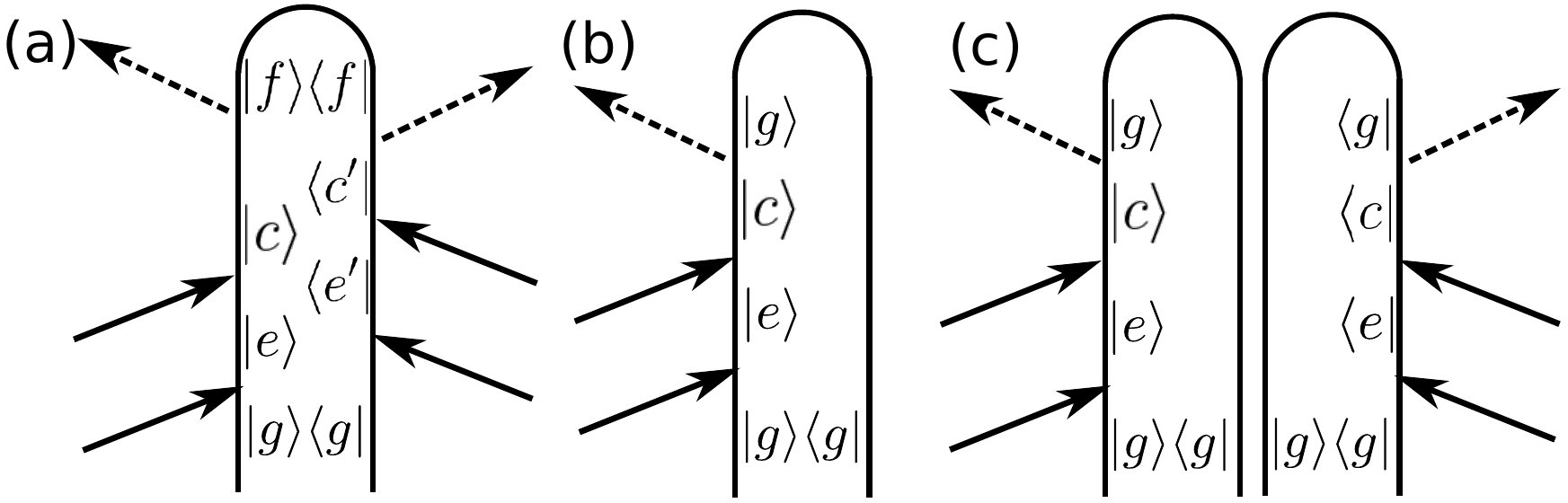}
  \caption{Loop diagrams for the SFG signals. Panels a, b and c represent the spontaneous incoherent, the stimulated and the spontaneous coherent SFG signals respectively.
\label{sfg1}}
\end{figure}

In the SFG process, Fig. \ref{sfg1}, the molecule interacts once with a pump field with frequency $\omega_\text{pu}$, once with a probe field at frequency $\omega_\text{pr}$ and finally emits a photon at the sum frequency $\omega_\text{pu}+\omega_\text{pr}$. In XSFG, $\omega_\text{pu}$ is resonant with a valence excitation and $\omega_\text{pr}$ is either a resonant or off-resonant core excitation. 
The loop diagrams representing the resonant SFG signal are given in Fig. \ref{sfg1}.
SFG experiments can be carried out in the frequency or in the time-domain. In the frequency domain, two plane waves with frequency $\omega_1$ and $\omega_2$ and wavevectors $\bm k_1$ and $\bm k_2$ are incident in a material and generate a wave at the sum frequency $\omega_s = \omega_1 + \omega_2$ in the phase matching direction $\bm k_s = \bm k_1 + \bm k_2$.

As discussed in Appendix \ref{appendixOBS}, the homodyne-detected signal can be split into coherent and incoherent components.
The incoherent signal originates from single molecules and thus scales as the number of molecules $N$.
The starting point for the signal is given by Eq. \ref{homoINC}.
In order to generate an observable photon, i.e. create a diagonal elements in the density matrix of the detected photon mode, both the bra and the ket of the molecule must undergo an SFG process, as displayed in diagram \ref{sfg1}a. 
To second order in the incoming fields, we get:
\begin{multline}
S_\text{SFG}(\omega_s, \Gamma) = \frac{2}{\hbar^6} N \mathcal E_s^2  \text{Re}\int dtdt' dt_2 dt_1 dt_2' dt_1' \ e^{i\omega_s (t-t')}\\ 
\langle 
\bm \mu_R(0)\cdot \bm E_1(t'-t_2'-t_1')
\bm \mu_R(t_1')\cdot \bm E_2(t'-t_2')
\bm \mu_R^\dagger(t_2')
\bm \mu_L(t_2)
\bm \mu_L^\dagger(t_1)\cdot \bm E_2(t-t_2)
\bm \mu_L^\dagger(0)\cdot \bm E_1(t-t_2-t_1)
\rangle \label{XSFGeq}
\end{multline}

The incoherent signal, Eq. \ref{XSFGeq}, is given in the time domain. 
The frequency domain signal is obtained by using plane wave for the fields $\bm E_1$ and $\bm E_2$: $\bm E_2(t-t_2) = \bm E_2 \ e^{-i\omega_2(t-t_2)}$ and $\bm E_1(t-t_2-t_1) = \bm E_1 \ e^{-i\omega_1(t-t_2-t_1)}$.

We now turn to the coherent signal which is phase-matched, scales as $N^2$, and can be recast as a modulus square of an amplitude.
\begin{multline}
S_\text{SFG}(\omega_s,\Gamma) = \frac{2}{\hbar^6}N^2 \text{Re}
|\int dt dt_2 dt_1 
\langle \bm \mu_L(t_2)
\bm \mu_L^\dagger(t_1)
\bm \mu_L^\dagger(0)\rangle 
\bm E_s^*(t) \bm E_2(t-t_2) \bm E_1(t-t_2-t_1)
|^2 \label{XSFGeq2}
\end{multline}
\noindent where $\bm E_s^*(t) = \mathcal E_s \bm e_s \ e^{i \omega_s t}$ is the spontaneous field emitted from the SFG process in the direction $\bm k_s = \bm k_1 + \bm k_2$.
Finally, the heterodyne SFG signal is given by:
\begin{multline}
S_\text{SFG}^\text{het}(\omega_s) = \frac{2}{\hbar^6} \text{Im}
\int dt dt_2 dt_1 
\langle \bm \mu_L(t_2)
\bm \mu_L^\dagger(t_1)
\bm \mu_L^\dagger(0)\rangle 
\bm E_\text{het}^*(t) \bm E_2(t-t_2) \bm E_1(t-t_2-t_1) \label{XSFGeq3}
\end{multline}

The SOS expression for Eq. \ref{XSFGeq2} is
\begin{equation}
S_\text{SFG}^\text{hom}(\omega_1,\omega_2) = \frac{2}{\hbar^6} N^2 E_s^2 E_2^2 E_1^2 \text{Re} 
|\sum_{e,c} \frac{
(\bm \mu_{gc}(\bold k_s)\cdot \bm \epsilon_s)
(\bm \mu_{ce}(\bold k_2)\cdot \bm \epsilon_2)
(\bm \mu_{eg}(\bold k_1)\cdot \bm \epsilon_1)
}{(\omega_1+\omega_2 - \omega_{cg}+i\Gamma_{cg})(\omega_1-\omega_{eg}+i\Gamma_{eg})}|^2
\end{equation}

The nonlinear response function introduced in Eqs. \ref{XSFGeq} to \ref{XSFGeq3} vanishes in an isotropic achiral ensemble.
Rotational averaging of the third rank response function in Eqs. \ref{XSFGeq2} and \ref{XSFGeq3} is obtained by contraction with the $I^{(3)}$ tensor
\begin{equation}
\langle \bm \mu_{i_1} \bm \mu_{i_2} \bm \mu_{i_3}\rangle_\Omega = \frac{1}{6} \epsilon_{i_1i_2i_3}\epsilon^{\lambda_1 \lambda_2 \lambda_3} \bm \mu_{\lambda_1} \bm \mu_{\lambda_2} \bm \mu_{\lambda_3}
= \frac{1}{6} \epsilon_{i_1i_2i_3} \ \bm\mu\cdot(\bm \mu \times \bm \mu)
\label{rotavSFG}
\end{equation}
\noindent where we have omitted the time arguments for brevity. The contraction of the Levi-Civita tensor and the response tensor give $\epsilon^{\lambda_1 \lambda_2 \lambda_3}
\bm \mu_{\lambda_1} \bm \mu_{\lambda_2} \bm \mu_{\lambda_3}=\bm\mu\cdot(\bm \mu \times \bm \mu)$, where the indices $i_1, i_2$ and $i_3$ get contracted with the electric field components. 
Thanks to the triple product, this averaging leads to a pseudo-scalar that is thus sensitive to chirality in isotropic media.
Orientationally-averaged second order signals provide an excellent probe for molecular chirality. They are relatively strong compared to other chiral-sensitive signals since they are of low order both in the perturbative expansion in the exciting fields and in the multipolar expansion.

Using Eq. \ref{rotavSFG} in Eq. \ref{XSFGeq3} allows to write the signal as an overlap integral in time-domain between chiral matter and field responses:
\begin{equation}
S_\text{SFG}(\omega_s) = \frac{2}{\hbar^6} \text{Im}
\int dt dt_2 dt_1 
\langle \bm \mu_L(t_2)\cdot
(\bm \mu_L^\dagger(t_1)\times
\bm \mu_L^\dagger(0))\rangle 
\bm E_\text{het}^*(t)\cdot( \bm E_2(t-t_2) \times\bm E_1(t-t_2-t_1)) \label{XSFGeq4}
\end{equation}

All even-order chiral techniques similarly involve an overlap between chiral response functions and a chiral field that does not vanish in the electric dipole approximation.
This observation has led  to the introduction
\cite{koroteev1993novel,koroteev1995biocars,
fischer2000three,fischer2005nonlinear}
of nonlinear chiral techniques in the frequency domain and in the optical regime.
Since SHG is forbidden even in optically active liquids because of the symmetry of the susceptibility over its last two indices, one must rather consider SFG and DFG spectroscopies\cite{koroteev1993novel}.
Higher even order nonlinear signals have been considered\cite{shkurinov1993second} as well.

Recently, Smirnova et. al\cite{ayuso2019synthetic,ayuso2018locally,ordonez2018generalized} 
have proposed an approach that makes further use of the even order nonlinear susceptibilities to detect molecular chirality.
Their approach goes as follow:
the rotational averaging in Eq. \ref{rotavSFG} introduces a Levi-Civita symbol acting on the incoming fields.
This means that, at second-order perturbation in the electric dipole Hamiltonian, the field tensor reduces to a triple product of the electric field, which is a pseudoscalar.

The  field correlation functions appearing in Eq. \ref{XSFGeq4} were introduced as a hierarchy of field pseudo-scalars $h^{(n)}$. 
For example, in the frequency domain, the field correlation in Eq. \ref{XSFGeq2} is
\begin{equation}
h^{(3)}(\omega_1,\omega_2) = 
\bm E_s^*(\omega_1 + \omega_2)\cdot(\bm E_2(\omega_2) \times\bm E_1(\omega_1))
\label{h3}
\end{equation}
The $h^{(3)}$ field correlation function appears in the chiral-sensitive $\chi^{(2)}$ signals.
Since $\chi^{(2)}(2\omega, \omega,\omega)$ which is responsible for SHG vanishes, the SFG or DFG processes are mandatory to use $h^{(3)}$ as a probe of molecular chirality.
Smirnova et al. proposed to used higher even-order response functions for which the HHG process linked to $h^{(2n+1)}$ no longer vanish.
For example, the lowest non-vanishing order probing the $\chi^{(4)}$ nonlinear susceptibility is $h^{(5)}(2\omega,\omega)$ field correlation function given by:
\begin{equation}
h^{(5)}(2\omega,\omega) = 
\bm E^*(2\omega)\cdot(\bm E(\omega) \times\bm E(\omega))(\bm E(\omega) \cdot\bm E(\omega))
\label{h5}
\end{equation}

\begin{figure}[!h]
  \centering
  \includegraphics[width=0.4\textwidth]{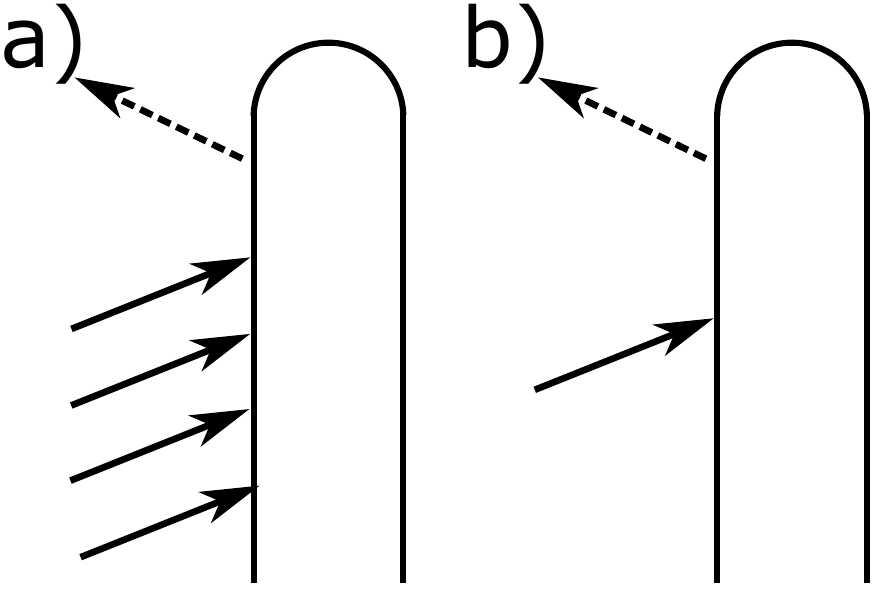}
  \caption{Loop diagrams of the two interfering pathways leading to the observation of the $\chi^{(4)}$ signal.
\label{sfg2}}
\end{figure}

A simple experimental setup was proposed to probe these chiral nonlinear responses by using two non-collinear fields, each of them made of fields with central frequency $\omega$ and $2\omega$ incident on a gas target of chiral molecules\cite{ayuso2019synthetic}.
The practical implementation of this excitation scheme requires the handedness of the field to be maintained within the interaction region thus introducing constraints on the phases of the incoming pulses.
Probing a chiral sample with a varying field ellipticity within the interaction region leads to a vanishing of the chiral signal.
The setup demonstrated by Ayuso et al.\cite{ayuso2019synthetic} uses two noncollinear pulses and thus generate an interference grating within the sample. 
Both the intensity and the ellipticity of the interference field are then modulated.
By a careful control of the incoming field phases, both gratings can be superimposed to ensure that intensity maxima correspond to the same ellipticity.
This has been coined as locally and globally chiral field\cite{ayuso2019synthetic}.
Spontaneous and coherent signals then originate from an interference term between the even order chiral amplitudes and a lower odd order achiral amplitude, see Fig. \ref{sfg2}a and b: $|\chi^{(4)} \bm E^*(\omega)\times \bm E(\omega)(\bm E(\omega)\cdot \bm E(\omega))+ \chi^{(1)} \bm E(2\omega)|^2 \sim \chi^{(4)} \chi^{(1)} h^{(5)}$.
This process can also be considered within the context of HHG emission (see section \ref{sectionCHHG}) where the incoming strong fields generate all even-order responses in parallel.
This approach has led to the measurement of giant asymmetry ratio (up to 200\%) when combined with an HHG measurement (chiral HHG is discussed in section \ref{sectionCHHG}).

So far, chiral SFG has been limited to visible or IR incoming pulses.
HHG detection has extended these concepts from this field into the EUV regime but little has been done with EUV or X-ray incoming pulses.
X-rays can offer a window to local chirality through its element sensitivity.

\subsection{Photo-electron circular dichroism}

In the previous section, we have demonstrated how the observation of even-order harmonics in isotropic ensembles is a signature of molecular chirality.
These signals originate from purely electric dipole coupling which makes them strong compared to their magnetic counterparts.
Another possibility to construct chiral signals within the electric dipole approximation is by using complex valued electric dipole transitions, that are typically present in ionization processes\cite{epifanovsky2013new,guo2021retrieval}.

As the incident photon energy is increased, it becomes possible to photoionize one of the core electrons via a bound to continuum transition involving a single or multiple photons.
Multiphoton ionization processes can be induced by visible or UV circularly polarized light\cite{barth2011nonadiabatic,lehmann2013imaging}. 
Single photon photoionization occur in the deep UV or soft X-ray regime and offer element-selectivity as routinely used in X-ray Photoelectron Spectroscopy (XPS)\cite{van2011x,schillinger2007probing}.
Using synchrotron or tabletop light sources, it has been shown that Angular Resolved PhotoElectron Spectroscopy (ARPES) displays an asymmetry between two enantiomers or ionization of chiral or aligned molecules with left and right polarized light\cite{dubs1985circular,dubs1986circular,garcia2003circular}.
The asymmetry of ARPES is the basis for the PhotoElectron Circular Dichroism (PECD) technique\cite{janssen2014detecting,lux2012circular}. 
Numerical calculations demonstrated that around 10\% asymmetry ratio, Eq. \ref{eq:pecdasym}, could be reached\cite{powis2008photoelectron}.
Much higher asymmetry ratios up to 100\% have also been reported\cite{fehre2021fourfold}.


A detailed review of PECD at synchrotrons has been written by Powis\cite{powis2008photoelectron}.
Here, we only survey the fundamental aspect of the theory and then introduce recent developments in time-resolved measurements.
The PECD signal is defined by \cite{powis2008photoelectron,harvey2018general}
\begin{equation}
S_\text{PECD}(E,\theta) = \frac{S_\text{PE}(E,\theta,\bold e_L)-S_\text{PE}(E,\theta,\bold e_R)}{S_\text{ARPES}(E,\theta)}
\label{eq:pecdasym}
\end{equation}
\noindent where $E$ is the ionized electron energy and $\theta$ is its  scattering angle with respect to the incident photon. 
Similarly to CD and ROA, PECD is a good signature of molecular chirality when considering randomly-oriented ensemble.
Otherwise, the signal may not vanish in oriented achiral samples and can also be of interest.
This normalization can lead to important fluctuation in angular region with low ARPES signals and, sometimes $ \text{max}(S_\text{ARPES}(E,\theta))$ is preferred.

Before discussing PECD, we briefly summarize how photoemission signals can be computed. 
So far, signals in this review were defined by the integrated change of photon number, see appendix B. However, the observable current is linked to the integrated change of electron number on the detector, that is:
\begin{equation}
S_\text{PE}(E,\theta,\bold e_L) = \int dt \ \langle \frac{d}{dt} N_p \rangle 
\end{equation}
\noindent where $N_p$ is the electron number operator for electron with momentum $\bm p$.
The procedure used in appendix \ref{appendixOBS} to express the signal in terms of a matter observable can be repeated by using the minimal coupling interaction Hamiltonian:
\begin{equation}
H_\text{int}(t) = - 
\bm p \cdot \bm A_\text{x}(t)
\end{equation}
\noindent where we use the $\bm p\cdot \bm A$ coupling and $\bm p = \sum_j \bm p_j$ is the many-body momentum operator.
Since the free-electron field is initially in the vacuum state, the photoelectron signal requires expansion to one additional order in the interaction Hamiltonian and can be expressed as a homodyne coherent signal:
\begin{eqnarray}
S_\text{PE}(\bm p,\bold e) &=&  \frac{2}{\hbar^2} 
\int dt dt' \bm A(t)^* \bm A(t') \langle \bm p(t) \bm p(t') \rangle_\Omega\label{PEdef}\nonumber\\
&=&  \frac{2}{\hbar^2}| \int dt  \ \bm A(t)^* 
\langle ^{N-1}\Psi(t)\otimes \Psi_\text{el}^p(t)|\bm p|^{N}\Psi(t)\rangle_\Omega|^2
\end{eqnarray}
\noindent where $\Omega$ indicates rotational averaging over the molecular degrees of freedom and $\bm e$ is the incoming electric field polarization.
Simulation of PE signals requires the evaluation of matrix elements $\langle ^{N-1}\Psi\otimes \Psi_\text{el}^p|\bm p(t)|^{N}\Psi\rangle$  between the non-ionized molecular states and the ionized states.
The ground state is a standard linear combination of Slater determinants over molecular orbitals.
The final state is a direct product of the $N-1$ molecular eigenstate and the quasi-free electron wavefunction $\Psi_\text{el}^p$.

Dyson orbitals\cite{melania2007dyson,oana2009cross} can be effectively used to represent these matrix elements. Expressing explicitly the matrix element as a function of the many-body coordinates, we have:
\begin{multline}
\langle ^{N-1}\Psi\otimes \Psi_\text{el}^p|\bm p|^{N}\Psi\rangle
= \sum_{ij} \frac{(-1)^{i+1}}{\sqrt N} \int d\bm x_1 ... d \bm x_N 
\ ^{N-1}\Psi(x_1...\{\bm x_i\} ... \bm x_N)\Psi_\text{el}^p(\bm x_i) \bm p_j \ ^{N}\Psi(x_1..\bm x_N)\\
= \sum_{ij,i\neq j}\frac{(-1)^{i+1}}{\sqrt{N}} 
\int d\bm x_1 ... d \bm x_N \ ^{N-1}\Psi_\text{I}(\bm x_1...\{\bm x_i\} ... \bm x_N)^*\Psi_\text{el}^p(\bm x_i)^* \bm p_j \ ^{N}\Psi(\bm x_1..\bm x_N)\\
+  \sum_{ij,i= j}\frac{(-1)^{i+1}}{\sqrt{N}} 
\int d\bm x_1 ... d \bm x_N \ ^{N-1}\Psi_\text{I}(\bm x_1...\{\bm x_i\} ... \bm x_N)^*\Psi_\text{el}^p(\bm x_i)^* \bm p_i \ ^{N}\Psi(\bm x_1..\bm x_N)
\label{matrixelem}
\end{multline}
In Eq. \ref{matrixelem}, the sum for $i\neq j$ corresponds to an Auger process: the light-matter interaction occurs on the orbital $j$ through the matrix elements of $\bm p_j$ and then orbital $i$ is ionized. We neglect this contribution in the following and only consider the second term for $i=j$ for photoemission: the ionized orbital is the one that interacts with the incoming field.
Using the Dyson orbitals defined as
\begin{equation}
\Phi^d(\bm x) =\sqrt{N}\int d\bm x_1... \{d\bm x\}...d\bm x_N \  ^{N-1}\Psi_\text{I}(\bm x_1...\{\bm x\} ... \bm x_N)^* \ ^N\Psi(\bm x_1...\bm x_N), 
\end{equation}
\noindent the matrix element can be recast as a single electron matrix element between the ionized electron wavefunction and the molecular Dyson orbital:
\begin{equation}
\label{dyson_eq}
\langle ^{N-1}\Psi\otimes \Psi_\text{el}^p|\bm p|^{N}\Psi\rangle = 
\int d\bm x \ \Psi_\text{el}^p(\bm x)^* \bm p \Phi^d(\bm x ) = \langle \Psi_\text{el}^p|\bm p|\Phi^d\rangle 
\end{equation}
When completely neglecting the molecular potential for the free electron, its wavefunction becomes a plane wave $e^{i \bm p\cdot \bm x}$ and the photoemission signal becomes the square of the Fourier transform of the Dyson orbital. 
However, approximating the ionized electron as a plane wave does not capture the chiral response of the molecular ensemble and a corrected approximation must be used in order to take in account the disturbance due to the chiral molecular potential\cite{gozem2015photoelectron}.
For an electron with momentum $\bm p$ in a spherical potential, the Schrodinger equation can be exactly solved and the eigenfunction are:
\begin{equation}
\Psi_p^\text{CW}(\bm r)= \sum_{lm} i^l R_{pl}(pr)\ Y^*_{lm}(\hat p) Y_{lm}(\hat r)
\end{equation}
\noindent where $R_{pl}(pr)$ is the radial Coulomb wave function that can be expressed as an $_1 F_1$  hypergeometric function\cite{gozem2015photoelectron}.
To account for the non spherical potential, the expansion is written in a similar manner:
\begin{equation}
\Psi_p(\bm r) = \sum_{lm} i^l \ Y^*_{lm}(\hat p) f_{lm}^{\bm p}(\bm r)
\end{equation}
The matrix element $\bm A(t)\cdot \langle \Psi_p|\bm p|\Phi^d\rangle$ now assumes the form
\begin{equation}
\bm A(t)\cdot \langle \Psi_p|\bm p|\Phi^d\rangle = A(t) \bm \epsilon \cdot \sum_{lm} i^l Y_{lm}^*(\hat p)\bm{\mu}_{lm}^p
\end{equation}
\noindent where
\begin{equation}
\bm{\mu}^{plm} = \int d\bm r f_{lm}^{\bm p}(\bm r) (i\hbar \nabla)\Phi^d(\bm r)
\end{equation}
All quantities can be expressed in the irreducible basis and the rotational averaging is then carried using the Wigner $\mathcal D$ matrices, leading to:
\begin{equation}
\langle\bm A(t)\cdot \langle \Psi_p|\bm p|\Phi^d\rangle\rangle_\Omega = \sum_{\sigma, lm,\mu s}(-1)^\sigma \epsilon_{-\sigma} i^l Y^*_{l\mu}(\hat p) \mu_s^{plm} \mathcal D_{m\mu}^{l*}(\Omega)\mathcal D_{\sigma s}^{1}(\Omega)
\end{equation}
By integrating over all angles and using some angular algebra, we can express the signal in the following form:
\begin{equation}
S_\text{PE}(\bm p,\bm e_i) = -\frac{4\pi}{\hbar^2}\sum_{j=0}^2 B_{pij}  \ P_j(\cos \theta_p)  
\end{equation}
\noindent where $\theta_p$ is the angle between the photoionized electron momentum and the $z$ axis, $P_j$ is a Legendre polynomial and the coefficients $B_{pij}$ are defined by:
\begin{equation}
B_{pij} = \sum_{\substack{lms\\ l'm's'}}\delta_{s-s',m-m'} 
(-1)^{1-s'-m+l+l'+j} i^{l-l'}
\frac{\sqrt{(2l+1)(2l'+1)}}{2j+1}
C_{1-i1i}^{j0}C_{1-s'1s}^{js-s'}C_{l'm'l-m}^{jm'-m}
\mu_s^{plm}\mu_{s'}^{p'l'm'}
\label{eq:defB}
\end{equation}
Using the symmetry properties of the $C_{1-i1i}^{j0}$ Clebsch-Gordan coefficient, we obtain the following relationships:
\begin{eqnarray}
B_{p01} &=& 0\\
B_{p11} &=& -B_{p-11}\\
B_{p\pm 12} &=& -B_{p02}
\end{eqnarray}
This indicates that 1) no $P_1(\cos \theta_p)$ term can be observed with linear polarization, 2) the coefficient of $P_1(\cos \theta_p)$ changes sign for opposite circular polarization and 3) the coefficients of $P_2(\cos \theta_p)$ have the same value for opposite circular polarization.

It can also be shown that $B_{p11}\neq 0 \longrightarrow  \mu_s^{plm} \neq \mu_{-s}^{plm}$, indicating that $B_{p11}$ is non vanishing only for non-centrosymmetric systems, i.e. chiral molecules.
Hence, the numerator of the PECD signals depends only on $S_\text{PE}(\bm p,\bm e_L) - S_\text{PE}(\bm p,\bm e_R) = 2 B_{p11} P_1(\cos \theta_p)$ and the denominator becomes $(S_\text{PE}(\bm p, \bm e_L)+S_\text{PE}(\bm p, \bm e_R))/2 = (2 B_{p10} P_0(\cos \theta_p)+2 B_{p12} P_2(\cos \theta_p))/2$. The normalized PECD signal can be finally written as:
\begin{equation}
S_\text{PECD}(\bm p) = \frac{2 B_{p11} P_1(\cos \theta_p)}{B_{p10} P_0(\cos \theta_p)+B_{p12} P_2(\cos \theta_p)}
\label{eq:pecd1}
\end{equation}


Molecular chirality is embedded in the coefficient $B_{p11}$, Eq. \ref{eq:defB}, and appears both in the spatial profile of the Dyson orbital $\Phi^d(\bm r)$ and in the multipolar expansion of the ionized electron wavefunction.
Thus, the asymmetry in PECD signals originate both from the bound and the continuum states chirality.
At high momenta, the photoelectron escapes quickly the molecular ion potential and is not impacted by its asymmetry.
This consideraction indicates that PECD should vanish at high energy, and different mechanisms have been proposed to account for PECD observations in the X-ray domain\cite{hartmann2019recovery,suzuki2020plane}.
Numerous demonstrations of PECD have been reported both in the EUV\cite{hergenhahn2004photoelectron,turchini2013conformational, ferre2015table,turchini2017conformational} and soft X-ray\cite{ulrich2008giant,ilchen2017emitter,kaiser2020angular} regime.

Time-resolved PECD (tr-PECD) with X-rays is now becoming possible with the availability of XFEL sources with polarization control. 
Ultrafast measurements of PECD are also being developed using sources from the IR to the EUV regimes\cite{beaulieu2016probing}.
The asymmetry in tr-PECD can originate from the nuclear chiral geometry and from electronic chiral currents\cite{ordonez2021propensity}.
tr-PECD has recently been measured on photoexcited fenchone at the C K-edge at FERMI\cite{facciala2020core}.
It has also been used to investigate the chiral fragment of trifluoromethyloxirane upon photolysis with an X-ray pump at 698eV\cite{schmidt2020ultrafast,ilchen2021site}.
While tr-PECD demonstrations are still scarce, many ultrafast achiral photoelectron spectroscopies have been reported\cite{al2015ultrafast,chergui2017photoinduced,suzuki2019ultrafast} and tr-PECD with femtosecond resolution is going to become increasingly available.

There are three ways to measure the dichroism by ultrafast time-resolved PECD: the differential photoemission can be obtained on the ionizing probe polarization (like in static PECD), on the pump polarization (photoemission of the asymmetrically populated excited state), or both.
\begin{equation}
S_\text{tr-PECD,I}(\bm p, T) = 
\frac{S_\text{tr-PE}(\bm p,T, \bm e_\text{pr}=\bm e_L, \bm e_\text{pu})-S_\text{tr-PE}(\bm p,T, \bm e_\text{pr}=\bm e_R, \bm e_\text{pu})}
{(S_\text{tr-PE}(\bm p,T, \bm e_\text{pr}=\bm e_L, \bm e_\text{pu})+S_\text{tr-PE}(\bm p,T, \bm e_\text{pr}=\bm e_R, \bm e_\text{pu}))/2}
\label{eq:pecdI}
\end{equation}
\begin{equation}
S_\text{tr-PECD,II}(\bm p, T) = 
\frac{S_\text{tr-PE}(\bm p,T, \bm e_\text{pr}, \bm e_\text{pu}=\bm e_L)-S_\text{tr-PE}(\bm p,T, \bm e_\text{pr}, \bm e_\text{pu}=\bm e_R)}
{(S_\text{tr-PE}(\bm p,T, \bm e_\text{pr}, \bm e_\text{pu}=\bm e_L)+S_\text{tr-PE}(\bm p,T, \bm e_\text{pr}, \bm e_\text{pu}=\bm e_R))/2}
\label{eq:pecdII}
\end{equation}
\begin{equation}
S_\text{tr-PECD,IIIa}(\bm p, T) = 
\frac{S_\text{tr-PE}(\bm p,T, \bm e_\text{pr}=\bm e_L, \bm e_\text{pu}=\bm e_L)-S_\text{tr-PE}(\bm p,T, \bm e_\text{pr}=\bm e_R, \bm e_\text{pu}=\bm e_R)}
{(S_\text{tr-PE}(\bm p,T, \bm e_\text{pr}=\bm e_L, \bm e_\text{pu}=\bm e_L)+S_\text{tr-PE}(\bm p,T, \bm e_\text{pr}=\bm e_R, \bm e_\text{pu}=\bm e_R))/2}
\label{eq:pecdIIIa}
\end{equation}
\begin{equation}
S_\text{tr-PECD,IIIb}(\bm p) = 
\frac{S_\text{tr-PE}(\bm p,T, \bm e_\text{pr}=\bm e_L, \bm e_\text{pu}=\bm e_R)-S_\text{tr-PE}(\bm p,T, \bm e_\text{pr}=\bm e_R, \bm e_\text{pu}=\bm e_L)}
{(S_\text{tr-PE}(\bm p,T, \bm e_\text{pr}=\bm e_L, \bm e_\text{pu}=\bm e_R)+S_\text{tr-PE}(\bm p,T, \bm e_\text{pr}=\bm e_R, \bm e_\text{pu}=\bm e_L))/2}
\label{eq:pecdIIIb}
\end{equation}
For an achiral system in the ground state, scheme I (Eq. \ref{eq:pecdI}) vanishes if the pump pulse is linearly polarized and it is thus more suitable for measuring the dynamics in chiral systems.
On the other hand, the use of a chiral pump (scheme II and III, Eqs. \ref{eq:pecdII}-\ref{eq:pecdIIIb}) is adequate to study achiral systems that acquire chirality in their excited state.
Chiral pumping favors the excitation of a given enantiomers in the excited PES.  

The combination of PECD measurements with quantum control schemes is another intriguing possibility.
Goetz et al.\cite{goetz2019quantum} have demonstrated the possibility to optimize the PECD asymmetry ratio in the UV regime.
Such scheme can also be envisioned in the time-domain by using an optimization routine on a pump pulse followed by a tr-PECD as a probe.

\subsection{Chiral High-Harmonic-Generation}
\label{sectionCHHG}

High Harmonic Generation is a physical process in which a photoionized electron gets accelerated by the ionizing field and then recollide with the parent ion, thus generating photons at high harmonics of the initial ionizing photon.
It has been a very prolific field in the past decades\cite{li2020attosecond} and can be used as a source of XUV light. 
HHG sources can reach the attosecond regime and have the advantage of being tabletop techniques.
Their energy limitation is constantly evolving with sources now reaching the carbon ($\sim$300 eV) and oxygen ($\sim$500eV) K-edges\cite{schmidt2018high,li201753}.
These sources can readily be used as an incident field for many of the ultrafast techniques discussed in this review.
One limitation is the generation of CPL HHG light necessitates the use of elliptically polarized driving pulses which dramatically reduces the HHG efficiency\cite{moller2012dependence}.
Circularly polarized HHG pulses have been successfully  produced and constant progress is being made towards higher quality and higher fluences of CPL \cite{kfir2016helicity,heslar2019conditions,
mauger2016circularly,bandrauk2016circularly,
chen2019circularly,he2022dynamical}.

Recently, it has been observed that the HHG process itself has different yields for opposite enantiomers
\cite{smirnova2015opportunities,neufeld2019ultrasensitive,
harada2018circular,wang2017high,neufeld2018optical,
cireasa2015probing,ayuso2021giant}.
The yield difference in the high harmonics intensities allows to define another observable of molecular chirality, chiral HHG (cHHG).
cHHG often relies on a bi-chromatic $\omega-2\omega$ strong field to drive the HHG process introduced in section \ref{sec:xsfg}, and displayed in Fig. \ref{sfg2}.
Using the same type of driving field, the Cohen group has analyzed the HHG from molecules with multiple stereogenic center with a deep-learning algorithm\cite{neufeld2022detecting}.

We first briefly review the standard approach to calculate achiral HHG signals, give generalized expression for HHG from molecules and then discuss the aspects specific to molecular chirality.
Spontaneous coherent HHG signals can be expressed as an amplitude squared:
\begin{equation}
S_\text{HHG}(\omega) \propto |\int dt e^{i\omega t}\langle \mu(t)\rangle|^2
\label{eq:defHHG}
\end{equation}

In the three-step process model propose by Lewenstein et al., the material correlation function is expanded to first order in the incoming field.
This interaction corresponds to the ionization of an electron while the final transition electric dipole corresponds to its recombination.
\begin{figure}[!h]
  \centering
  \includegraphics[width=0.6\textwidth]{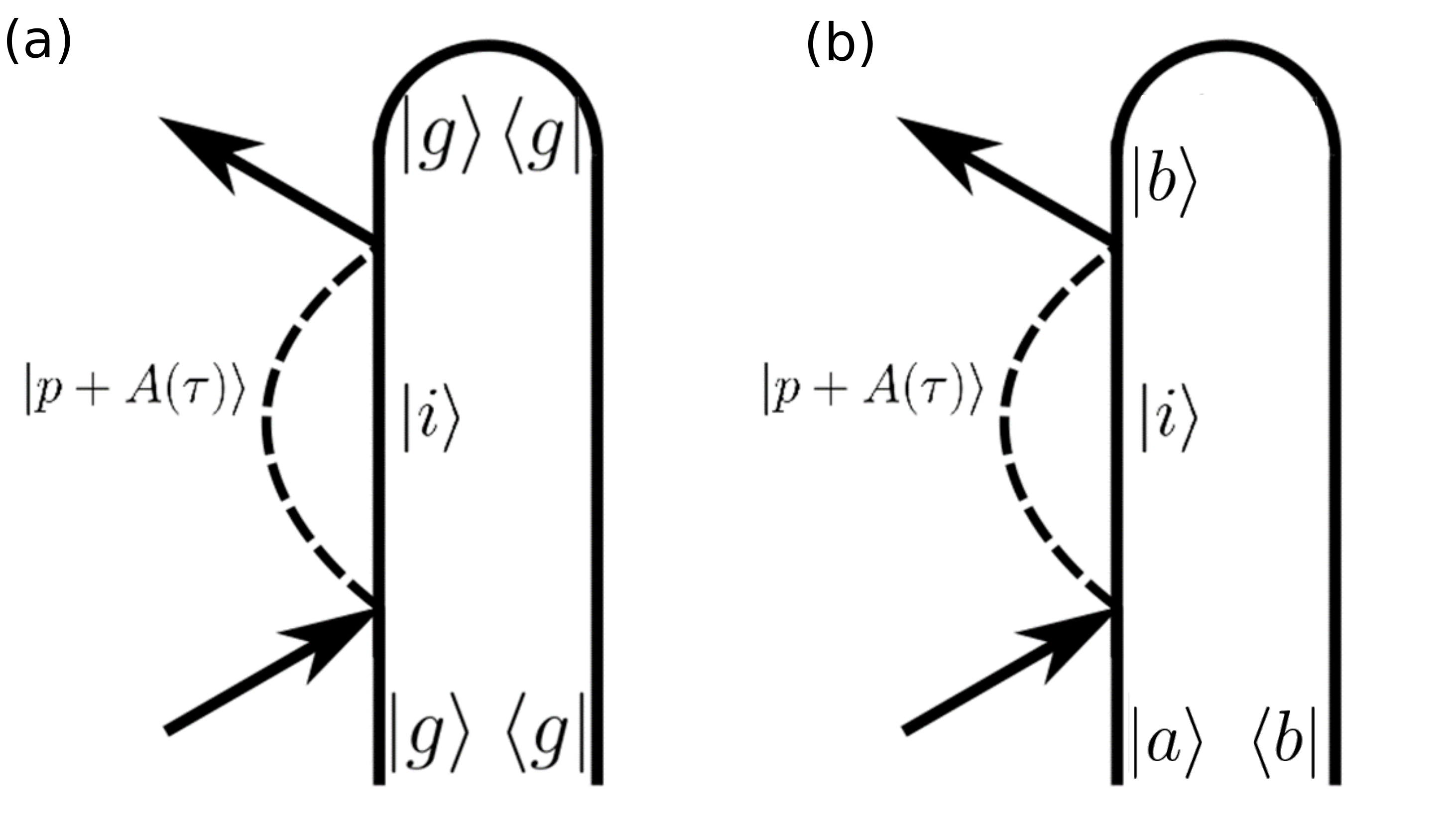}
  \caption{Loop diagrams for the HHG process. a) Diagram for an atom or molecule initially in the ground state. b) Diagrams for a molecule in a excited population ($a=b \neq g$) or coherence ($a\neq b$) prior to interaction with the pulse triggering the HHG process. 
\label{hhg1}}
\end{figure}

In HHG the electron between the two events is driven by the strong ionizing pulse.
Thus, we introduce the field-free propagator $U_0(t',t_0)$ and the strong field propagator $U_v(t,t')$.
Following the diagram in Fig. \ref{hhg1}, the matter correlation function becomes:
\begin{equation}
\langle \bm \mu(t) \rangle = 2\text{Im}\int_{t_0}^t dt'
\langle\psi(t_0)|U_0(t,t_0)^\dagger \bm \mu U_v(t,t')(-\mu\cdot \bm E(t'))U_0(t',t_0)
|\psi(t_0)\rangle
\label{hhg_atom}
\end{equation}
The propagation with $U_v(t,t')$ can be solved numerically or analytically if some simplifying assumptions are made.
Further analytical developments are achieved within the strong field approximation that neglects the atomic or molecular potential on the free electron and using a plane wave basis for it. 
$U_v(t,t')$ can be calculated as a Volkov propagator\cite{lewenstein2008principles}:
\begin{equation}
U_v(t,t')|\bm p + \bm A(t')\rangle = e^{-i\frac{1}{2}\int_{t'}^t d\tau (\bm p + \bm A(\tau)^2)} |\bm p + \bm A(t)\rangle
\end{equation}
\noindent where $\langle r |\bm p + \bm A(t)\rangle = e^{i(\bm p + \bm A(t))\cdot \bm r}$.

Assuming that the atom is initially in its ground state $g$ and introducing the ionization energy $I_p = E_g^N - E_g^{N-1}$, the signal can be recast as
\begin{equation}
S_\text{HHG}(\omega) \propto 
\Big| 2\text{Im}\int_{t_0}^t dt'\int d\bm p dt e^{i \omega t}e^{-i S(p,t,t')}
\langle\psi_g| \bm \mu| \bm p + \bm A(t)\rangle  
\langle \bm p + \bm A(t')|\bm \mu|g\rangle \cdot \bm E(t')
\Big|^2 
\end{equation}
\noindent where $S(\bm p,t,t') = 1/2\int_{t'}^t d\tau (\bm p + \bm A(\tau)^2) + I_p(t-t')$.
This highly oscillatory integral can prove challenging to calculate numerically.
This problem has been solved by Lewenstein et al. by using a saddle point analysis\cite{lewenstein1994theory}.


In molecules, the ionized system has its own dynamics that potentially interacts with the ionized electron.
Additionally, the molecular wavefunction is now a many-body wavefunction and Eq. \ref{hhg_atom} becomes
\begin{equation}
\label{hhg_molecule1}
\langle \bm \mu(t) \rangle = 2\text{Im}\int_{t_0}^t dt'
\langle\psi^N(t_0)| U_0(t,t_0)^\dagger \bm \mu  U_I(t,t')\otimes U_v(t,t') \bm \mu U_0(t',t_0) |\psi^N(t_0)\rangle
\end{equation}
\noindent with $|\psi^N(t_0)\rangle = \sum_a | \chi_a^N(t_0)\rangle 
| \varphi_a^N\rangle$ and $|\psi^{N-1}(t)\rangle = \sum_i | \chi_i^{N-1}(t)\rangle | \varphi_i^{N-1}\rangle$ are the neutral and ionized molecule wavefunctions respectively. 
Note that $|\psi^N(t_0)\rangle$ is not necessarily the molecular ground state and generating a HHG process on a molecule excited by an actinic pump is an interesting possibility to optimize the HHG yield or the chiral asymmetry.
Eq. \ref{hhg_molecule1} is obtained within the strong field approximation for the free electron and thus neglects its interaction with the ionized molecules.
Expanding in the adiabatic states, see section \ref{sec:simulation}, the molecular HHG signal becomes
\begin{multline}
\langle \bm \mu(t) \rangle = 2\text{Im} \sum_{iab} \int_{t_0}^t dt'\int d\bm p
\langle \chi_b^N(t) \varphi_b^N |\bm \mu | \chi_i^{N-1}(t) \varphi_i^{N-1}\bm p+\bm A(t)\rangle\\
\times e^{i S_{ab}(\bm p,t,t')}
\langle \chi_i^{N-1}(t')\varphi_i^{N-1} \bm p + \bm A(t')| \bm \mu | \chi_a^N(t') \varphi_a^N\rangle\cdot \bm E(t')
\end{multline}
\noindent where $S_{ab}(\bm p,t,t') = 1/2\int_{t'}^t d\tau (\bm p + \bm A(\tau)^2) + (E_b^N - E_i^{N-1})t- (E_a^N - E_i^{N-1})t'$.
Finally, by writing explicitly the integrals over the normal coordinates and using the definition of the Dyson orbitals, see Eq. \ref{dyson_eq}, we get 
\begin{multline}
\langle \bm \mu(t) \rangle = 2\text{Im} \sum_{iab} \int_{t_0}^t dt'\int d\bm p d\bm R d\bm R'
\chi_b^{N*}(\bm R,t)\chi_i^{N-1}(\bm R,t)
\chi_i^{N-1,*}(\bm R',t')\chi_a^N(\bm R',t')\\
\times
\langle  \Phi_{bi}^d(\bm R) |\bm \mu(\bm R) | \bm p+\bm A(t)\rangle e^{i S_{ab}(\bm p,t,t')}
\langle \bm p + \bm A(t')| \bm \mu |  \Phi_{ia}^d(\bm R')\rangle\cdot \bm E(t')
\label{eq:hhgcalc}
\end{multline}
Eqs. \ref{eq:defHHG} to \ref{eq:hhgcalc} can be used to compute HHG spectra from a molecular system.
In particular, Eq \ref{eq:hhgcalc} is suitable for molecules experiencing nonadiabatic nuclear dynamics.
Using circular polarization for the driving strong field, opposite enantiomers display an asymmetry in the HHG yield.
Similarly to PECD, the asymmetry can take its origin in the bound state geometry through the Dyson orbitals and in the ionized electron propagator. 
The former is accounted for in the discussed formalism while the latter would require a more accurate propagator for the electron than the Volkov one and would include the asymmetric Coulombic potential of the parent ion.

\section{Chiral signals based on the orbital angular momentum of light}
\label{minimalPART}

\begin{figure}[!h]
  \centering
  \includegraphics[width=0.5\textwidth]{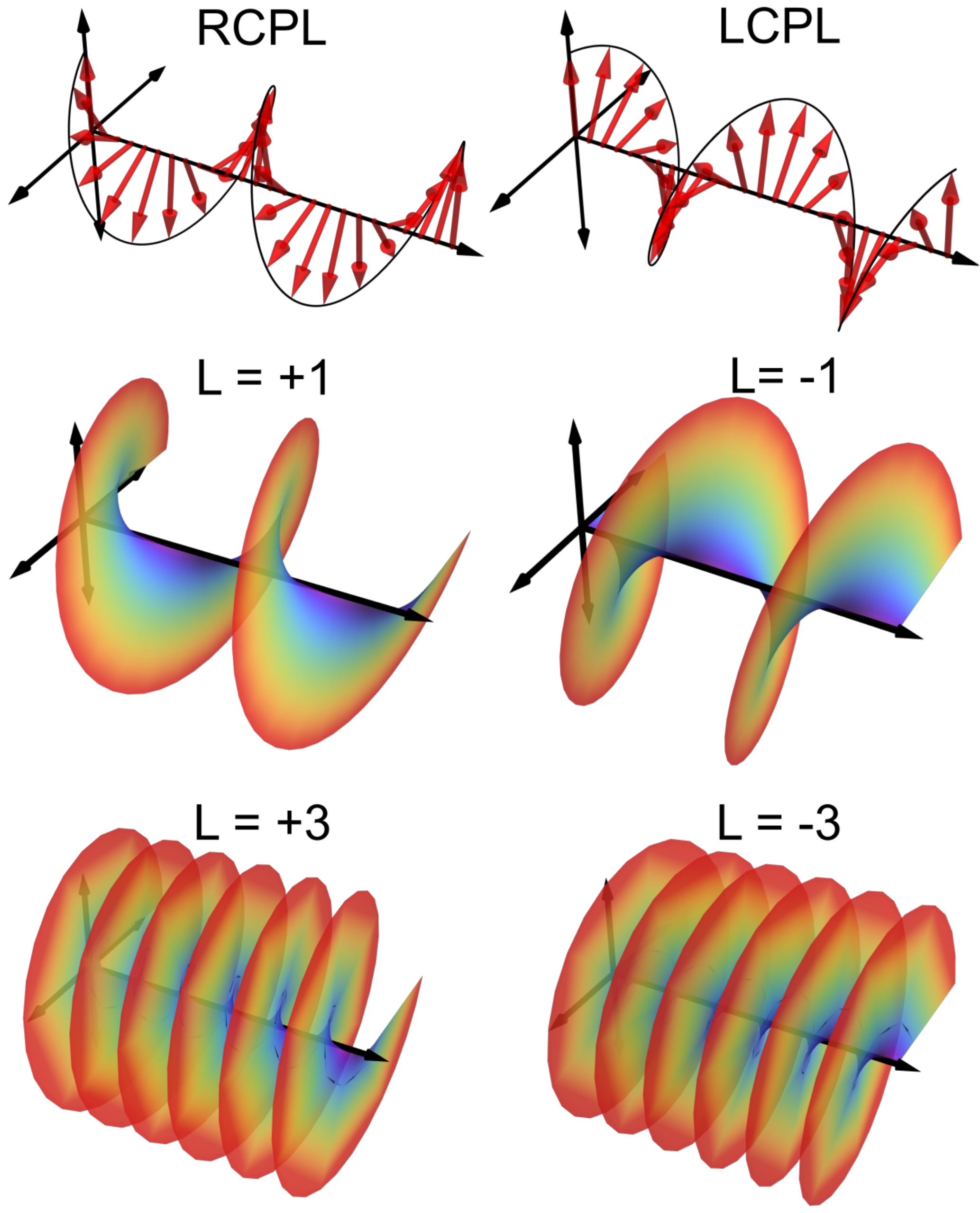}
  \caption{Top row: representation of the electric field polarization in right and left circular polarized light (RCPL and LCPL respectively). Middle and bottom rows: helical wavefronts of OAM beams with L = +1, -1, +3 and -3.
\label{fig:oambeam}}
\end{figure}

Over the past few decades, the control of the spatial profile of laser beams have greatly improved.
This control encompasses both the variation of the amplitude and the polarization in space.
Many spatial shaping techniques have already been implemented in the X-ray regime.
This has led to the development of a large variety of beams such as vector beams\cite{morgan2020free} (e.g. radial and azimuthal polarizations) and vortex beams\cite{peele2003x,seiboth2019refractive}.
The latter, also named twisted beams\cite{hernandez2017twist} or OAM (Orbital Angular Momentum) beams\cite{liu2020orbital}, possess a chirality that can interact with and discriminate chiral molecules.
Vortex beams are defined by a screw-type wavefront that twists along the beam propagation as displayed in Fig. \ref{fig:oambeam}.
They are eigenstates of the OAM operator $L_z = -i \frac{\partial}{\partial z}$ where $z$ is the axis of propagation of the beam.
The eigenvalue of the beam is called the topological charge $l$ and the incoming field assumes the form:
\begin{equation}
\bm E(\bold r, l) = \bm E(r, z) e^{i l \varphi}
\end{equation}
\noindent where $(r, \varphi, z)$ are the cylindrical coordinates and the explicit form of $\bm E(r, z)$ depends on the vortex beam generation process.
Examples of vortex beams include the Laguerre-Gauss (LG)\cite{bahrdt2013first,sasaki2008proposal} and the Hypergeometric-Gaussian (HGG)\cite{rosner2017high,kotlyar2006diffraction,ribivc2017extreme} beams.
A more general OAM-carrying beam can be constructed as a linear combination of eigenstates of $L_z$ and can display complicated wavefronts.

Different strategies have been adopted to generate X-ray vortex beams.
Spiral Fresnel zoneplates (FZP) act as diffractive X-ray lense and can generate vortex beams\cite{sakdinawat2007soft,vila2014characterization,
zhang2010focusing,ribivc2017extreme,rosner2017high} with nanofocusing capabilities.
Other diffractive optics geometries such as fork dislocation gratings\cite{lee2019laguerre} have also been used to generate OAM in X-ray beams.
Alternatively, transmission optics such as spiral phase plates\cite{seiboth2019refractive} have also been recently implemented at energies as high as 8.2 keV but remain limited to low OAM values.
Finally, the use of deformable mirrors\cite{spiga2013x} is another technique to spatially shape X-ray wavefronts, although it has not been used to generate vortices yet.
For the latter, the OAM-inducing optics have been designed in order to sustain the high-peak power of XFEL radiation.

Given the vortex beam novelty, considerable research has been dedicated to their generation and characterization at various frequency regimes.
Early applications include telecommunication (by using the OAM as a physical layer in multiplexing\cite{lee2017orbital}) and laser-assisted surface processing\cite{kohmura2018nano,syubaev2017direct}.
More general information on optical vortex beams and their applications can be found in \cite{shen2019optical,omatsu2019new}.
The interaction between the OAM of light and molecular chirality has only gained interest recently.
It was noticed by Cohen et al.\cite{tang2010optical} in 2010 that spatially-shaped chiral light can engage efficiently with molecular chirality.
However, the effect was poised by the difficulty to generate a relevant phase twist at optical wavelength with respect to typical molecular sizes.
The diffraction limit of X-ray beams allows for a focusing down to \aa ngstr{\"o}m-sized beams and helical wavefront can be generated within few nanometers which make them most promising  for local probes of chirality. Core-resonant excitations are also highly localized spatially.

\subsection{General considerations on vortex beams}

Various quantities originating from geometrical or physical considerations can characterize the twisted nature of a vortex beam.
The geometric quantities are the vorticity $\bm \omega$ and the helicity $H$ defined by
\begin{eqnarray}
\bm \omega(\bm r) = \nabla \wedge \bm E(\bm r)\\
H = \int d\bold r \ \bm E\cdot \bm \omega(\bm r)\label{eq:helicity}
\end{eqnarray}
Vorticity and helicity of the magnetic field and the vector potential can be defined in a similar way and are connected to the electric field ones through the Maxwell equations.
These quantities which can be defined for any vector fields and have long been used in fluid mechanics.

From physical considerations, the total angular momentum of a beam is given by
\begin{eqnarray}
\bold J &=& \epsilon_0 \int d\bold r \bm E\wedge \bm B\\
&=& \epsilon_0 \int d\bold r (E_i (\bold r\wedge \nabla) A_i + \bm E \wedge \bm A) = \bm L + \bm S
\end{eqnarray}
\noindent where $\bm L = \epsilon_0 \int d\bold r (\bm E_i (\bold r\wedge \nabla)\bm A_i$ is the OAM of the beam\cite{yao2011orbital,allen1999iv} and $\bm S  = \epsilon_0 \int d\bold r \bm E \wedge \bm A$ is its Spin Angular Momentum (SAM). $\bm E$ and $\bm A$ are the gauge-invariant transverse component of these fields. 
For plane waves, the SAM can be recast as $S = (\epsilon_0 /(2\omega)) \int d\bm r (|E_L|^2 - |E_R|^2$ where $E_L$ and $E_R$ are the amplitude of left and right circularly polarized light. Thus, $S/\hbar = n_L -n_R$ is a measure of the difference between the number of left ($n_L$) and right($n_R$) circularly polarized photons within the beam.

Starting with Maxwell's equations, Lipkin\cite{lipkin1964existence} has shown that the following chiral local quantity is conserved:
\begin{equation}
C = \frac{\epsilon_0}{2} \bm E \cdot \nabla\wedge \bm E + \frac{1}{2\mu_0} \bm B \cdot \nabla\wedge \bm B
\label{eq:defC}
\end{equation}
This quantity known as the optical chiral density defines a pseudo-scalar.
Being conserved, the chiral density satisfies a continuity equation:
\begin{equation}
\frac{\partial C}{\partial t} + \nabla \Phi_C = 0
\end{equation}
\noindent where
\begin{equation}
\Phi_C = \frac{1}{2 \mu_0} (\bm E\wedge(\nabla \wedge \bm B) - \bm B\wedge(\nabla \wedge \bm E))
\end{equation}
\noindent is the chiral density flux\cite{bliokh2011characterizing}.
The chiral density $C$ is the local version of the helicity $H$, Eq. \ref{eq:helicity} summed over electric and magnetic contributions.

Finally, vortex beams can also be characterized by their topological charges $q$\cite{kotlyar2020topological,berry2004optical} which measures the phase increase of the field around a closed loop $\mathcal C$ enclosing the vortex:
\begin{equation}
q = \lim_{r \to +\infty}\frac{1}{2\pi}\oint_{\mathcal C} d\varphi \frac{\partial \Phi(r,\varphi)}{\partial \varphi}
\end{equation} 
\noindent where $\Phi(r,\varphi)$ is the incoming field phase. 
For the common case of light vortices with a complex amplitude proportional to $e^{\pm i l \varphi}$ for the angular component ($\Phi(r,\varphi) = \pm l \varphi$), the topological charge is $q = \pm l$.
The topological charge is closely linked to the OAM of light and it can be shown that OAM normalized to the beam power is equal to the topological charge of the beam.

The first part of this review was dedicated to the use of the SAM of X-rays to probe chiral molecules.
Their OAM offers new avenues for the design of chiral signals.
The spatial variation of the incoming fields can either be taken as a whole or expanded in multipoles.
In the former case, spectroscopic signals can be recast in terms of transition matrix elements of current and charge densities (Eqs. \ref{hetmcj}-\ref{homoCOmc}) using the minimal coupling Hamiltonian (Eq. \ref{hintminimal}).
This level of theory has various merits over the multipolar Hamiltonian. 
First, the transition matrix elements bear a clear physical meaning in real space using simple operators unlike transition multipoles that are a mathematical constructions obtained by a Taylor expansion and become increasingly more elaborate at higher orders.
Second, the transition current and charge densities implicitly contain all multipoles via nonlocal response functions and avoid the tedious sums over the contributing multipoles to the chiral signals.

Indeed, this description becomes increasingly more advantageous over the multipolar approach when the incoming field is strongly varying across the molecule.
Finally, the minimal coupling description retains the full spatial profile of the matter multipoint correlation function and the incoming electromagnetic fields.
The signals are recast as an overlap integral over space, suggesting the possibility to optimize this quantity by spatial shaping of the incoming field.

The price to pay for these advantages is that the matter transition matrix elements are now tensor fields instead of simple tensors. 
The extraction of transition current and charge densities from ab initio quantum chemistry packages requires some extra efforts and the signal calculations require a higher numerical cost.

Before discussing specific applications of the non local-formalism, we show how the minimal coupling Hamiltonian introduces the spatial variations of the field into the signal expressions.
To first order in the incoming fields, one gets:
\begin{equation}
S_\text{abs}(\Gamma) = \frac{2}{\hbar^2} \text{Re} \int dt dt' d\bold r d\bold r' \ \bold A^*(\bold r,t)\bold A(\bold r',t') 
\langle \bold j_L(\bold r)\mathcal G(t-t') \bold j_-(\bold r')\rangle
\label{linabsMC}
\end{equation}
At third order in the incoming field, we obtain:
\begin{multline}
S_\text{abs}(\Gamma) = \frac{2}{\hbar^4} \text{Re} \int dt dt_3 dt_2 dt_1 d\bold r d\bold r_3 d\bold r_2 d\bold r_1  
\bold A^*(\bold r,t)\bold A(\bold r_3,t_3) \bold A(\bold r_2,t_2) \bold A(\bold r_1,t_1)\\ 
\times \langle \bold j(\bold r)|
\mathcal G(t_3) \bold j_-(\bold r_3)
\mathcal G(t_2) \bold j_-(\bold r_2)
\mathcal G(t_1) \bold j_-(\bold r_1)\rangle
\label{4wmMC}
\end{multline}

The spatial variation of the beam now enters as overlap integrals over space between the current densities and the vector potential envelopes.

\subsection{Enhanced dichroism with optimized fields}

Tang and Cohen\cite{tang2010optical,tang2011enhanced} have generalized CD, Eq. \ref{CDdef}, to include beams of opposite parity instead of restricting it to left and right circular polarization.
Within the electric dipole - magnetic dipole approximation, they were able to link the observed dissymmetry factor $S_\text{chir}$ to the field chirality $C$, Eq. \ref{eq:defC}, and the matter linear chiral response, Eq. \ref{eq:Gchirrep}:
\begin{equation}
S_\text{chir}(\Gamma) = \frac{A_+ - A_-}{\frac{1}{2}(A_+ + A_-)} = \frac{2 C \text{Im}G}{\omega U_e\text{Im}\alpha}
\label{eq:cohen_results}
\end{equation}
\noindent where $S_\text{chir}$ is normalized by the achiral contribution and $U_e = \epsilon_0|E|^2/2$ is the local time-averaged electric energy density. $A_+$ and $A_-$ are the absorption at frequency $\omega$ measured using beams with opposite parities.
The numerator of Eq. \ref{eq:cohen_results} clearly shows that the dissymmetry factor increases with the field and the matter chirality.
It immediately follows that minimizing the incoming field intensity leads to larger asymmetry ratio, at the expense of an overall weaker total signal.
The vortex beams can be used to maximize the dissymmetry factor by maximizing the field chirality $C$ at a fixed incoming energy density.

In the optical regime, one can safely truncate the multipolar expansion at low orders given the small molecular size compared to the incoming wavelength.
This is also what limits the interaction of optical vortices with chiral molecules since most molecules within a large optical beam do not experience the twisted wavefront over their spatial extension.
Alternatively, one can rely on near-field phenomena to generate sub-wavelength vortices\cite{schouten2004optical}.
Tighter focus can be achieved in the X-rays and it will be interesting to describe the vortex complex amplitude as a single entity.
Thanks to the short wavelength of X-ray photons ($\sim$ 0.1 to 10 nm), they can achieve focusing and spatial variations at the molecular scale\cite{matsuyama2016nearly,bergemann2003focusing}.
The minimal coupling Hamiltonian, Eq. \ref{hintminimal} retains the complete spatial profiles of the exciting fields.

Alternatively, one can consider schemes involving multiple pulses whose interferences generate the desired chiral spatial variation.
For example, Cohen et al.\cite{tang2010optical} proposed to use two counter-propagating circular polarized light to enhance chirality.
This allows to maximize the dissymmetry ratio, but it does so mostly for molecules located in regions where the field intensity is weak.
In this configuration, the maxima of $C$ are located in places where the field intensity is minimal.
Additionally, this scheme relies on the use of a partially reflecting mirror which is difficult to implement in the X-ray regime where grazing incidence is typically used for mirrors.

In previous work\cite{rouxel2016non}, we demonstrated how the minimal coupling Hamiltonian allows to generalize Tang and Cohen result, Eq. \ref{eq:cohen_results}, for spatially varying beams.
We restrict our description to resonant interactions where the $\sigma \bm A^2$ term in Eq. \ref{hintminimal} can be neglected.
The molecular linear response, Eq. \ref{linabsMC}, then only involves current density operators through the interaction Hamiltonian $H_\text{int} = \int d\bm r \ \bm j(\bm r)\cdot \bm A(\bm r)$. 
The current density is an operator, Eq. \ref{eq:j}, that can be readily computed by ab initio calculations.
It can be partitioned into divergence-free (transverse) $\bm j_\perp$ and curl-free (longitudinal) $\bm j_\parallel$ components using the Helmholtz decomposition:
\begin{equation}
\bm j(\bm r) = \bm j_\parallel(\bm r) + \bm j_\perp(\bm r)
\end{equation}

The transverse part can be written as the curl of an auxiliary field $\bm j_\perp(\bm r) = \nabla \wedge \bm a(\bm r)$.
The chiral contribution to Eq. \ref{linabsMC} then originates from crossterms involving $\bm j_\parallel(\bm r)$ and $\bm a(\bm r)$:
\begin{multline}
S_\text{chir}(\Gamma) = \frac{2}{\hbar^2} \text{Re} \int dt dt' d\bold r d\bold r' \ \bold A^*(\bold r,t) \nabla\wedge\bold A(\bold r',t') 
\langle \bold j_{L,\parallel}(\bold r)\mathcal G(t-t') \bold a_{-}(\bold r')\rangle \\
+ \nabla\wedge\bold A^*(\bold r,t) \bold A(\bold r',t') 
\langle \bold a_{L}(\bold r)\mathcal G(t-t') \bold j_{-,\parallel}(\bold r')\rangle
\label{linabsMCchir}
\end{multline}

One can readily notice that the chiral contribution to the signal is an overlap spatial integral over a chiral field tensor $\bm A^*(\bm r,t)\otimes\nabla \wedge \bm A(\bm r',t')$ and a chiral matter tensor $\langle \bold a_{L}(\bold r)\otimes\bold j_{-,\parallel}(\bold r')\rangle$.
The chiral field tensor is reminiscent to Lipkin's chiral invariant $C$, Eq. \ref{eq:defC}.
In order to optimize chiral signals with spatially shaped pulses, one must then aim at maximizing this overlap, i.e. tune the field local chirality to match the one of the matter.
In order to calculate signals from a molecular ensemble, one must rotationally average the matter tensor field, which can be a demanding task in the most general case\cite{rouxel2018translational} that can be tackled numerically\cite{parrish2019ab}.

Other approaches to chirality based on the polarization $\bm P$ and the magnetization $\bm M$ densities are also possible.
These involve the electric and magnetic field through the following coupling Hamiltonian:
\begin{equation}
H_\text{int} = -\int d\bm r \ \bm P(\bm r)\cdot \bm E(\bm r) + \bm M(\bm r)\cdot \bm B(\bm r)
\end{equation}
\noindent where we have neglected the diamagnetic contribution\cite{babiker1983derivation,andrews2018perspective} for simplicity.
This coupling is obtained from the minimal coupling one, Eq. \ref{hintminimal},
by using the Power-Zienau-Woolley canonical transformation\cite{cohen1997photons} $e^{iS/\hbar}$ with $S = \int d\bm r \bm P(\bm r)\cdot \bm A(\bm r)$ and using that $\bm j(\bm r,t) = \frac{\partial \bm P(\bm r,t)}{\partial t} + \nabla \wedge \bm M(\bm r,t)$.
This transformation has been extensively used, especially as a starting point for a multipolar expansion.
However, if one wants to include all multipoles and use the complete vector fields, the polarization and the magnetization densities are not trivially calculated at the ab initio level since their expression are not uniquely defined.
The current density operator, in contrast, can be calculated in a simple manner from the many-body molecular eigenstates expanded in a molecular orbital basis.
Additional studies are called for to clarify the nature of the Lipkin's chiral density $C$ expressed in the various formulations of electromagnetism.

\subsection{X-ray Helical Dichroism}

So far, we have discussed how OAM-carrying beams can engage with molecular chirality in a different way than SAM-carrying beams.
Elaborate beam profiles can be created to maximize the overlap between the matter and the field chiralities.
The simplest approach, readily achievable experimentally, is to use twisted beams carrying a single OAM value.
Transfer of OAM of light to trapped ions have been studied\cite{peshkov2016absorption,schmiegelow2016transfer} and it was demonstrated that the OAM interacts with bound electrons.
By analogy to circular dichroism which observes the absorption difference for opposite SAM, the difference in absorption for opposite OAM, $+l$ and $-l$, named Helical Dichroism (HD) has been proposed\cite{brullot2016resolving}.
Various measurements on nanosystems have been reported \cite{brullot2016resolving,kerber2018orbital,ni2021giant} in the optical range but none on molecules.
Optical HD on molecules is difficult due to the necessity to generate a phase vortex relevant on molecular sizes.
X-rays can readily overcome this difficulty thanks their ability to be focused down to a few nanometer spot size\cite{bajt2018x}.
A demonstration of X-ray HD has been made at the synchrotron SLS on Fe-diMe(tpy)$_3$ at the Fe K-edge\cite{rouxel2022hardhd}.
In the X-ray, spiral Fresnel zoneplates were used to generate vortex beams with OAM $L=\pm1$ and $\pm3$ and the X-ray absorption spectrum (XAS) was measured both in transmission and in Total Fluorescence Yield mode (TFY).
Further experimental efforts at the carbon K-edge are ongoing, making the technique a promising chiral X-ray spectroscopy.

\begin{figure}[!h]
  \centering
  \includegraphics[width=0.8\textwidth]{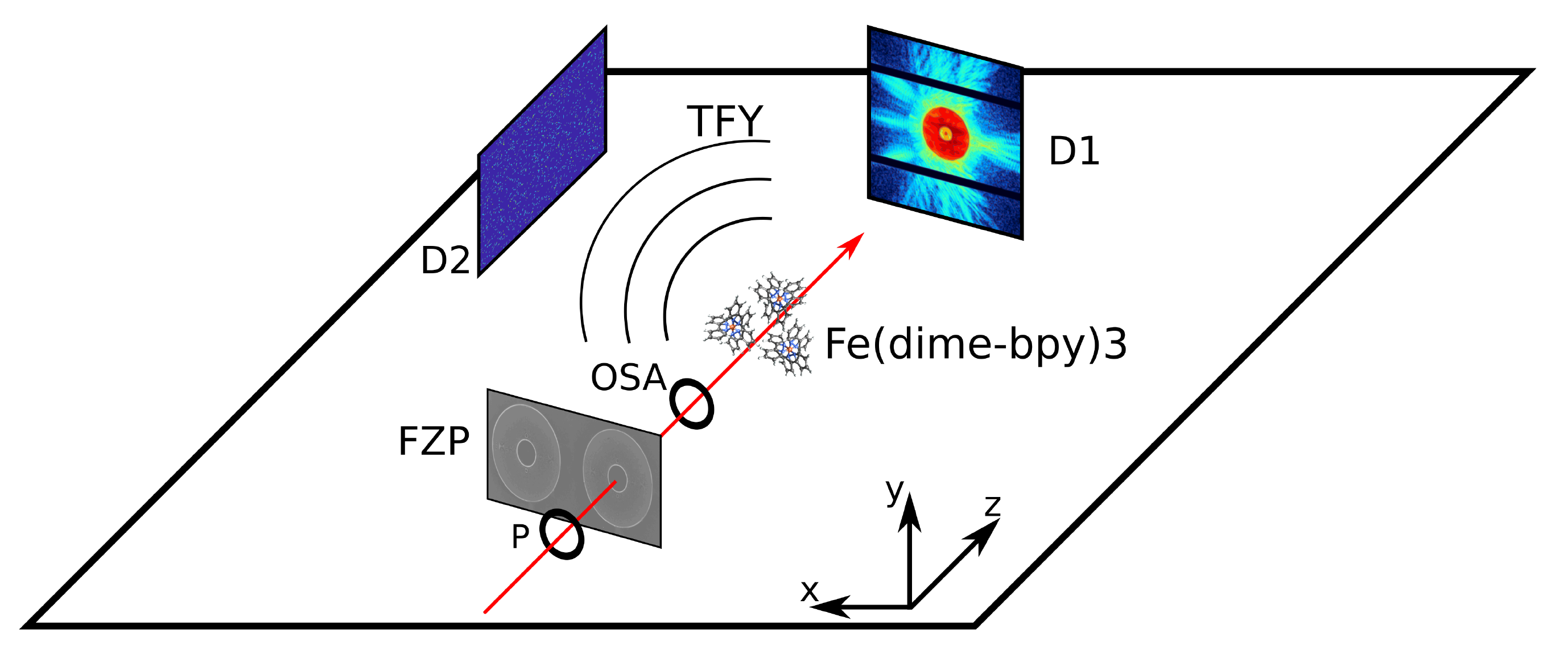}
  \caption{Setup to measure X-ray HD\cite{rouxel2022hardhd}: spiral Fresnel zoneplates are used to generate helical wavefront. The absorption signal can then be detected in transmission (detector D1) or in Total Fluorescence Yield (TFY, detector D2). The pinhole (P) and the Order Sorting Aperture (OSA) ensure the quality of the generated OAM beams.
\label{hd1}}
\end{figure}

Since an important transverse spatial phase vortex of the electromagnetic field is required, the electric dipole approximation is not sufficient to describe the interaction of matter with the OAM.
We shall use the minimal coupling Hamiltonian to derive the HD signal.
The minimal coupling Hamiltonian fully captures the spatial variation of the incoming beam.
A multipolar approach is also possible\cite{forbes2018optical} but its truncation at the quadrupolar order does not account for possibly large field variations.
The HD signal is defined as the differential XAS between $+l$ and $-l$ OAM beams normalized by the average of the two XAS:
\begin{equation}
S_\text{HD}(l,\omega) = \frac{A_{+l}(\omega)-A_{-l}(\omega)}{\frac{1}{2}(A_{+l}(\omega)+A_{-l}(\omega))}
\label{eq:defHD}
\end{equation}
Unlike for the SAM of light, the OAM value are not restricted to $\pm 1$ values and the HD signal now depends on $l$ that can take an arbitrary value among the natural integers.

The absorption signals in Eq. \ref{eq:defHD} can be expressed using Eq. \ref{linabsMC} in the frequency domain beams and by summing over molecular eigenstates:
\begin{multline}
A(\Gamma) = \frac{2}{\hbar^2} \text{Re}\sum_c \int d\bold r d\bold r' d\omega \ (\bold A^*(\bold r,\omega) \bold A(\bold r',\omega)) 
i (\bold j_{gc}(\bold r)\mathcal G_{cg,cg}(\omega)\bold j_{cg}(\bold r') - \bold j_{cg}(\bold r)\mathcal G_{gc,gc}(\omega)\bold j_{cg}^*(\bold r')
\end{multline}
Assuming monochromatic beams and using the definition of the Liouville space Green function, Appendix \ref{appendix:perturbation}, the signal can be recast as
\begin{equation}
A_{\pm l}(\omega) = -\frac{2}{\hbar^2} \text{Im}\sum_c \frac{\langle d_{gc}^{\pm l} d_{cg}^{\pm l}\rangle_\Omega}{\omega - \omega_{cg} + i\Gamma_{cg}} - \frac{\langle d_{cg}^{\pm l} d_{cg}^{\pm l,*}\rangle_\Omega}{\omega - \omega_{gc} + i\Gamma_{gc}}
\end{equation}
\noindent where the $\bm d$ transition matrix elements are given by the integrated transition current multiplied by the vector potential:
\begin{equation}
d_{ij} = \int d\bm r \bold j_{ij}(\bm r) \cdot \bm A(\bm r, \pm l)
\end{equation}
This signal is similar to standard absorption calculated in the electric dipole approximation but the transition matrix elements are now replaced by overlap integrals between the transition current density matrix elements and the vortex beams spatial profiles.
The HD signal, $S_\text{HD}(l,\omega)$, is chirality-induced, i.e. it vanishes in achiral media. 
It can be shown as follow.
When taking the difference between left and right OAM, the difference in absorption for each  state can be expressed as:
\begin{multline}
A_{+l}(\omega)-A_{-l}(\omega) \propto \int d\bm r d\bm r' 
\langle j_{gc}(\bm r) j_{cg}(\bm r')\rangle_\Omega (\bm A^*(\bm r, l) \bm A(\bm r', l) - \bm A^*(\bm r, - l) \bm A(\bm r', - l))\\
\propto \int d\bm r d\bm r' 
\langle j_{gc}(\bm r) j_{cg}(\bm r')\rangle_\Omega \bm A^*(\bm r, l) \bm A(\bm r', l) - \langle j_{gc}(-\bm r) j_{cg}(-\bm r')\rangle_\Omega\bm A^*(-\bm r, - l) \bm A(-\bm r', - l))
\end{multline}
Using the symmetry $\mathcal P \bm A(\bm r, - l) = \bm A(-\bm r, - l) = \bm A(\bm r, l)$, we obtain:
\begin{equation}
A_{+l}(\omega)-A_{-l}(\omega) \propto  \int d\bm r d\bm r' \bm A^*(\bm r, l) \bm A(\bm r', l) \Big(\langle j_{gc}(\bm r) j_{cg}(\bm r')\rangle_\Omega - \langle j_{gc}(-\bm r) j_{cg}(-\bm r')\rangle_\Omega\Big)
\end{equation}

The matter quantity $\langle j_{gc}(\bm r) j_{cg}(\bm r')\rangle_\Omega - \langle j_{gc}(-\bm r) j_{cg}(-\bm r')\rangle_\Omega = (1 - \mathcal P)\langle j_{gc}(\bm r) j_{cg}(\bm r')\rangle_\Omega$ vanishes for achiral molecules but is finite for chiral molecules. HD is thus induced by chirality.

The explicit computation of HD signal on a molecules requires to define the beam profile of the vortex beams.
The LG and the HG beams are two basis sets commonly used and corresponding to experimentally generated beams.
The spatial profiles of LG beams propagating along the $z$ axis are given by
\begin{multline}
\text{LG}_{lp}(r,\phi,z) =\sqrt{\frac{2 p!}{\pi(p+|l|)!}} \frac{w_0}{w(z)} 
\bigg(\frac{\sqrt{2} r}{w(z)} \bigg)^{|l|} 
e^{-\frac{r^2}{w^2(z)}} e^{-ik \frac{r^2}{2 R(z)}}\\
\times L_p^{|l|}\big(\frac{2r^2}{w^2(z)}\big)
 e^{i(2p+|l|+1)\arctan(\frac{z}{z_R})}
e^{i l \phi}e^{-ikz}
\end{multline}
The transverse spatial extent of the wave is described by the beam width $w(z) = w_0 \sqrt{1+ z^2/z_R^2}$ with $w_0$ being the beam waist. $R(z) = z + z_R^2/z$ is the radius of curvature, $z_R = \pi w_0^2/\lambda$ is the Rayleigh length and $(2p+|l|+1)\text{atan}(z/z_R)$ is the Gouy phase.
HG beams generated by Fresnel zoneplates can be obtained from a Fraunhofer diffraction and are given near focus by
\begin{multline}
\text{HG}_{l}(r,\phi,z) = \frac{(-i)^{l+1}}{(l+2)l!}\frac{k R^2}{f}(\frac{kR r}{2f})^l 
{}_1 F_2(\frac{l+2}{2},\frac{l+4}{2},l+1; -\frac{kRr}{2f})
e^{i l \phi}e^{-ikz}
\end{multline}
\noindent where $f$ is the focal length of the Fresnel zoneplate and $R$ is the beam radius on focus.\\

An alternative approach to HD using the multipolar expansion has been introduced by Andrews et al.\cite{andrews2004optical,forbes2018optical}.
They had shown that the magnetic transition dipole does not contribute to the HD signal by calculating the absorption cross-section in the multipolar coupling, truncated at the quadrupolar term. 
They showed that the OAM of light has an observable effect only when the light carries also a SAM.
Nonetheless, the HD signal may be finite without the use of the SAM of light at higher orders in the multipolar expansion.
The electric dipole - electric quadrupole contribution to the absorption spectrum is given by:
\begin{multline}
A(\omega) = -\frac{2}{\hbar^2} \text{Im}\sum_c 
\frac{\langle \bm\mu_{gc}\cdot \bm E^*  \bm q_{cg}\cdot \nabla \bm E\rangle_\Omega}{\omega - \omega_{cg} + i\Gamma_{cg}}
-\frac{\langle \bm\mu_{cg}\cdot \bm E^* \bm q_{cg}\cdot \nabla \bm E\rangle_\Omega}{\omega - \omega_{gc} + i\Gamma_{gc}}\\
+\frac{\langle \bm q_{gc}\cdot \nabla\bm E^*  \bm \mu_{cg}\cdot \bm E\rangle_\Omega}{\omega - \omega_{cg} + i\Gamma_{cg}}
-\frac{\langle \bm q_{cg}\cdot \nabla\bm E^* \bm \mu_{cg}\cdot  \bm E\rangle_\Omega}{\omega - \omega_{gc} + i\Gamma_{gc}}
\end{multline}
Upon computing the gradient in cylindrical coordinates, terms of the form $il\bm{e_\varphi}$ appear in the transition amplitude and lead to an $l$-dependant signal.

A more general use of vortex beams can employ beams carrying both an OAM and a SAM.
The resulting circular-helical dichroism (CHD) signal defined by:
\begin{equation}
S_\text{CHD}(l,\omega) = \frac{A_{L,+l}(\omega)-A_{R,-l}(\omega)}{\frac{1}{2}(A_{L,+l}(\omega)+A_{R,-l}(\omega))}
\end{equation}
\noindent where $A_{L,+l}(\omega)$ is the absorption of a left-polarized $+l$ OAM beam and $A_{R,-l}(\omega)$ of a right-polarized beam with $-l$ OAM.

\begin{figure}[!h]
  \centering
  \includegraphics[width=0.6\textwidth]{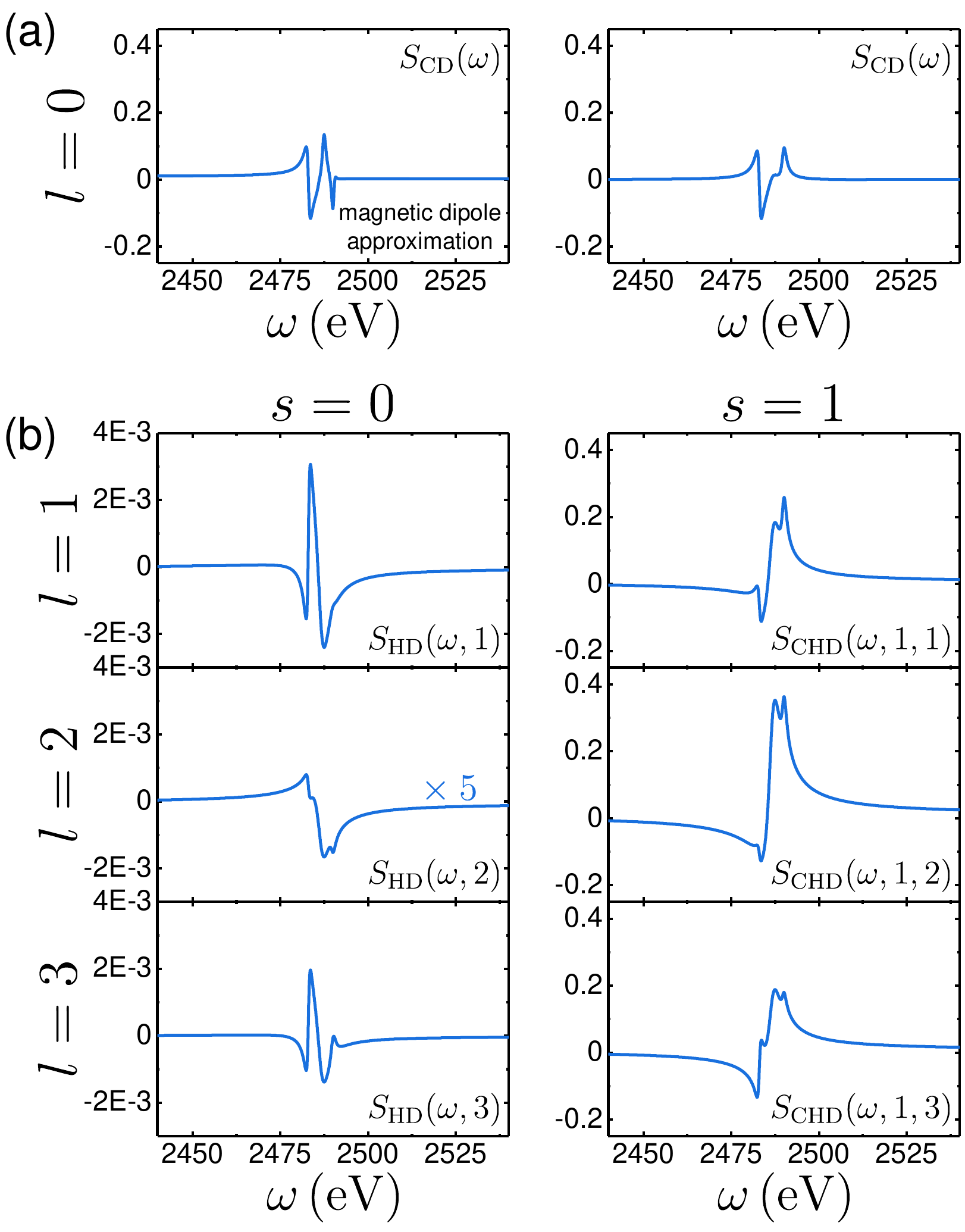}
  \caption{Simulation of X-ray CD, HD and CHD spectra at the S K-edge on L-cystein. Reprint from Ye et al\cite{ye2019probing}.
\label{hd2}}
\end{figure}

Simulated X-ray CD, HD and CHD spectra on L-cysteine at the sulfur K-edge (2.48 keV)\cite{ye2019probing} are displayed in Fig. \ref{hd2}.
This work highlights the necessity to create a phase vortex relevant at molecular sizes.
The asymmetry ratio, Eq. \ref{eq:defHD}, is maximum for molecules situated at the center of the twisted beam.
For large spot sizes, most molecules are off the beam axis and experience a lower asymmetry ratio which weakens the HD signal upon averaging over all molecules within the interaction volume.

Finally, we note that time-resolved HD (tr-HD) can be formally defined in a similar way as tr-CD, Eq. \ref{trcddef}:
\begin{equation}
S_{\text{tr-HD}}(l, \omega,\tau) = 
\frac{A(+l,\omega, \tau)-A_{-l, \text{R}}(\omega, \tau)}
{\frac{1}{2}(A(+l,\omega, \tau)+A(-l,\omega, \tau))}
\label{trhddef}
\end{equation}
tr-HD has not been reported yet and constitute an interesting future extension of the HD technique.

\subsection{Chiral X-ray diffraction}

As a last technique of this review, we discuss chiral-sensitive opportunities using X-ray Diffraction (XRD) schemes.
It is well-known that static XRD has an ambiguity that cannot distinguish experimentally randomly-oriented enantiomers\cite{flack2008use}.
In structure reconstruction, this limitation is often circumvented by combining the XRD measurement with other chiral-sensitive signals, e.g. CD, or but using oriented samples.
The advent of ultrafast time-resolved XRD (tr-XRD) at XFEL is offering new avenues to probe chirality with X-rays in a diffraction setup.
For example, Giri et al. have recently shown\cite{giri2021imaging} how charge migration in chiral epoxypropane can be monitored using tr-XRD.
Additionally, the use of OAM beams combined with XRD could generate new approaches to design chiral sensitive XRD signal.
It this section, we recall why beams without OAM cannot generate a static chiral-sensitive XRD and how the OAM beams can generate a chiral version of XRD.

The interaction Hamiltonian, Eq. \ref{hintminimal}, simplifies far from resonance where we can only retain the $\sigma \bold A^2$ contribution.
X-ray Diffraction (XRD) is well described by that coupling and the use of a scattered monochromatic beam leads to the following expression for the XRD signal:
\begin{equation}
S_\text{XRD}(\bm q) \propto |\bm \epsilon_\text{X}\cdot\bm \epsilon_s|^2 |\int d\bm r \sigma(\bm r) e^{-i\bm q\cdot \bm r}|^2
\end{equation}
\noindent where $q = \bm k_s - \bm k_\text{X}$ is the momentum transfer between the incident X-ray wavevector $\bm k_\text{X}$ and the scattered one $\bm k_s$. $|\bm \epsilon_\text{X}\cdot\bm \epsilon_s|^2$ is the Lorentz polarization factor.
Since the coupling Hamiltonian contains $\bm A_\text{X}\cdot \bm A_s$, the incoming $\bm \epsilon_\text{X}$ and scattered $\bm \epsilon_s$ field polarizations do not engage with the scalar charge density and XRD can not have a chiral component originating from the SAM of light.
Additionally, usual X-ray beams for XRD are assumed to be plane wave and thus do not carry an OAM.

OAM beams offer the potential to design chiral XRD techniques.
With spatially-varying beam, the XRD signal is given by:
\begin{multline}
S_{\text{XRD}}(\bold q)=\frac{1}{\hbar^2}\Big(\frac{e}{2m}\Big) \Big(\frac{\hbar}{2\epsilon_0\omega_s}\Big)^{2} \Re
\int d\bold r dt d\bold r' dt' \\
\langle \sigma_L(\bold r,t)\sigma_R(\bold r',t')\rangle
\bm{\mathcal A}_\text{X}^*(\bold r,t)\cdot
\bm{\mathcal A}_\text{X}(\bold r',t')
e^{i(\bold k_s\cdot \bold r-\omega_s t)}
e^{-i(\bold k_s\cdot \bold r'-\omega_s t')}.
\label{XRD1}
\end{multline}

When the coherent signal is simulated for an assembly of coherent scatterers, the two-point correlation function of charge densities can be factorized over molecules:
\begin{equation}
S_{\text{XRD}}(l,\bold q) \propto F_2(\bold q)| \int d\bm r \langle\sigma(\bm r)\rangle \bm A_\text{X}(\bm r, l) e^{i \bold q\cdot \bm r}|^2
\label{eq:xrd2molcoh}
\end{equation}
\noindent where $F_2(\bold q)$ is the structure factor of the lattice. $\langle\sigma(\bm r)\rangle$ is the expectation value of the charge density operator which reduce to $\sigma_{gg}(\bm r)$ for molecules in the ground state.
The vortex beam $A_\text{X}(\bm r, l)$ radial profile depends on the way the beam has been generated.
In the absence of long range order, e.g. molecules in liquid or gas phase, the structure factor in Eq. \ref{eq:xrd2molcoh} vanishes upon isotropic averaging over molecular orientations.
The XRD signal is then dominated by the single molecule term:
\begin{equation}
S_\text{XRD}(l,\bold q) \propto \int d\bm r d\bm r' \langle\sigma(\bm r)\sigma(\bm r')\rangle \bm A_\text{X}(\bm r, l) \bm A_X^*(\bm r', l) e^{i \bold q\cdot (\bm r - \bm r')}
\label{eq:xrd1molspon}
\end{equation}
For a molecule in the ground state, the expectation value reduces to $\langle\sigma(\bm r)\sigma(\bm r')\rangle = \sigma_{gg}(\bm r)\sigma_{gg}(\bm r')$.

OAM X-ray beams have the potential to add a chiral contribution to XRD signal and thus break Friedel's law which states that the XRD pattern is always centrosymmetric even if the sample is not.
It holds because XRD patterns are amplitudes squared of the Fourier transform of the electronic charge density $\sigma(\bm r)$, which is a real function.
The real nature of the charge density imposes the symmetry $\sigma(\bold q) = \sigma^*(-\bold q)$ and the XRD signal can then be written as $S_{\text{XRD}}(\bold q) = |\sigma(\bold q)|^2 = \sigma(\bold q)\sigma^*(\bold q) = \sigma^*(-\bold q)\sigma(-\bold q) = S_{\text{XRD}}(-\bold q)$.

This simple statement has profound consequences on the determination of chiral structures since XRD can not trivially distinguish between opposite enantiomers. 
The added chiral contribution from X-ray OAM could allow XRD to discriminate enantiomers in structure determination.

\section{Simulation strategies for X-ray chiral signals }
\label{sec:simulation}

Most signals presented in this review are expressed in terms of multipoint correlation functions that can be expanded in molecular eigenstates.
Transition matrix elements of the electric dipoles, magnetic dipoles, electric quadrupoles and current density operators must be computed to simulate a signal on a given molecule.

In this section, we discuss ab initio computational strategies commonly used to compute X-ray chiral signals.
These signals require the computation of core-excited states and their transition multipole moments.
The available methods include multi-configurational self-consistent field (MCSCF) approaches such as RASSCF (Restricted Active Space Self Consistent Field)\cite{helmich2021simulating, casanova2022restricted}, density density functional theory such as TDDFT (Time-Dependent Density Function Theory)\cite{autschbach2002chiroptical, autschbach2002chiropticalII, lopata2012linear, zhang2014nonlinear}, pertubation theories including the ADC (Algebraic-Diagrammatic Construction methods) schemes\cite{} and coupled pair theories such as EOM-CC (Equation-of-Motion Coupled-Cluster)\cite{vidal2019new, nanda2020stay, nanda2020cherry,faber2020magnetic, andersen2022probing}.
Numerical strategies for optical chiral signals have been extensively discussed by Crawford et al.\cite{crawford2007current} while the computation of core-excited states have recently been reviewed by Norman et al. \cite{norman2018simulating}.

Some specific care must be used to calculate higher order multipoles with acceptable precision.
In any ab initio calculation involving various multipoles, one must keep in mind that only the first non-vanishing transition multipole, typically the electric dipole, is origin invariant.
This origin invariance is recovered when calculating observables such as spectroscopic signals that typically involve multipoint correlation functions of transition matrix elements\cite{ruud2002gauge}.
However, the basis truncation needed to numerically simulate signals is another source of origin variance that must be addressed.
More explicitly, in the multipolar Hamiltonian, also known as the length gauge and labelled by the superscript $r$, the origin variance of the first three multipoles is given by
\begin{eqnarray}
\bm \mu^r_{ij}(\bold O + \bold a) &=& \bm \mu^r_{ij}(\bold O)\\
\bm m^r_{ij}(\bold O + \bold a) &=& \bm m^r_{ij}(\bold O)+ (\frac{1}{2} \bold a \times \bold p)_{ij}\\
\bm q^r_{ij}(\bold O + \bold a) &=& \bm q^r_{ij}(\bold O) - \frac{3}{2} (\bold a \otimes \bm \mu_{ij}+ \bm \mu_{ij}\otimes\bold a) + \bold a \cdot \bm \mu_{ij} \mathds{1}
\end{eqnarray}
\noindent where $\bm O$ is an arbitrary origin of coordinates, usually taken within the molecule, and $\bm a$ is a translation from that origin.

The transition matrix elements involved in the computation of X-ray signals can be divided into  bound to bound and bound to continuum type. 
The former dominate the few eV regime around the absorption edges. 
This includes pre-edge transitions as well as the main absorption peaks at and above the edge.
Such transitions are well treated by keeping an orbital basis set to describe the core excited states. 
However, many-electron effects play an important role in these highly excited states and multi-determinantal wavefunctions are usually needed to achieve accurate eigenstates.

Time-Dependent Density functional theory (TDDFT) approaches have been widely used to compute XCD spectra\cite{alagna1994random, kimberg2007calculation,jiemchooroj2007near,
villaume2009circular,takahashi2015theoretical}.
Kimberg et al.\cite{kimberg2007calculation} have carried out a comparative study of Hartree-Fock and DFT approaches for various amino-acids in the soft X-ray regime (C, N and O K-edges).
The CAM-B3LYP exchange-correlation functional has been used to compute XCD spectra on L-alanine\cite{jiemchooroj2007near} .
Takahashi et al.\cite{takahashi2015theoretical} have studied the basis set and gauge dependence of DFT methods for serine and alanine in the soft X-ray regime.

For post Hartree-Fock methods, the ground state and the valence excited manifold can be calculated at the MCSCF level (multi-configurationnal self consistent field method) after geometry optimization of the molecular structure.
At this stage, it is important to select an active space that is large enough to capture all the occupied and unoccupied orbitals that can be involved in both the valence and the core transitions over the desired energy range.
A consistent active space of the valence excited manifold and the core excited one should be kept in calculating transition matrix elements between the many-body states.
Thus, the core orbitals involved in the considered transitions are translated into the active space and frozen to double occupancy.
Once the valence excited manifold has been calculated, the computation can be repeated with the core orbital frozen with a core hole.
Typically, for K-edge calculations, the 1s orbital is rotated in the active space used for the previous valence excited state calculation and frozen to single occupancy. For L-edge calculations that involve the 2p orbitals, the three orbitals are rotated into the active space and their occupancy restricted to 5 electrons.

Finally, Coriani et al. \cite{andersen2022probing} have recently shown that EOM-CC and CC-RSP (CC Response Theory) using a restricted EOM-CC operator\cite{vidal2019new} can produce XCD molecular spectra. 
They observed a significant gauge origin dependence. 
A good agreement with other theoretical approaches such as CC-RSP (Response Theory) or TDDFT was achieved but comparison with experimental spectra still proved challenging.
The amount of available experimental spectra is still sparse, even for the simplest technique, XCD, and this is a bottelneck for the improvement of computational techniques.

\section{Conclusions and outlook}

Chirality is a fundamental property of molecular structures which is further connected to important applications.
Spectroscopic techniques specifically sensitive to chirality provide useful structural information. 
Chiral techniques have long been used in the infrared and visible regimes, with great success but have a weak magnitude. 
Considerable efforts have been made to increase their strength.
Intense, ultrashort X-ray sources have undergone a rapid development over the past decades and further progress is on the horizon.
X-rays have a long history of resolving atomistic structures, and their use to investigate molecular chirality on ultrafast timescales is a recent development.

In this review, we have broadly surveyed chiral techniques involving X-ray or EUV pulses.
We have classified them into three categories based on the required X-ray/molecule coupling: signals based on magnetic dipole and electric quadrupole interactions, signals solely based on the electric dipole interaction, and signals based on interaction with the OAM of light.
The field of X-ray chiral spectroscopy is still at an early stage with some of the simplest signals such as XCD only implemented quite recently on few systems.
This review aims at providing closed form expressions for most of the discussed signal.
Further simulation work on a variety of systems should provide guidelines for the design of exciting experiments.

X-rays can selectively probe specific elements within a molecule and thus gain information on how the non-centrosymmetry of the electron distribution is localized in space.
This element-selectivity further allows to study how fast a chiral disturbance propagates within a molecule. 
The ultrashort time-duration of EUV and X-ray pulses down to the attosecond regime helps monitor ultrafast electron dynamics happening within a much shorter time frame that the subsequent nuclear dynamics.
X-rays have also long been used for structure determination and the development of chiral sensitive X-ray diffraction techniques could lead to absolute configuration determination.

The field of X-ray chirality, and even more so of ultrafast X-ray chirality, is quite recent. Nonetheless, numerous experimental realizations offer the perspective of a rich future making use of the variety of tehcniques presented in this review.
Among multipolar signals, experiments have so far used XCD as an observable\cite{turchini2004core, izumi2009measurement, hussain2012circular, izumi2013characteristic, izumi2014nitrogen, auvray2019time, oppermann2019ultrafast}. Deep UV experiments \cite{oppermann2019ultrafast} have reached the femtosecond with a high level sensitiviy, counteracting the low signal magnitudes ($\sim 10^{-4}$ mOD). 
Extensions to higher energy regimes for femtosecond tr-XCD will make a great use of the polarization control available at FELs.
Signals in the electric dipole approximation have also experienced a great leap forward over the last few years, both those relying on high harmonics\cite{cireasa2015probing, wang2017high, harada2018circular, neufeld2018optical, ayuso2019synthetic, beaulieu2018photoexcitation,neufeld2022detecting, ayuso2022strong,neufeld2022detecting} or on photoionization \cite{janssen2014detecting,lux2012circular, beaulieu2016probing, turchini2013conformational, ilchen2021site}. 
Giant asymmetry ratios, possiblity reaching 100\% have been reported. 
Minimal coupling signals, relying on the incoming beam OAM, also offer an exciting avenue\cite{ribivc2017extreme, de2020photoelectric, rouxel2022hardhd} to probe molecular chirality, with asymmetry ratios in the 1-10\% range. 
They are made possible by the spatial coherence of X-ray sources and progress in the making of diffractive optics that generate them. 
The extra control of the  X-ray OAM offer an extra parameters to observe higher asymmetry ratios that is typically observed in XCD measurements.
More interestingly, the signal dependence on the OAM values contains new physical information.
Finally, the detection of photoelectron OAM is also an intriguing and recently proposed direction\cite{planas2022strong}.

\begin{appendices}

\section{Rotational averaging of tensors}
\label{appendixROTAV}
A $n$-th rank tensor $T$ is rotationally averaged using the averaging tensor $I^{(n)}$\cite{craig1998molecular}.
Here, we give the tensor expressions of the rotationally averaged tensors up to rank 5. Expressions for arbitrary ranks can be found in references\cite{craig1998molecular}.
The averaging procedure can be seen as a projection over a basis of isotropic tensors.
Rank 2 and 3 averaging tensors are simple since the basis of isotropic tensors is one-dimensional and are built from the Kronecker $\delta_{ij}$ and the Levi-Civita symbols $\epsilon_{ijk}$ giving
\begin{equation}
I^{(2)} = \frac{1}{3} \delta_{i_1i_2}\delta_{\lambda_1 \lambda_2}
\end{equation}
\begin{equation}
I^{(3)} = \frac{1}{6} \epsilon_{i_1i_2i_3}\epsilon{\lambda_1 \lambda_2 \lambda_3}
\end{equation}
For higher rank tensors, there are more linearly independent isotropic tensors. $I^{(4)}$ and $I^{(5)}$ are given by
\begin{equation}
I^{(4)} = \frac{1}{30}
\left( \begin{array}{c}
\delta_{i_1i_2}\delta_{i_1i_2}   \\
\delta_{i_1i_3}\delta_{i_2i_4}  \\
\delta_{i_1i_4}\delta_{i_2i_3}  \end{array} \right)^T
\left( \begin{array}{ccc}
4  & -1 & -1 \\
-1 & 4  & -1 \\
-1 & -1 &  4 \end{array} \right)
\left( \begin{array}{c}
\delta_{\lambda_1\lambda_2}\delta_{\lambda_1\lambda_2}   \\
\delta_{\lambda_1\lambda_3}\delta_{\lambda_2\lambda_4}  \\
\delta_{\lambda_1\lambda_4}\delta_{\lambda_2\lambda_3}  \end{array} \right)
\label{tensorI4}
\end{equation}
\begin{equation}
I^{(5)} = \frac{1}{30}
\left( \begin{array}{c}
\epsilon_{i_1i_2i_3}\delta_{i_4i_5}   \\
\epsilon_{i_1i_2i_4}\delta_{i_3i_5}  \\
\epsilon_{i_1i_2i_5}\delta_{i_3i_4}  \\
\epsilon_{i_1i_3i_4}\delta_{i_2i_5}  \\
\epsilon_{i_1i_3i_5}\delta_{i_2i_4}  \\
\epsilon_{i_1i_4i_5}\delta_{i_2i_3}  \end{array} \right)^T
\left( \begin{array}{cccccc}
3   & -1 & -1 &  1 & 1  & 0\\
-1  &  3 & -1 & -1 & 0  & 1\\
-1  & -1 &  3 &  0 & -1 & -1\\
1   & -1 &  0 &  3 & -1 & 1\\
1   &  0 & -1 & -1 & 3  & -1\\
0   &  1 & -1 &  1 & -1 & 3 \end{array} \right)
\left( \begin{array}{c}
\epsilon_{\lambda_1\lambda_2\lambda_3}\delta_{\lambda_4\lambda_5}   \\
\epsilon_{\lambda_1\lambda_2\lambda_4}\delta_{\lambda_3\lambda_5}  \\
\epsilon_{\lambda_1\lambda_2\lambda_5}\delta_{\lambda_3\lambda_4}  \\
\epsilon_{\lambda_1\lambda_3\lambda_4}\delta_{\lambda_2\lambda_5}  \\
\epsilon_{\lambda_1\lambda_3\lambda_5}\delta_{\lambda_2\lambda_4}  \\
\epsilon_{\lambda_1\lambda_4\lambda_5}\delta_{\lambda_2\lambda_3}  \end{array}\right)
\label{tensorI5}
\end{equation}
\noindent where $\delta$ and $\epsilon$ are the Kronecker and Levi-Civita symbols respectively.

For example, the 4th-rank tensor involving one magnetic interactions and three electric ones is averaged as 
\begin{equation}
(\langle \bm \mu\bm \mu\bm \mu\bm m\rangle_\Omega)_{i_1 i_2 i_3 i_4} = (I^{(4)})^{\lambda_1\lambda_2\lambda_3\lambda_4}_{i_1i_2i_3i_4}\langle \bm \mu\bm \mu\bm \mu\bm m\rangle_{\lambda_1\lambda_2\lambda_3\lambda_4}
\end{equation}
\noindent where we used Einstein summation convention for the Cartesian in indices $i_1, i_2, i_3, i_4$, $\langle ...\rangle_\Omega$ stands for rotational averaging.

\section{Spectroscopic signals}
\label{appendixOBS}
We now review the first principles calculation of spectroscopic observables. 
A photon detection event can always be viewed as the change of photon number in the photon modes that the detector is sensitive to \cite{roslyak2009unified}. 
Since camera shutters are typically slow compared to ultrafast molecular dynamics, signals are given by the time-integrated photon number change:
\begin{equation}
S(\Gamma) = \int dt \ \langle \frac{d}{dt} N_s \rangle
\label{ALLSIG}
\end{equation}
\noindent where $\Gamma$ represents all the experimental control knobs that can be varied in a specific experiment. 
The time integration can run to infinity since we assume to detect all photon during the ultrafast process.
We have assumed an ideal detector sensitive to a single photon mode $s$.
For real detectors, one has to convolute this response to a single mode with the instrument response function in $k$ space (finite detector size), frequency domain (bandwidth of detection of the detector pixel), in polarization (polarization resolve detection or not)\cite{bennett2014time}.
In many cases, this gating procedure can be reduced to a few variables depending on the experimental setup.

We next show how the heterodyne, the homodyne coherent and the homodyne incoherent signals can be obtained from Eq. \ref{ALLSIG} for the three Hamiltonians given in Eq. \ref{hintEdip} to \ref{hintMULTI}.
We start with the simplest dipolar Hamiltonian.
The expectation value of the photon number change can be calculated using the Heisenberg equation:
\begin{eqnarray}
\langle \frac{d}{dt} N_s \rangle &=& \frac{i}{\hbar} \langle[H_\text{int}(t),a_s^\dagger a_s]\rangle\nonumber\\
&=& \frac{i}{\hbar}\int d \bold r \langle[-\bm \mu(\bold r)\cdot \bold E_s^\dagger(\bold r,t)-\bm \mu^\dagger(\bold r) \cdot \bm E_s(\bold r,t),a_s^\dagger a_s]\rangle\nonumber\\
&=& -\frac{2}{\hbar} \text{Im} \int d\bm r \langle \bm \mu(\bm r) \cdot \bm E_s^\dagger(\bm r,t) \rangle
\end{eqnarray}

The photon signal thus becomes:
\begin{equation}
S(\Gamma) = -\frac{2}{\hbar} \text{Im} \int dt d\bm r \langle \bm \mu(\bm r) \cdot \bm E_s^\dagger(\bm r,t) \rangle
\label{ALLSIG2}
\end{equation}

Various detection modes can be described starting with this expression.
If a classical field is present in the $s$ mode, the expectation value over the quantum operator $\bm E_s^\dagger$ can be expressed simply as a function of the classical field: $\langle\bm E_s^\dagger\rangle = \bm E_\text{het}^*(\bm r,t)$. 
In this heterodyne detection mode, the photon emitted from the process with the matter interfere with an external local oscillator and gives the heterodyne signal:
\begin{equation}
S_\text{het}(\Gamma) = -\frac{2}{\hbar} \text{Im} \int dt d\bm r \bm E_\text{het}^*(\bm r,t)\cdot \langle \bm \mu(\bm r,t)  \rangle
\end{equation}
\noindent where the time variable on $\langle \bm \mu(\bm r,t)  \rangle$ arises by taking the expectation value at time $t$.

Homodyne detection is achieved by detecting the spontaneously emitted photons into initially unoccupied modes.
In this case, the Fock state of the electromagnetic field is the vacuum before the experiment and the interaction populates the $s$ mode that can be detected.
It is then clear that the signal in Eq. \ref{ALLSIG2} vanishes at this order and one must perform a perturbative expansion to first order in the spontaneous mode:
\begin{equation}
S_\text{hom}(\Gamma) = \frac{2}{\hbar} \text{Im} \frac{i}{\hbar} \int dt d\bm r dt' d\bm r' \langle \bm \mu_\text{left}(\bm r,t) \cdot \bm E_{s,\text{left}}^\dagger(\bm r,t)
\bm \mu_\text{right}^\dagger(\bm r',t') \cdot \bm E_{s,\text{right}}(\bm r',t') \rangle
\label{homo1}
\end{equation}
\noindent where $\text{left}$ and $\text{right}$ stand for left and right interaction on the density matrix\cite{mukamel1999principles}. They are used to ensure that a photon population in the mode $s$ is created.

The electric dipole operator is localized on each molecule. Hence, the spatial dependence of $\bm \mu(\bm r)$ for equivalent molecules is 
\begin{equation}
\bm \mu (\bm r) = \sum_\alpha \bm \mu \delta(\bm r - \bm r_\alpha)
\end{equation}
\noindent where $\alpha$ denotes molecules and $\bm r_\alpha$ is the position of molecule $\alpha$.
The correlation function in Eq. \ref{homo1} contains terms involving a single molecule (homodyne incoherent signals) or two molecules (homodyne coherent signals).
\begin{eqnarray}
S_\text{hom,inc}(\Gamma) &=& \frac{2}{\hbar^2} \mathcal E_s^2 N \text{Re} \int dt dt' e^{i\omega_s(t-t')} \langle \bm \mu_L(t) \cdot \bm \mu_R^\dagger(t')\rangle \label{homoINC}\\
S_\text{hom,co}(\Gamma) &=& \frac{2}{\hbar^2} \mathcal E_s^2 \text{Re} F_2(\Delta \bold k) |\int dt e^{i\omega_s t} \langle \bm \mu_L(t)\rangle|^2 \label{homoCO}
\end{eqnarray}
\noindent where $\mathcal E_s^2 = \hbar \omega_s / 2 \epsilon_0$ and $N$ is the number of scatterer in the process that can be directly related to the molecular concentration. 
To obtain Eq. \ref{homoINC} and \ref{homoCO}, we have summed over all polarizations of the spontaneous field.
The $F_2(\Delta \bm k)$ function is the structure factor and will be discussed in the next section.

We now provide the signal expressions for these three detection modes for the other two interaction Hamiltonian.
For the multipolar Hamiltonian, we have:
\begin{equation}
S_\text{het}(\Gamma) = -\frac{2}{\hbar}N \text{Im} \int dt \bm E_\text{het}^*(t)\cdot \langle \bm \mu(t) \rangle + \bm B_\text{het}^*(t)\cdot \langle \bm m(t)\rangle + \nabla\bm E_\text{het}^*(t)\cdot \langle \bm q(t)\rangle \label{hetmulti}
\end{equation}
\begin{multline}
S_\text{hom,inc}(\Gamma) = \frac{2}{\hbar^2} \mathcal E_s^2 N \text{Re} \int dt dt' e^{i\omega_s(t-t')}\Big( \langle \bm \mu_L(t) \cdot \bm \mu_R^\dagger(t')\rangle 
+ \frac{1}{c}\langle \bm m_L(t) \cdot \bm \mu_R^\dagger(t')\rangle\\
+ \frac{1}{c}\langle \bm \mu_L(t) \cdot \bm m_R^\dagger(t')\rangle
-\frac{i}{c}\langle (\bm{\hat k_s}\cdot\bm q_L(t)) \cdot \bm \mu_R^\dagger(t')\rangle 
 + \frac{i}{c}\langle \bm \mu_L(t) \cdot (\bm{\hat k_s} \cdot \bm q_R^\dagger(t'))\rangle\Big)\label{homoINCmulti}
\end{multline}
\begin{equation}
S_\text{hom,co}(\Gamma) = \frac{2}{\hbar^2} \mathcal E_s^2 (\text{Re} F_2(\Delta \bold k)) |\int dt e^{i\omega_s t} 
\langle \bm \mu_L(t)\rangle 
+ \frac{1}{c}\langle \bm m_L(t)\rangle
+ \frac{1}{c}\langle -i\bold{\hat k_s}\cdot\bm q_L(t)\rangle|^2 \label{homoCOmulti}
\end{equation}

Finally, we give the signal computed by the minimal coupling Hamiltonian. 
In this coupling, the final interaction can be either with the charge density or with the current density. 
The latter leads to signal expression similar to the electric dipole coupling but takes into account the full spatial extension of the incoming beam and thus avoids using a high order multipolar expansion. 
The former contributes mostly off-resonance and leads to all the diffraction-like observable.
\begin{equation}
S_\text{het}^{\bold j}(\Gamma) = -\frac{2}{\hbar}N \text{Im} \int dt d\bold r \bm A_\text{het}^*(\bold r, t)\cdot \langle \bm j(r, t) \rangle \label{hetmcj}
\end{equation}
\begin{equation}
S_\text{het}^{\sigma}(\Gamma) = -\frac{2}{\hbar}N \text{Im} \int dt d\bold r \bm A_\text{het}^*(\bold r, t)\cdot\bm A_\text{in}(\bold r, t) \langle \bm \sigma(r, t) \rangle 
\label{hetmcs}
\end{equation}
\noindent where $\bm A_\text{in}(\bold r, t)$ is the diffracted incoming beam.
It is worth noting that heterodyne-detected diffraction has not yet been implemented and that this detection scheme would solve the phase-problem occurring in structure reconstruction\cite{shen1998solving}.
The one-molecule signals are given by:
\begin{equation}
S_\text{hom,inc}^{\bold j}(\Gamma) = \frac{2}{\hbar^2} \mathcal A_s^2 N \text{Re} \int dt dt' d\bold r d\bold r' e^{i\omega_s(t-t')} \langle \bm j_L(\bold r, t) \cdot \bm j_R^\dagger(\bold r', t')\rangle\label{homoINCmcj}
\end{equation}
\begin{equation}
S_\text{hom,inc}^{\sigma}(\Gamma) = \frac{2}{\hbar^2} \mathcal A_s^2 N \text{Re} \int dt dt' d\bold r d\bold r' e^{i\omega_s(t-t')} 
\bm e_s^*\cdot \bm A_\text{in}(\bold r, t) 
\bm e_s\cdot \bm A_\text{in}^*(\bold r', t')
\langle \sigma_L(\bold r, t) \sigma_R^\dagger(\bold r', t')\rangle\label{homoINCmcs}
\end{equation}

Finally, the two-molecule homodyne detected signals are
\begin{equation}
S_\text{hom,co}^{\bold j}(\Gamma) = \frac{2}{\hbar^2} \mathcal A_s^2 \text{Re} F_2(\Delta \bold k) |\int dt d\bold r e^{i\omega_s t} \ \langle \bm j_L(\bold r, t)\rangle|^2\label{homoCOmcj}
\end{equation}
\begin{equation}
S_\text{hom,co}^{\sigma}(\Gamma) = \frac{2}{\hbar^2} (\frac{e}{2mc})^2\mathcal A_s^2 \text{Re} F_2(\Delta \bold k) |\bm \epsilon_s \cdot \bm \epsilon_X|^2 |\int dt d\bold r e^{i\omega_s t}\langle \bm \sigma_L(\bold r, t)\rangle|^2 \label{homoCOmc}
\end{equation}

\section{Perturbative expansion of spectroscopic signals}
\label{appendix:perturbation}
In appendix B, we have reviewed the observables relevant for the calculation of most spectroscopic signals.
The signals usually require the calculation of the expectation value of an operator $X$ defined as
\begin{equation}
\langle X(t) \rangle = \langle \psi(t) | X | \psi(t) \rangle = \text{Tr}(X \rho(t)) = \langle \langle X | \rho(t)\rangle \rangle
\end{equation}
The first equality is the standard expectation value calculated from the wavefunction in Hilbert space. For a given order $n$ of perturbation, one has to expand the bra at order $m$, the ket at order $n-m$ and them sum over $m$.
The second equality provide the expectation given as a function of the density matrix. The density matrix formulation allows a proper treatment of relaxation and avoid summing over the bra and the ket perturbative order. This comes at the expense of nested commutators from the Liouville-von Neumann equation that dictates the time evolution of the density matrix.
Finally, the last term gives the expectation value in Liouville space, i. e. the Hilbert space spawned by the density matrix seen as a vector.

The time evolution of the wavefunction is given by the following propagators:
\begin{eqnarray}
|\psi(t)\rangle &=& e^{-\frac{i}{\hbar}\int_{t_0}^t d\tau H(\tau)} |\psi(t_0)\rangle\\
\rho(t) &=& e^{-\frac{i}{\hbar}\int_{t_0}^t d\tau [H(\tau),\bullet]} \rho(t_0)\\
|\rho(t)\rangle\rangle &=& e^{-\frac{i}{\hbar}\int_{t_0}^t d\tau \mathcal L(\tau)} |\rho(t_0)\rangle\rangle
\end{eqnarray}

The exponential operators are a symbolic notation defined by their Dyson series:
\begin{eqnarray}
e^{-\frac{i}{\hbar}\int_{t_0}^t d\tau H(\tau)}|\psi(t_0)\rangle&=&
1 + \sum_{n=1}^{\infty} (\frac{-i}{\hbar})^n \int_{t_0}^t d\tau_n \int_{t_0}^{\tau_n} d\tau_{n-1} ... \int_{t_0}^{\tau_2} d\tau_{1} H(\tau_n) H(\tau_{n-1}...H(\tau_1)|\psi(t_0)\rangle \nonumber\\
e^{-\frac{i}{\hbar}\int_{t_0}^t d\tau [H(\tau),\bullet]} \rho(t_0) &=& 
1 + \sum_{n=1}^{\infty} (\frac{-i}{\hbar})^n 
\int_{t_0}^t d\tau_n \int_{t_0}^{\tau_n} d\tau_{n-1} ... \int_{t_0}^{\tau_2} d\tau_{1} [H(\tau_n),[H(\tau_{n-1},...[H(\tau_1),\rho(t_0)]...]]\nonumber\\
e^{-\frac{i}{\hbar}\int_{t_0}^t d\tau \mathcal L(\tau)} |\rho(t_0)\rangle\rangle &=& 
1 + \sum_{n=1}^{\infty} (\frac{-i}{\hbar})^n 
\int_{t_0}^t d\tau_n\int_{t_0}^{\tau_n} d\tau_{n-1} ... \int_{t_0}^{\tau_2} d\tau_{1} 
\mathcal L(\tau_n) \mathcal L(\tau_{n-1}... \mathcal L(\tau_1)|\rho(t_0)\rangle\rangle
\end{eqnarray}

To $n$-th order, a given perturbative expansion will be the sum over $2^n$ terms. Moreover, the electric field contributes with either positive or negative frequencies at each interaction. It is thus convenient to introduce a diagrammatic representation for each term contributing to the signal.
For a given signal, diagrams can be written in the wavefunction formalism or in the density matrix formalism. The former are represented by loop diagrams, taking their origins from the Keldysh loop formalism, and the latter are given by ladder diagrams.
Ladder diagrams have the advantage to be fully time-ordered and to naturally be able to include relaxation by tracing out the bath degrees of freedom in the total density matrix.
On the other hand, loop diagrams are more suitable for numerical propagation in between perturbative interaction since they rely on an Hilbert space picture.
For example, such numerical propagation are better suited for molecules experiencing nuclear dynamics or evolving in a non-perturbative strong field.

We now summarize the rules to translate a diagram into a sum over state expressions.

\begin{figure}[!h]
  \centering
  \includegraphics[width=0.5\textwidth]{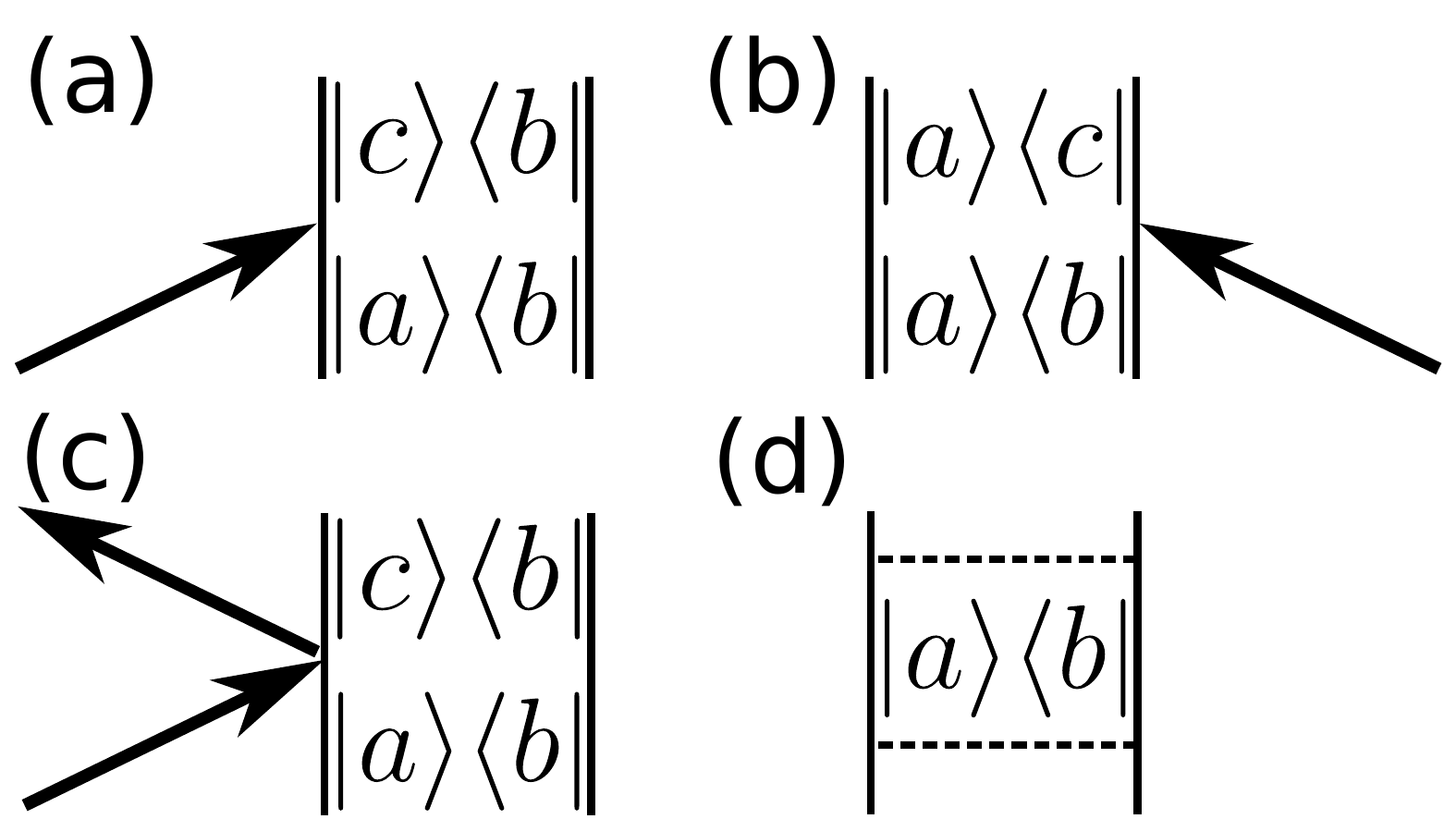}
  \caption{Diagrams.
\label{Fig:appendixC1}}
\end{figure}

\textbf{Ladder diagram in time domain:}
\begin{enumerate}
\item Time goes from bottom to top, the two straight lines represent the ket and the bra of the molecular density matrix.
\item Fields are split in two contributions $\bm F(r,t) = \int d\omega d\bm k \ \bm F(\bm k, \omega) e^{i(\bm k\cdot \bm r - \omega t)}$ and $\bm F^*(r,t) = \int d\omega d\bm k \ \bm F(\bm k, \omega)^* e^{-i(\bm k\cdot \bm r - \omega t)}$. Arrows pointing to the right are interaction with $\bm F(r,t)$ and arrows pointing to the left are interactions with $\bm F(r,t)^*$. In the dipolar coupling, $\bm F = \bm E$, in the multipolar coupling $\bm F = \bm E, \bm B$ or $\nabla\bm E$, in the minimal coupling $\bm F = \bm A$.
\item Each interaction vertex gives an interaction with a transition matrix element: electric dipole $\bm\mu_{ij}$, magnetic dipole $\bm m_{ij}$, electric quadrupole $\bm q_{ij}$, current density $\bm j_{ij}$ or current density $\sigma_{ij}$.
\item Each interaction on the left introduces a factor $i/\hbar$ and a factor $-i/\hbar$ on the right.
\item Intervals between interaction introduces a Liouville-space propagator $\mathcal G(t-t_0) = \theta(t-t_0) e^{-\frac{i}{\hbar}\int_{t_0}^t d\tau \mathcal L_0(\tau)} $ where $\mathcal L_0(\tau)$ is the non-interacting Liouvillian. Matrix elements of $\mathcal G(t-t_0)$ (shown in Fig. \ref{Fig:appendixC1}(d)) are $\mathcal G_{ab}(t-t_0) = \theta(t-t_0) e^{-i\omega_{ab}(t-t_0)-\Gamma_{ab}(t-t_0)}$ where $\Gamma_{ab}$ describe the relaxation of $|a\rangle \langle b|$.
\end{enumerate}

\textbf{Ladder diagram in frequency domain:}
\begin{enumerate}
\item Similar rules to the time-domain case applied except for the following ones.
\item Pointing left arrows are $\bm F^*(\bm k, \omega)$ and pointing right arrows are $\bm F(\bm k, \omega)$.
\item Propagators are $\mathcal G(\omega) = 1/(\omega - \mathcal L_0/\hbar +i \Gamma_{ab})$ and $\mathcal G_{ab}(\omega) = 1/(\omega - \omega_{ab} +i \Gamma_{ab})$.
\item Frequency variable are cumulative along the diagram.
\end{enumerate}

For example, in Fig.\ref{Fig:appendixC1} and in the dipolar coupling, diagram (a) gives a contribution $(i/\hbar)\bm \mu_{ca}\cdot \bm E(r,t)$ and diagram (b) gives $(-i/\hbar)\bm \mu_{cb}\cdot \bm E^*(r,t)$. In the minimal coupling, the $\sigma\bm A^2$ interaction interacts twice with the field at a given vertex and diagram (c) gives $(i/\hbar)\sigma_{ca}(\bm r,t) \bm A_{i}(\bm r,t)\cdot \bm A_{d}^*(\bm r,t)$. 

Finally, we introduce loop diagrams rules based on a wavefunction approach. More details can be found in the literature\cite{mukamel2010ultrafast}.

\textbf{Loop diagram in time domain:}
\begin{enumerate}
\item The left branch represents the evolution of the ket and the right branch the one on the bra. Unlike ladder diagrams, the interactions are not time-order on different branches but are time-ordered within a branch.
\item Arrows pointing to the right are interaction with $\bm F(r,t)$ and arrows pointing to the left are interactions with $\bm F^*(r,t)$.
\item Rules 2, 3, and 4 of time-domain ladder diagrams are kept.
\item Intervals between interactions introduces an Hilbert space propagator 
$G(s) = \theta(s) e^{-\frac{i}{\hbar}\int_{t_s}^{t_s'} d\tau \mathcal H_0(\tau)} $ where $H_0(\tau)$ is the non-interacting Hamiltonian and $s = t_s - t_s'$ is the propagation time interval between $t_s$ and $t_s'$. 
Matrix elements of $G(s)$ over field-free molecular eigenstates are $G(s) = \theta(t-t_0) e^{-i E_{a}s/\hbar}$. Alternatively, the propagator $G(s)$ can be treated numerically for the propagation.
\item The last interaction at the observation time $t$ is conventionally chosen to be occurring from the left.
\end{enumerate}

\textbf{Loop diagram in frequency domain:}
\begin{enumerate}
\item Rules 1, 2, 3 and 5 of time-domain loop diagrams are kept.
\item Propagation on the left branch provides $iG(\omega)$ propagator. The right branch provides a $-iG^\dagger(\omega)$. The field-free propagator is given by $G(\omega) = (\omega - H_0/\hbar + i \epsilon)^{-1}$.
\item  The frequency arguments of the propagators in a branch are
cumulative. The ground state frequency $E_g/\hbar$ is added to all propagators' arguments.
\end{enumerate}

\end{appendices}

\renewcommand{\abstractname}{Acknowledgements}
\begin{acknowledgement}
The work covered in this review was primarily supported by the Chemical Sciences, Geosciences, and Biosciences division, Office of Basic Energy Sciences, Office of Science, U.S. Deparment of Energy through Award No.DE-FG02-04ER15571 (J.R.R) and Award No DE-SC0019484 (S.M) and by the National Science Foundation (Grant No. CHE-1953045). J.R.R. was supported by the LABEX MANUTECH-SISE (ANR-10-LABX-0075) of the Université de Lyon, within the program 'Investissements d’Avenir' (ANR-11-IDEX-0007) operated by the French National Research Agency (ANR).
\end{acknowledgement}

\section*{Biographies}

Jérémy Rouxel is an associate professor at the Laboratoire Hubert Curien, Saint-Etienne. He received his Ph.D. from Troyes Technological University and from Nanyang Technological University. He then joined University of California, Irvine and the Ecole Polytechnique Fédérale de Lausanne as a postdoctoral fellow. His research interests include nonlinear spectroscopies, ultrafast molecular dynamics, molecular chirality in the X-ray regime and the development of novel X-ray ultrafast techniques.

Shaul Mukamel, currently Distinguished Professor of Chemistry and of Physics \& Astronomy at the University of California, Irvine, received his Ph.D. in 1976 from Tel Aviv University. 
Following postdoctoral appointments at MIT and the University of California, Berkeley, he has held faculty positions at Rice University, the Weizmann Institute, and the University of Rochester, before joining UCI at 2003. 
His research interests broadly span the  areas of ultrafast multidimensional spectroscopy of molecules from the infrared to the X-ray regimes, excitons in chromophore aggregates  and photosynthetic  complexes, nonadiabatic conical intersection dynamics,polaritons in cavities, and molecular electrodynamics.

\bibliographystyle{plain}

\bibliography{biblio_all}

\end{document}